\documentclass[useAMS,usenatbib]{mn2e}
\usepackage{graphicx}
\usepackage{amsmath}
\usepackage{placeins}

\newcommand{\ha}{H$\alpha$} 
\newcommand{\hbeta}{H$\beta$}
\newcommand{\hc}{H$\gamma$}
\newcommand{\hdelta}{H$\delta$}

\newcommand{\neon}{[Ne~{\sc iii}]}
\newcommand{\sulfurvi}{[S~{\sc vi}]}
\newcommand{\neoniv}{[Ne~{\sc iv}]}
\newcommand{\neonv}{[Ne~{\sc v}]}

\newcommand{\argon}{[Ar~{\sc iv}]}

\newcommand{\argonV}{[Ar~{\sc v}]}

\newcommand{\argoniii}{[Ar~{\sc iii}]}
\newcommand{\helium}{He~{\sc i}}
\newcommand{\heliumb}{He~{\sc ii}}

\newcommand{\oiii}{[O~{\sc iii}]~5007~\AA}

\newcommand{\nitrogen}{[N~{\sc ii}]}

\newcommand{\nitrogena}{[N~{\sc i}]}
\newcommand{\oxygeniii}{[O~{\sc iii}]}
\newcommand{\oxygeniiib}{O~{\sc iii}}
\newcommand{\oxygeni}{[O~{\sc i}]}
\newcommand{\oxygenii}{[O~{\sc ii}]}

\newcommand{\sulfur}{[S~{\sc iii}]}
\newcommand{\sulfurt}{[S~{\sc ii}]}
\newcommand{\chloro}{[Cl~{\sc iii}]}

\newcommand{\ironvi}{[Fe~{\sc vi}]}

\newcommand{\ironii}{[Fe~{\sc ii}]}
\newcommand{\ironi}{[Fe~{\sc i}]}
\newcommand{\kripto}{[K~{\sc iv}]}
\newcommand{\carboniv}{C~{\sc iv}}
\newcommand{\carboniia}{C~{\sc ii}}
\newcommand{\carboniib}{C~{\sc ii}}
\newcommand{\carboniii}{[C~{\sc iii}]}
\newcommand{\silicon}{[Si~{\sc ii}]}

\newcommand{\degree}{$^{\circ}$}
\def\vhel{\ifmmode{V_{{\rm HEL}}}\else{$V_{{\rm HEL}}$}\fi}
\def\vsys{\ifmmode{V_{\rm sys}}\else{$V_{\rm sys}$}\fi}
\def\kms{\ifmmode{~{\rm km\,s}^{-1}}\else{~km~s$^{-1}$}\fi}
\def\vlsr{\ifmmode{v_{\rm lsr}}\else{$v_{\rm lsr}$}\fi}

\title[Evidence for a Wolf--Rayet or WEL--type binary nucleus in the bipolar Planetary Nebula Vy~1--2]
{Evidence for a [WR] or WEL--type binary nucleus in the bipolar planetary nebula Vy~1--2}

\author[Evidence for a WR or WEL type binary nucleus in Vy~1--2] 
{S. Akras$^{1,2,3}$ $\thanks{e-mail:akras@astro.ufrj.br}$, 
P. Boumis$^{3}$, J. Meaburn$^{4}$, J. Alikakos$^{3}$, J. A. L\'opez$^{2}$, D. R. Gon\c calves$^{1}$\\
$^{1}$Observat\'orio do Valongo, Universidade Federal do Rio de Janeiro, Ladeira Pedro Antonio 43, 20080-090, Rio de Janeiro, Brazil\\
$^{2}$Instituto de Astronomia, Universidad National Autonoma de Mexico, Ensenada 22800, Baja California, Mexico.\\
$^{3}$IAASARS, National Observatory of Athens, I. Metaxa \& V. Pavlou, Penteli, GR--15236, Athens, Greece.\\
$^{4}$Jodrell Bank Centre for Astrophysics, School of Physics and Astronomy, University of Manchester, M13 9PL, UK\\}

\begin{document} 

\date{Received **insert**; Accepted **insert**}

\pagerange{\pageref{firstpage}--\pageref{lastpage}} \pubyear{2013}

\maketitle
\label{firstpage}

\begin{abstract} 

We present high--dispersion spectroscopic data of the compact planetary nebula Vy~1--2, where high 
expansion velocities up to 100~\kms\ are found in the \ha, \nitrogen\ and \oxygeniii\ emission lines. 
HST images reveal a bipolar structure. Vy~1--2 displays a bright ring--like structure with a size of 
2.4\arcsec$\times$3.2\arcsec\ and two faint bipolar lobes in the west--east direction. A faint pair of 
knots is also found, located almost symmetrically on opposite sides of the nebula at PA=305\degree. 
Furthermore, deep low--dispersion spectra are also presented and several emission lines are detected for the first time 
in this nebula, such as the doublet \chloro\ 5517, 5537~\AA, \kripto\ 6101~\AA, \carboniib\ 6461~\AA, the doublet 
\carboniv\ 5801, 5812~\AA. By comparison with the solar abundances, we find enhanced N, depleted C and solar O.  
The central star must have experienced the hot bottom burning (CN--cycle) during the $\rm{2^{nd}}$ 
dredge--up phase, implying a progenitor star of M$\geq$3~M$_{\odot}$. The very low C/O and N/O abundance ratios 
suggest a likely post--common envelope close binary system. 

A simple spherically symmetric geometry with either a blackbody or a H--deficient stellar atmosphere model is not able 
to reproduce the ionisation structure of Vy 1--2. The effective temperature and luminosity of its central star 
indicate a young nebula located at a distance of $\sim$9.7~kpc with an age of $\sim$3500~years. 
The detection of stellar emission lines, \carboniib\ 6461~\AA, the doublet \carboniv $\lambda\lambda$ 5801, 5812 
and \oxygeniiib\ 5592~\AA, emitted from a H--deficient star, indicates the presence of a late--type Wolf--Rayet or 
a WEL type central star.

\end{abstract}


\section{Introduction}

\begin{figure*}
\centering
\includegraphics[scale=0.80]{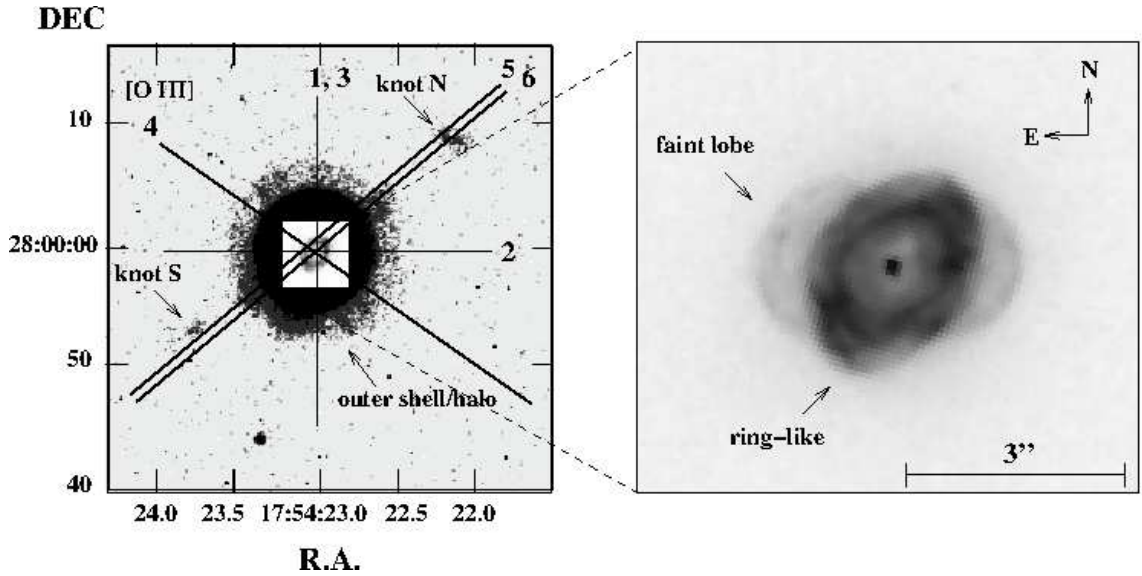}
\caption[]{Slit positions of the low--(1) and high--(2, 3, 4 and 5) dispersion spectra superimposed 
upon the high--resolution HST WFPC3 narrow--band \oxygeniii\ image (ID: 11657; PI: Stanghellini). 
Left panel reveals the presence of a pair of knots at P.A.=305$^{\circ}$. 
Right panel is in logarithmic scale and illustrates the ellipsoidal structure of the nebula with extended 
filamentary structures and the presence of two fainter lobes at P.A. $\sim$90$^{\circ}$.}
\label{fig1}
\end{figure*}

Planetary nebulae (PNe) are the consequent result of the interaction of stellar winds ejected from 
low--to--intermediate mass stars (1-8~M$_{\odot}$). A slow and dense wind ($10^{-4}$~M$_{\odot}$), 
emerges from an Asymptotic Giant Branch (AGB) star, is swept up into a dense spherically symmetric 
shell by a second, faster and tenuous wind expelled during the post-AGB phase (interacting stellar 
wind model; Kwok et al. 1978). This model can reasonably explain the formation and evolution of spherical PNe. However, 
most post--AGB stars and PNe show highly asymmetric circumstellar envelopes and complex morphologies 
(Manchado et al. 2006, Meixner et al. 1999, Parker et al. 2006, Boumis et al. 2003, 2006, Miszalski et al. 2008, 
Lagadec et al. 2011, Sahai 2011, Sabin et al. 2014). 

The deviation of PNe from spherically symmetric structures has its origin in an enhanced 
equatorial--to--polar density gradient in the AGB circumstellar envelopes resulting in the formation 
of aspherical and more complex nebulae. This equatorial density enhancement is interpreted as the 
result of common envelope evolution of close binary systems (Soker \& Livio 1994, Nordhaus \& Blackman 2006, De Marco 2009). 
Although, most of the post--AGB and PNe show complex morphologies, only a small number of them are known 
to contain a binary system (Miszalski et al. 2009, De Marco 2009). Recent work by Jones et al. (2014) 
confirmed the binary nature of Hen~2--11's nucleus, providing further evidence of a positive link 
between binary nuclei and aspherical PNe.

The compact planetary nebula Vy~1--2 (PNG 053.3+24.0), discovered by Vyssotsky (1942), had first been described as an 
elliptical PN ($\lq\lq$The IAC Morphological Catalogue of Northern Galactic Planetary Nebulae\rq\rq; Manchado et al. 1996), 
but recent high angular resolution HST images revealed a more complex structure (Sahai et al. 2011, Shaw  et al. in preparation). 
According to Winberger's catalogue of expansion velocities of Galactic PNe, Vy~1--2 shows 
$\rm{V_{[O {\sc\ III}]}}$ $>$ $\rm{V_{[N {\sc\ II}]}}$ (Weinberger 1989). This result is inconsistent with 
a simple homologous expansion law. A similar velocity trend has 
also been found for few other PNe, (e.g. BD +30\degree 3639, Akras \& Steffen 2012, Bryce \& Mellema 1999; 
see also Medina et al. 2006 for more objects). A detailed study of the velocity field as a function of distance 
from the central stars in PNe has revealed a more complex expansion law than the simple homologous--law, probably associated 
with the fast wind from the central stars (Gesicki \& Zijlstra 2003). Interestingly, 
the central stars of this rare group of PNe have been identified as Wolf--Rayet type ([WR]) or weak emission line 
stars (WELS) (Medina et al. 2006). Unfortunately, the spectral type of the central star of Vy~1--2 is still unknown. 
Barker (1978a,b), and more recently Wesson et al. (2005), as well as Stanghellini et al. 
(2006) studied the ionisation structure and chemical composition of this nebula. A comparison between these 
studies shows a significant discrepancy in the ionic and total chemical abundances (see Tables 
5 and 6). This may be related to the different values of T$_e$ and N$_e$ used by these authors. In particular, 
Wesson et al. (2005) estimated N$_e$ and T$_e$ using diagnostics line ratios from low to high excitation lines (see Table 4). 
The high difference in N$_e$ between \sulfurt\ and \oxygenii\ cannot be explained since they have similar 
ionisation potentials. Barker (1978a) gave an average value of 5000~cm$^{-3}$ using different diagnostics from 
low excitation lines (\sulfurt\ and \oxygenii). Therefore, the low $\rm{N_e}$ in \sulfurt\ of 1160~cm$^{-3}$ derived by Wesson 
et al. (2005), may not be correct.

\begin{table*}
\centering
\caption[]{Observation log}
\label{table5}
\begin{tabular}{llllllllllll}
\hline 
Slit & Dispersion & Date & Filter & wavelength (\AA ) & FWHM (\AA ) & P.A. ($^{\circ}$) & Exp. times (sec)\\   
\hline
& & & & 1.3m Skinakas Telescope & \\
1     &  low   &  2011 Sep 04 & $-$ &  4700--6800 & $-$ & 0   & 3600 and 300$\times$3 \\
\hline
& & & & 2.1m SPM Telescope $+$ MES &\\    

2     &  high  &  2011 May 15 & \ha +\nitrogen\ & 6550 & 90 & 90 & 1800\\
2     &  high  &  2011 May 15 & \oxygeniii\     & 5007 & 60 & 90 & 1800\\
3     &  high  &  2012 Mar 30 & \ha +\nitrogen\ & 6550 & 90 & 0  & 1200 \\   
4     &  high  &  2012 Apr 01 & \ha +\nitrogen\ & 6550 & 90 & 55 & 1200 \\   
5     &  high  &  2012 Apr 01 & \ha +\nitrogen\ & 6550 & 90 & 310 & 1800 \\   
5     &  high  &  2012 Apr 01 & \oxygeniii\     & 5007 & 60 & 310 & 1800 \\   
6$^a$ &  high  &  2002 Jul 23 & \ha +\nitrogen\ & 6550 & 90 & 310 & 1800 \\   
\hline
& & & & HST+WFPC3 &\\ 
$-$   & imaging & 2010 Feb 08 & F502N   & 5010.0 &   65.0  & $-$ &  600\\
$-$   & imaging & 2010 Feb 08 & F200LP  & 4883.0 & 5702.2  & $-$ &   11\\
$-$   & imaging & 2010 Feb 08 & F350LP  & 5846.0 & 4758.0  & $-$ &   11\\
$-$   & imaging & 2010 Feb 08 & F814W   & 8024.0 & 1536.0  & $-$ &   23\\
\hline
\end{tabular}
\medskip{}
\begin{flushleft}
$^a$ It was drawn from the SPM Kinematic Catalogue of Planetary Nebulae
\end{flushleft}
\end{table*}

The logarithmic \hbeta\ flux (erg s$^{-1}$ cm$^{-2}$) was estimated to be -11.52$\pm$0.01 (Barker 1978a; 
hereafter Bar78a) and -11.53 (Wesson et al. 2005; hereafter WLB05). Vy 1--2 is also a radio source with 1.4~GHz flux 
equal to $12.4\pm0.9$~mJy and extinction coefficient c(radio)=0.1, which is slightly lower than the optical value 
from WLB05 (c(\hbeta)=0.139). The coordinates of Vy~1--2 are given at RA $\rm{17^h 54^m 23^s}$ and Dec. $\rm{27\degr\ 59\arcmin\ 58\arcsec}$ 
for the optical, and RA $\rm{17^h 54^m 23^s.49}$ and Dec. $\rm{+27\degr\ 59\arcmin\ 58\arcsec.5}$ for the radio position (Condon \& Kaplan 1998). 

The first estimation of the effective temperature of the central star of Vy~1--2 was by Gurzadyan (1988), 
124~kK and 66~kK using the \heliumb/ \hbeta\ and \oxygeniii/ \heliumb\ line ratios, respectively. More 
recently, Stangellini et al. (2002) calculated $\rm{T_{eff}}$ = 119$\pm25$~kK and $\rm{log(L/L_{\odot}})$ 
= 3.275$\pm$0.285, for a distance of $7.24\pm0.13$~kpc, whereas Phillips (2003) estimated the \helium\ and \heliumb\ 
Zanstra temperatures of $75.4\pm10.2$~kK and $99.8\pm4.3$~kK, respectively. Its distance is uncertain. It has been estimated 
to range from 4.38~kpc to 7.6~kpc: 4.7~kpc (Maciel 1984); 5.7~kpc (Sabbadin et al. 1984); 4.38~kpc (Phillips 1998) and 7.6~kpc (Cahn \& Kaler 1971, 
Cahn et al. 1992). However, all the aforementioned authors used a radius of 2.4\arcsec\ for their distance 
estimation obtained from Cahn and Kaler (1971). No high--angular resolution of Vy~1--2 had been obtained until 1996 
(Manchado et al. 1996). More recently, Shaw et al. (in preparation) obtained the high--angular HST images 
that we present here. The value of 2.4\arcsec\ is almost $\sim$1.6 times larger than the radius find 
in this work using the HST images (see below). This implies that the previous reported distances of Vy 1--2 are likely underestimated.

The large differences in chemical abundances and effective temperature inspired us to study this nebula 
and provide more accurate determinations of stellar and nebular properties. We present high--resolution HST images 
of Vy~1--2 in the \oxygeniii\ emission line and broad--band filter F200LP, as well as new high-- and low--dispersion 
spectra in order to provide further insight into the kinematic structure and chemical composition of this compact 
nebula. In Section 2, we describe the observations and data reduction procedures. In Section 3, we present the results 
from the morpho--kinematic and chemical analysis. We also present the physical properties 
and chemical abundances derived from the ionisation correction factor (ICF; Kingsburgh \& Barlow 
1994) and from a photo--ionisation model performed with Cloudy (Ferland et al. 2013). Our results are 
discussed in Section 4 and conclusions are summed up in Section 5. 

\begin{figure*}
\centering
\includegraphics[scale=0.425]{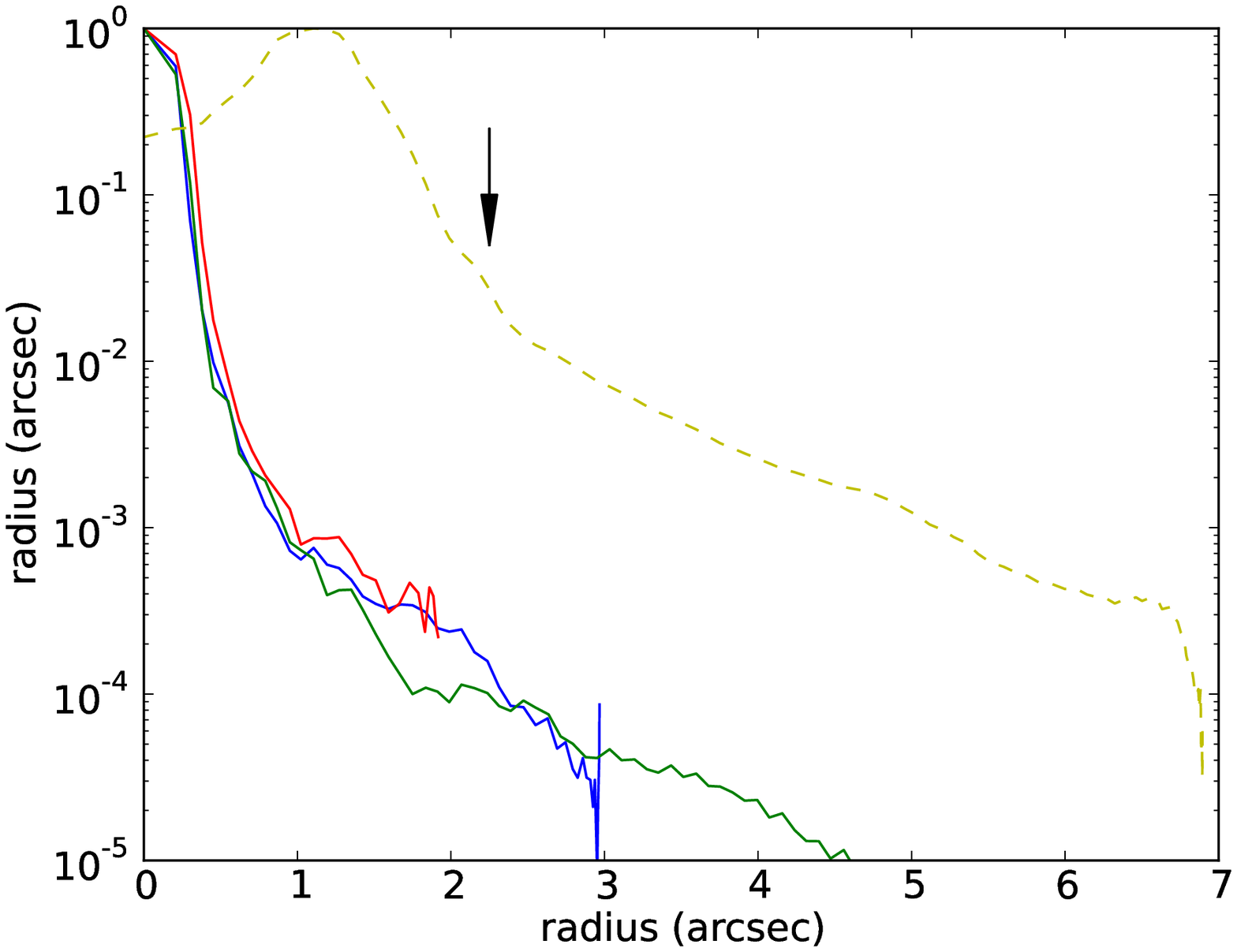}
\includegraphics[scale=0.425]{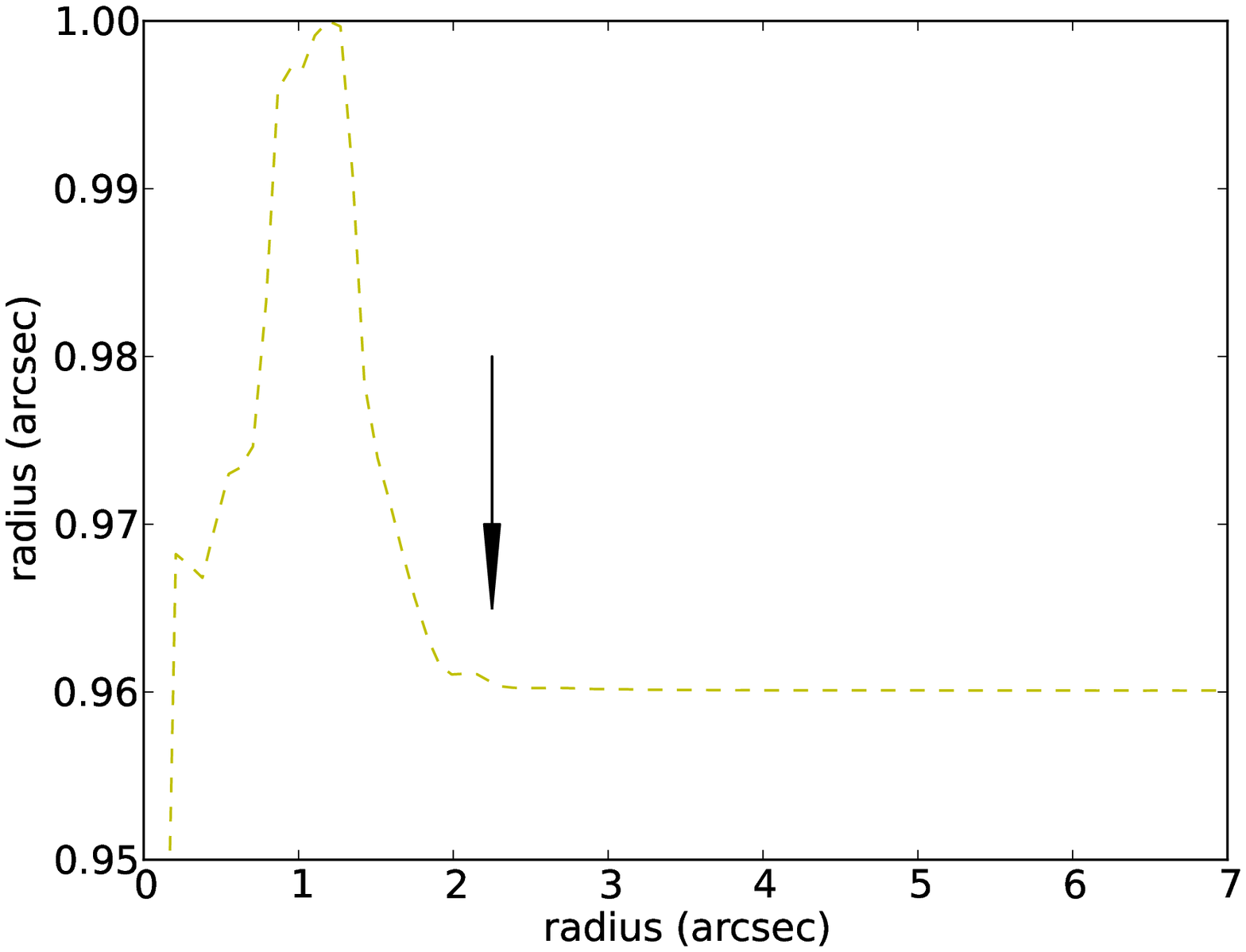}
\caption[]{Left panel: surface brightness profile of \oxygeniii\ emission in Vy~1--2 averaged over all angles 
(yellow dashed line). The bump at r$\sim$2.2~\arcsec--2.3~\arcsec, indicated by an arrow, corresponds to the 
contribution of the faint lobes along the east--west direction. The red and blue solid lines represent the radial 
brightness of two stars observed with the same camera and filter while the green solid line corresponds to the 
profile of a theoretical HST point spread function (PSF) created using the program Tiny Tim. Right panel: 
surface brightness profile of the convolved \oxygeniii\ image averaged over all angles, using the theoretical 
PSF and Leon Lucy's algorithm.}
\label{fig2}
\end{figure*}

\section{Observations}

\subsection{HST imaging}
High--resolution HST images of Vy~1--2 were drawn from the Mikulski archive for 
space telescopes (Shaw et al. in preparation). They were taken with the Wide Field Planetary 
Camera 3 (WFPC3) on February 8$^{\rm{th}}$ 2010 (GO program 11657, PI: Stanghellini L.). The narrow--band 
F502N filter ($\lambda$=5009.8 \AA) was used to isolate the \oxygeniii\ 5007 \AA\ emission line and determine 
the morphology of Vy~1--2 (Fig. 1), while the broad filters F200LP, F350LP and F814W 
($\lq\lq$I\rq\rq, see Table 1) were obtained in order to probe the presence of a cold 
binary companion. The exposure times were 600, 11, 11, and 23~sec, respectively.
A faint outer shell or halo as well as a pair of knots can be discerned in the \oxygeniii\ image (Fig. 1, 
left panel), while the F200LP (Fig 1, right panel) image is used to highlight the inner nebular 
components: the bright elliptical ring--like structure and the faint lobes.

We carefully investigate whether the outer shell/halo is a real nebular component or the result of scattered light from 
the bright inner nebula in the HST optics. In Figure~2, left panel, we display the radial profile of a theoretical PSF\footnote{The program Tiny Tim developed 
by Krist et al. (2011), was used to create the theoretical PSF for the \oxygeniii\ image.} (green solid line), of two 
field stars (observed with the same camera and filter; red and blue solid lines) and of Vy~1--2 (yellow dashed line).
The radial profiles between the two field stars and the theoretical PSF agree reasonably well. A discrepancy with 
the modelled PSFs can be seen at around 1.5--2\arcsec, though it is negligible compared to the Vy 1--2's profile. 
On the other hand, Vy~1--2 shows a maximum brightness at r=1.2\arcsec\ (at the position of the ring--like structure), 
while for r$>$2.2--2.3\arcsec\ (indicated by a vertical arrow in Fig.~2) shows a similar steep decline for the stars and 
theoretical PSF but two magnitude brighter. The theoretical \oxygeniii\ PSF was used to convolve the HST image using Leon Lucy's algorithm in {\sc STSDAS}. 
The azimuthally--averaged profile of the convolved image is presented in Figure~2, right panel. The bump at 
r=2.2--2.3\arcsec\ is indicated by a vertical arrow. For r$>$2.3\arcsec, the profile is a constant function, which means that the outer 
shell/halo is seen due to the PSF scattering. In particular, for a power-law density distribution of $\rho$(r)$\propto$r$^{-1}$, a r$^{-2}$ 
surface brightness is expected (Sahai et al. 1999), which we do not see here. Nevertheless, some contribution of scattering light from 
dust can not be excluded, given that Vy~1--2 has been found to be a strong IR emitter.

\subsection{High--dispersion longslit spectroscopy}

High--dispersion, long--slit spectroscopic data of Vy~1--2 in \ha +\nitrogen\ 
and \oxygeniii\ were obtained using the Manchester echelle 
spectrometer, (MES--SPM; Meaburn et al. 2003) on the 2.1~m telescope at the Observatorio 
Astron\'omico of San Pedro Martir in Baja California, Mexico, in its 
f/7.5 configuration. The observing run took place on May 2011 and April 2012 (see Table 1).
The MES-SPM was equipped with a Marconi CCD detector with 2048$\times$2048 
square pixels and each side of 13.5~$\mu$m ($\equiv$0.176\arcsec\ pixel$^{-1}$). 
A 90~\AA\ bandwidth filter was used to isolate the 87$^{th}$ order containing 
the H$\alpha$ and \nitrogen{} $\lambda \lambda ~$6548,~6584~\AA, and 
the 113$^{th}$ order containing the \oiii\ emission lines. 
Two-by-two binning was employed in both the spatial and spectral directions 
(a three--by--three binning was employed in both directions for the observing run in May). 
Consequently, 1024 increments, each 0\farcs352{} long, gave a projected slit 
length of 5\farcm12 on the sky. The slit used was 150~$\mu$m{} wide ($\equiv$~11~\kms\ and 
1\farcs9). During the observations the slit was oriented at different position angles. 
The integration times were 1800 sec, 1500 sec and 1200 sec (see Table 1 for details).
In order to avoid any contamination from the bright central star, the slits were slightly 
offset from it by $\sim$0.5~\arcsec, covering only the bright ring--like structure.

A faint continuum could be discerned within the very short spectral range of 
our single--order echelle spectra. The wavelength calibration was performed using 
a Th/Ar calibration lamp to an accuracy of $\pm$1~\kms\ when converted 
to radial velocity. The data reduction was performed using the standard {\sc IRAF}~\footnote 
{IRAF (Image Reduction and Analysis Facility) is distributed by the National Optical Astronomy Observatory, 
which is operated by he Association of Universities for Research in Astronomy (AURA) Inc., under cooperative agreement 
with the National Science Foundation.} and 
{\sc STARLINK} routines. Individual images were bias subtracted and cosmic rays removed.
All spectra were also calibrated to heliocentric velocity. The position--velocity 
(PV) diagrams and line profiles for the \ha, \nitrogen\ and \oxygeniii, and for all slit 
positions are presented in Figures 4--8 and Figures 9--12, respectively.

\begin{figure}
\centering
\includegraphics[scale=0.30]{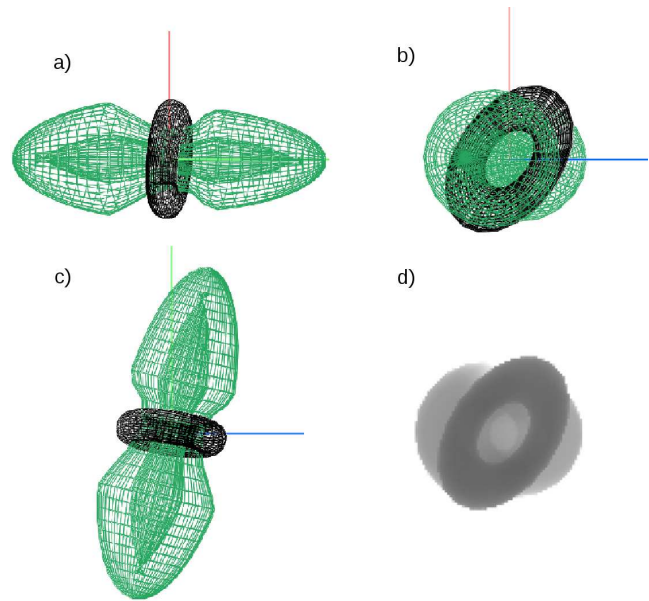}
\caption[]{Schematic diagrams of Vy~1--2 using the code {\sc SHAPE}. Panels a) to c) show the mesh {\sc SHAPE} images for three 
different orientations right, front and top views, respectively. Panel d) illustrates the final rendered 2--D image.}
\label{fig3}
\end{figure}

\subsection{Low--dispersion spectroscopy}

\begin{figure*}
\centering
\includegraphics[scale=0.65]{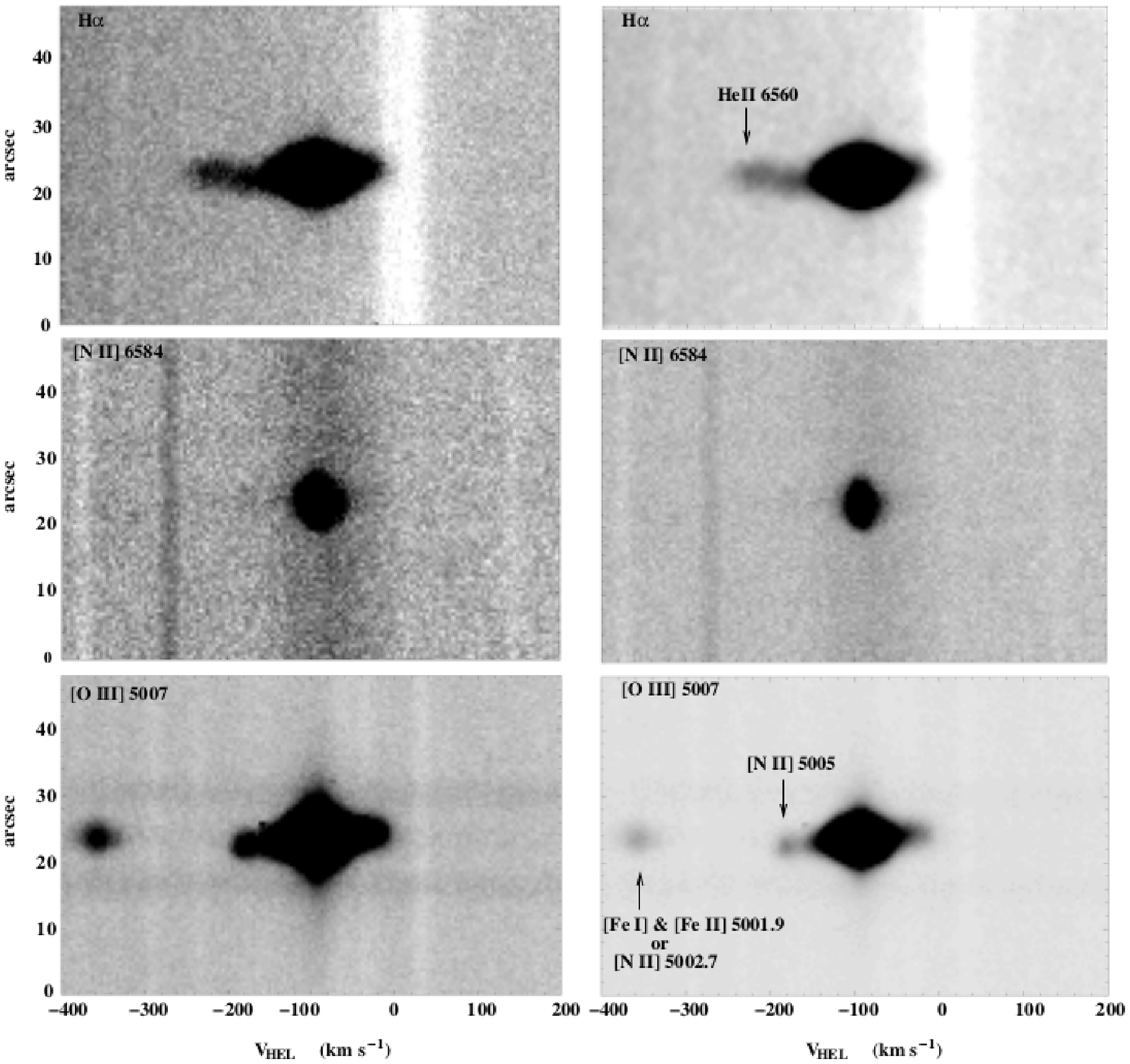}
\caption[]{\ha\ (upper panel), \nitrogen\ 6584~\AA\ (middle panel) 
and \oxygeniii\ 5007 \AA\ (lower panel) position--velocity (PV) diagrams of Vy~1--2 shown at two different contrast levels  
the for slit position 2 (P.A.=90$^{\circ}$).}
\label{fig4}
\end{figure*}

\begin{figure*}
\centering
\includegraphics[scale=0.65]{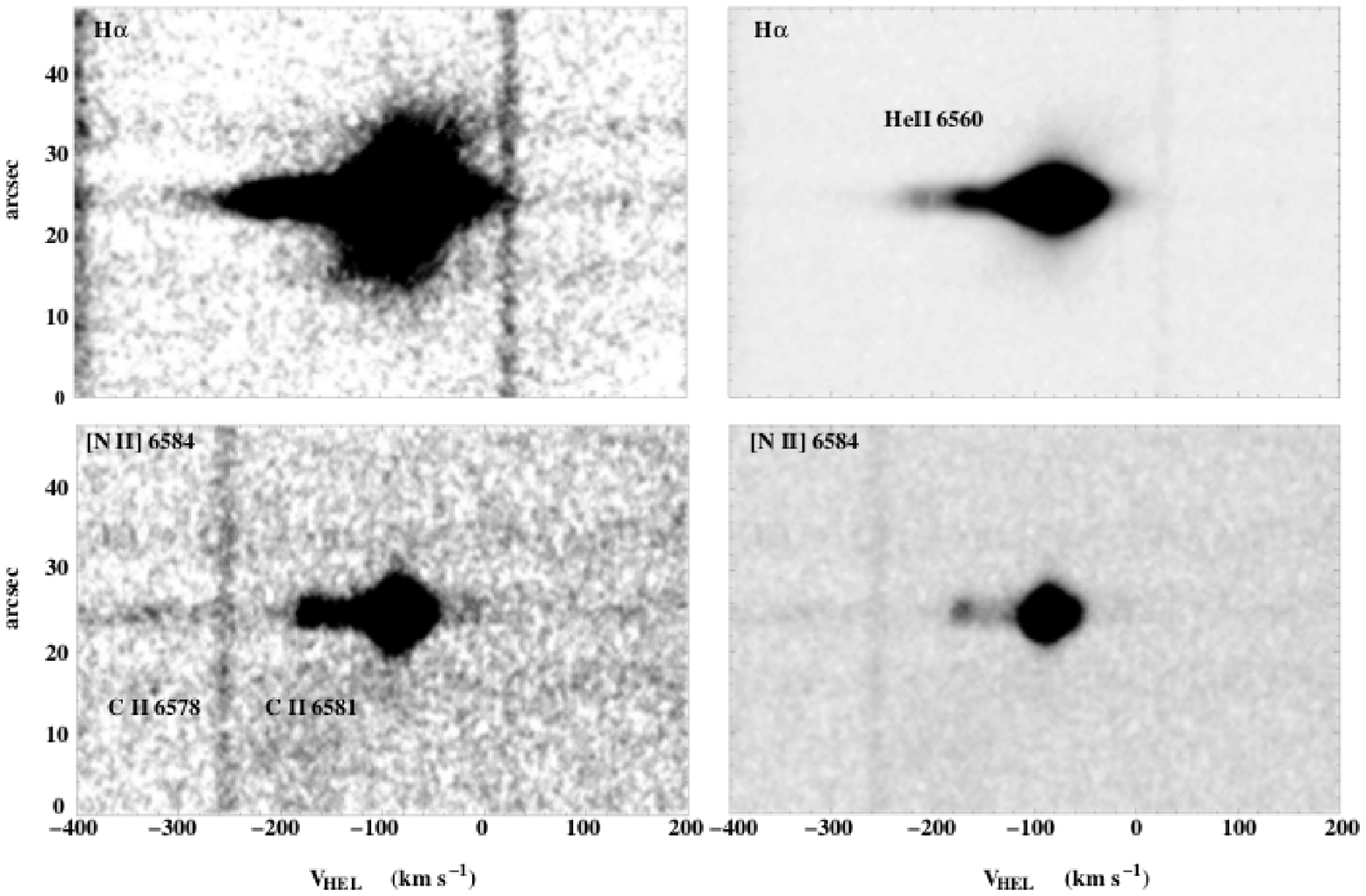}
\caption[]{\ha\ (upper panel) and \nitrogen\ 6584~\AA\ (lower panel) 
position--velocity (PV) diagrams of Vy~1--2 shown at two different contrast levels for the slit position 3 (P.A.=0$^{\circ}$).}
\label{fig5}
\end{figure*}

\begin{figure*}
\centering
\includegraphics[scale=0.65]{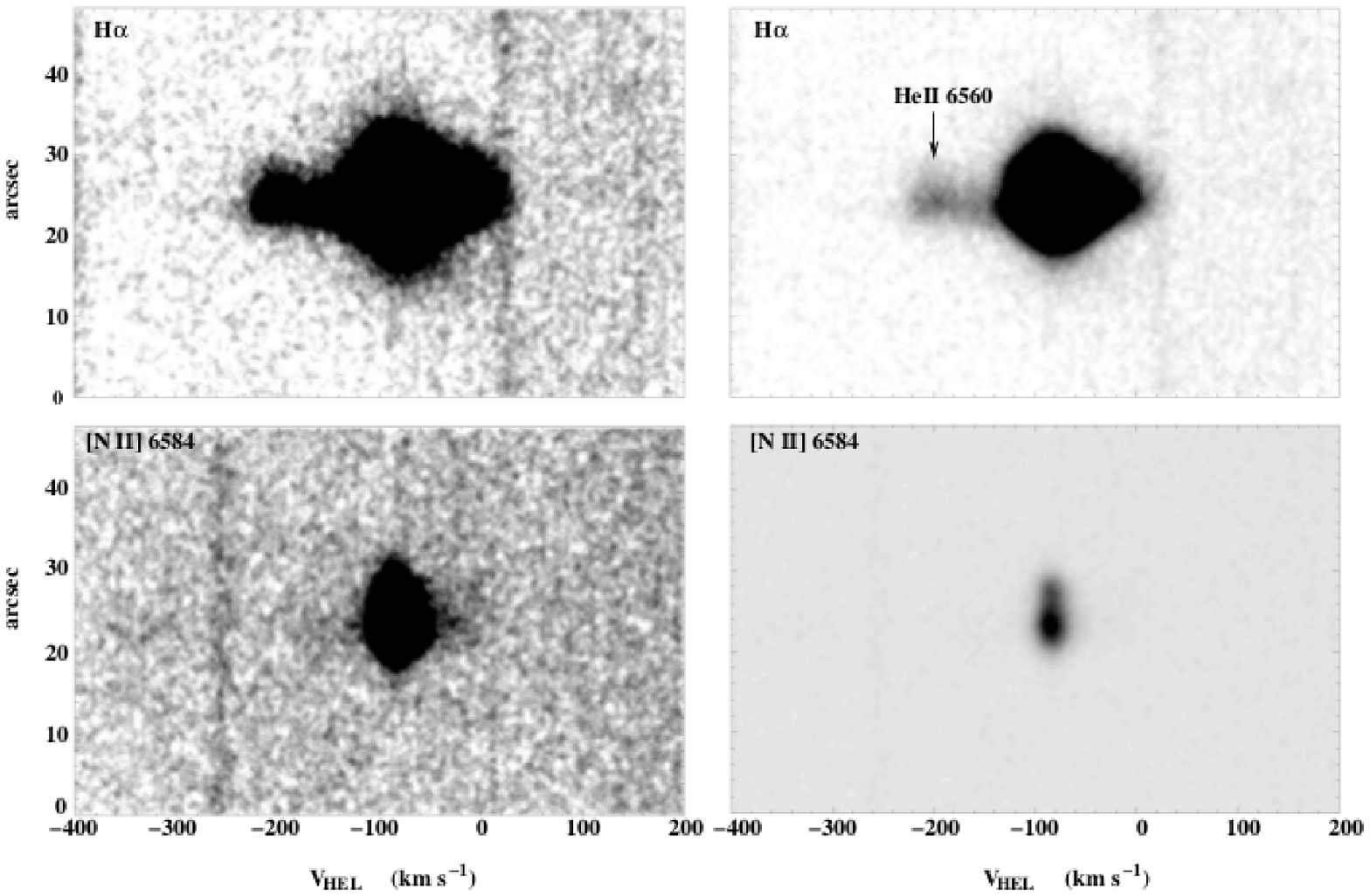}
\caption[]{\ha\ (upper panel) and \nitrogen\ 6584~\AA\ (lower panel) 
position--velocity (PV) diagrams of Vy~1--2 shown at two different contrast levels the for slit position 4 (P.A.=55$^{\circ}$).}
\label{fig6}
\end{figure*}

\begin{figure*}
\centering
\includegraphics[scale=0.65]{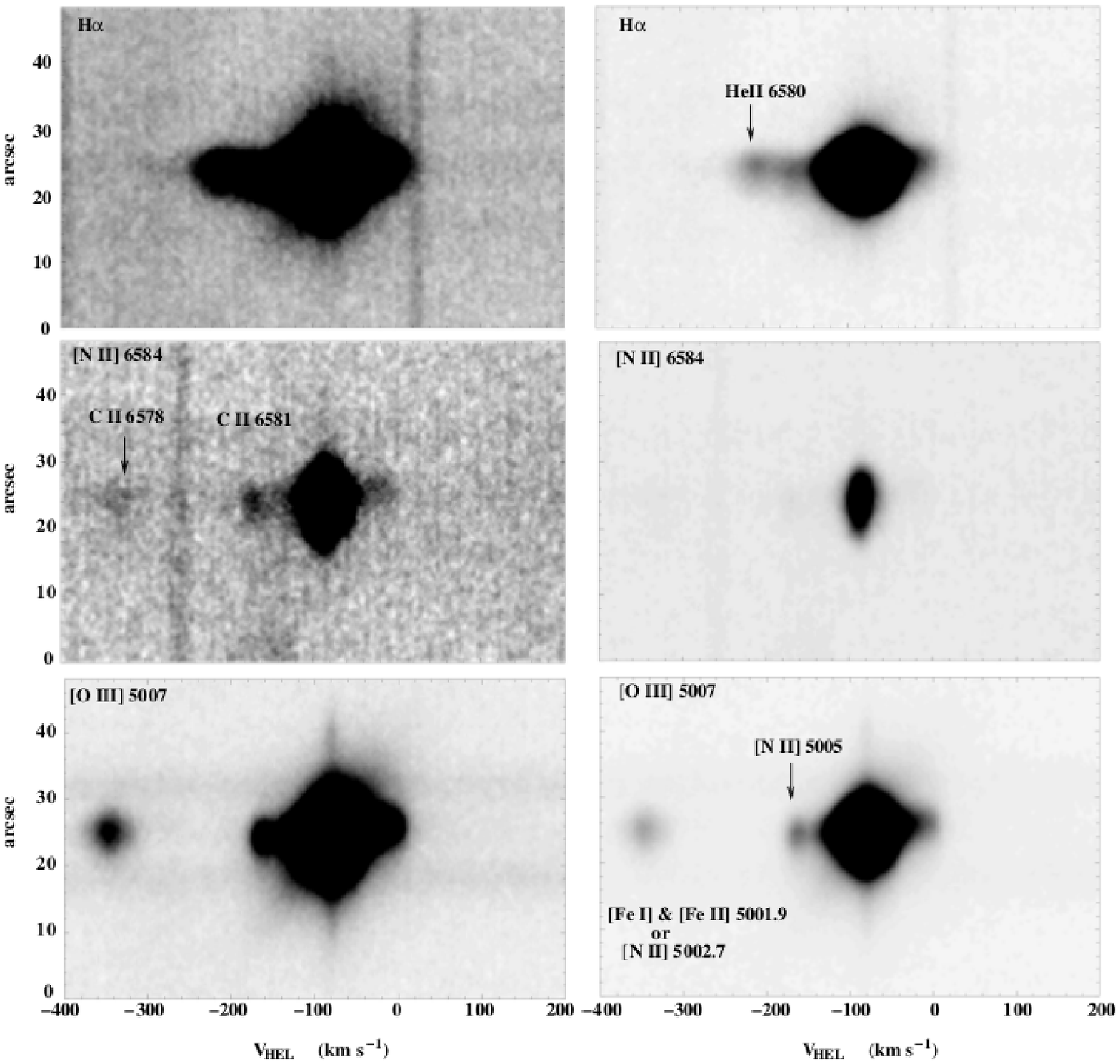}
\caption[]{\ha\ (upper panel), \nitrogen\ 6584~\AA\ (middle panel) and \oxygeniii\ 5007 (lower panel) 
position--velocity (PV) diagrams of Vy~1--2 shown at two different contrast levels for the slit position 5 (P.A.=310$^{\circ}$).}
\label{fig7}
\end{figure*}

\begin{figure*}
\centering
\includegraphics[scale=0.65]{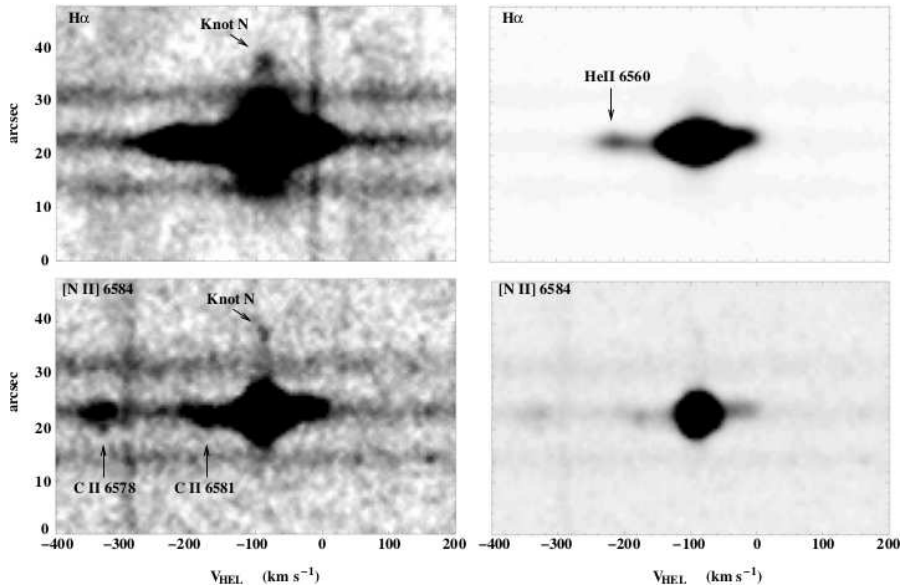}
\caption[]{\ha\ (upper panel) and \nitrogen\ 6584~\AA\ (lower panel) 
position--velocity (PV) diagrams of Vy~1--2 shown at two different contrast levels for the slit position 6 (P.A.=310$^{\circ}$), 
drawn from the SPM catalogue (L\'opez et al. 2012).}
\label{fig8}
\end{figure*}

Low--dispersion spectra of Vy~1--2 were obtained with the 1.3~m telescope at 
Skinakas Observatory in Crete, Greece, on September $\rm{4^{th}}$, 2011. A 1300 line
mm$^{-1}$ grating was used in conjunction with a 2000 x 800 SITe CCD 
(15 $\mu$m pixel), resulting in a scale of 1~\AA\ $\rm{pixel^{-1}}$ and 
covering the wavelength range of 4700--6800~\AA. The slit width used was 7.7\arcsec\ 
with a slit length of 7.9 \arcmin\ and oriented in the north--south direction 
(Slit position 1). Two different exposure times were taken in order to identify 
both the bright and faint emission lines: 3600~sec (saturated in \ha\ 
and \oxygeniii\ 5007~\AA\ lines, but detecting the much fainter lines) and 
3 x 300~sec. The spectro--photometric standard stars HR5501, HR7596, HR9087, HR718 and HR7950 
(Hamuy et al. 1992) were observed in order to flux calibrate the spectra, while the fluxes were 
also corrected for atmospheric extinction and interstellar reddening. 

The data reduction was performed using standard IRAF routines. 
Individual images were bias subtracted and flat--field
corrected using a series of twilight flat--frames. The sky--background
was also subtracted from each individual frame while no dark
frames were used as the dark current was negligible. Finally, the
wavelength calibration was performed using a Fe--He--Ne--Ar calibration lamp. 

\section{Results}

\subsection{Morphology}

Figure 1 displays the high--resolution HST \oxygeniii\ (left panel) and ultraviolet (F200LP, right panel) 
images of Vy~1--2, at two different scales, in order to highlight the fainter nebular components.
The left panel displays a faint outer shell/halo and two much fainter knots at a position angle (hereafter P.A.) of 305$^{\circ}$, located 
at 11.5\arcsec\ (knot S) and 14.5\arcsec\ (knot N) from the central star. The right panel shows an enlargement of the nebula, 
revealing a bright elliptical ring--like structure with a minor and major axis of 2.4\arcsec\ and 3.2\arcsec, respectively. 
This implies a mean radius of $\sim$1.4\arcsec. This value is $\sim$1.6 times lower than the value used previously for the distance 
estimation obtained from ground--based observations. Thus, we recalculated the distance of Vy 1--2 using our more accurate 
value of radius, and we obtain a distance between 7.0~kpc and 14.5~kpc. These distances clearly indicate that Vy 1--2 is a 
distant nebula. An average distance of 9.7~kpc is adopted here.

The ring--like structure has a non--uniform emission along its perimeter, with an apparent filamentary structure, which indicates 
a strong interaction of stellar winds in addition of an inner cavity. Its major axis is oriented towards to a P.A.$=$320$^{\circ}$,  
15$^{\circ}$ rotated from the knots' direction. This may be associated with a possible precession of the rotation axis of a single 
central star, or the orbital plane of a binary star at the centre of Vy~1--2. Two faint lobes are also apparent in the 
east--west direction (P.A. $\sim$90$^{\circ}$). Despite their very low surface brightness compared to the central 
ring-like structure, their morphology suggests a bipolar structure, like the Homunculus nebula and Hb~5 PN. 
The eastern and western bipolar lobes are found to be blue-- and red--shifted, respectively, while their different 
sizes imply a tilted nebula towards the line of sight. In Figure 3, we illustrate a schematic 3--D diagram of 
Vy~1--2 for different points of view, using the code {\sc SHAPE} (Steffen et al. 2011). This simple model reproduces fairly well 
the main components of Vy~1--2, such as the ring---like structure and the bipolar lobes as well as the orientation of the nebula 
(inclination and position angles). The inclination angle of the model is 13\degree, while the ring--like 
structure is best described with an ellipse. Any attempt to reproduce the ring--like structure assuming a round shape, 
resulted in a very high inclination angle of $\sim$50\degree\ that we do not get from the PV diagrams. 
Our 3-D structure of Vy 1–2 is only one of many other solutions that can be obtained just by varying the size 
of bipolar lobes and the orientation accordingly.

Moreover, in order to investigate the possible presence of a cold binary companion, the F200LP, F350LP and F814W broad--band 
images were studied. With the available data, we found no evidence of a nearby companion. If the central 
star is a binary system, then the projected separation must be smaller than 0.08\arcsec. 

\begin{table}
\centering
\caption[]{Expansion velocities of Vy~1--2}
\label{table5}
\begin{tabular}{llcccc}
\hline 
 slit & P.A.  & V(\ha)& V(\nitrogen) & V(\oxygeniii) & V(\heliumb)\\
      & ($^{\circ}$) & \kms\ &\ \kms\  & \kms\         &  \kms\      \\      
\hline
\multicolumn{5}{l}{\textbf{low velocity component}}\\
2     &  90  &  19 & 9 & 18 & 25\\
3     &   0  &  18 & 8 & $-$& $-$\\
4     &  55  &  19 & 9 & $-$ & 25 \\ 
5     & 310  &  19 & 9 & 18 & 23\\
6     & 310  &  19 & 8 & $-$ & 23\\
\\
\multicolumn{5}{l}{\textbf{high velocity component}}\\
2     &  90  &   90 & 40 & 90 & $-$\\
3     &   0  &  100 & 40 & $-$& $-$\\ 
4     &  55  &  100 & 40 & $-$& $-$\\ 
5     & 310  &   95 & 90 & 90 & $-$ \\
6     & 310  &  100 & 90 & $-$& $-$\\ 
\hline
\end{tabular}
\begin{flushleft}
\end{flushleft}
\end{table}

\subsection{Kinematics}

\begin{figure*}
\includegraphics[scale=0.25]{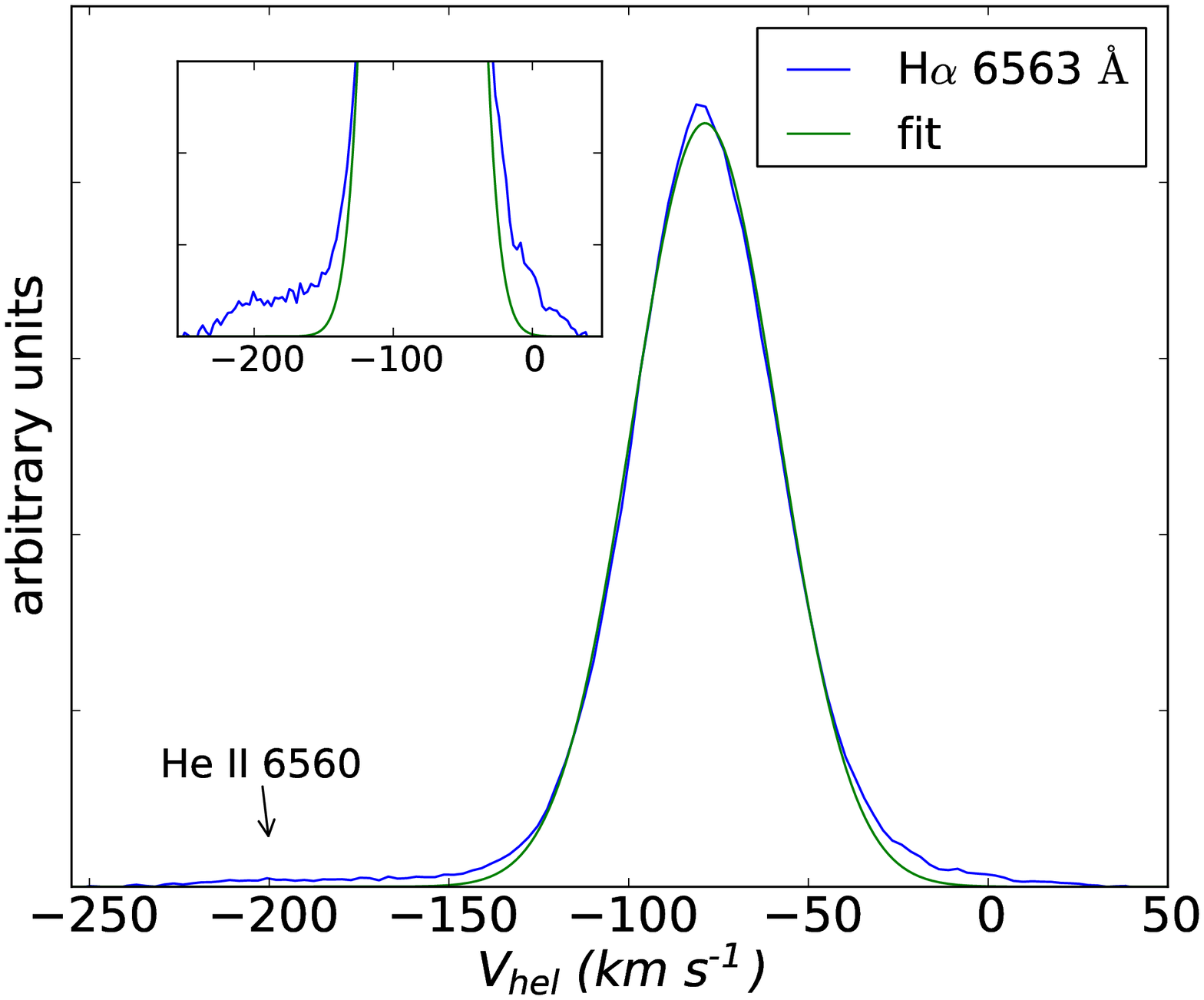}
\includegraphics[scale=0.25]{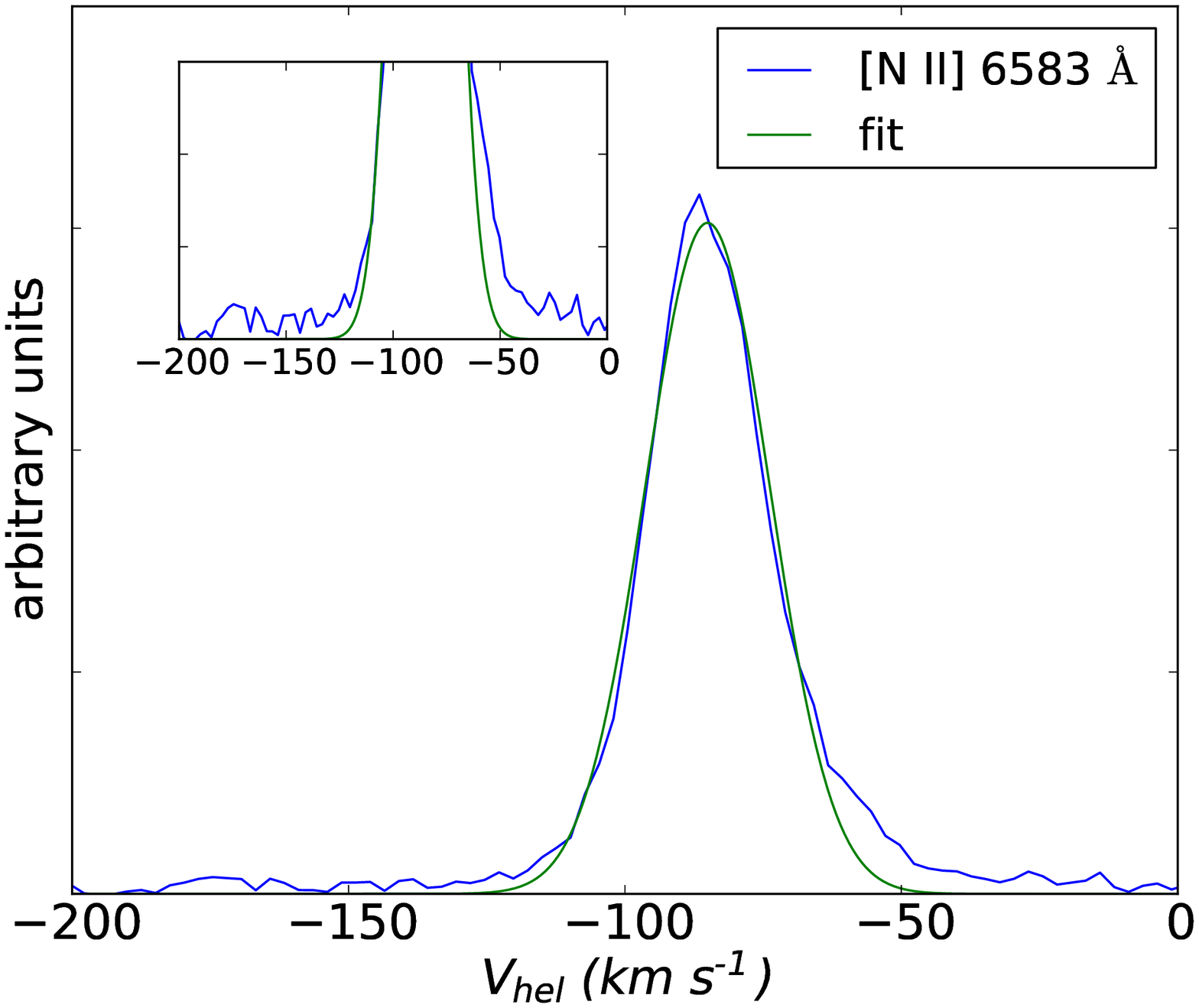}
\includegraphics[scale=0.25]{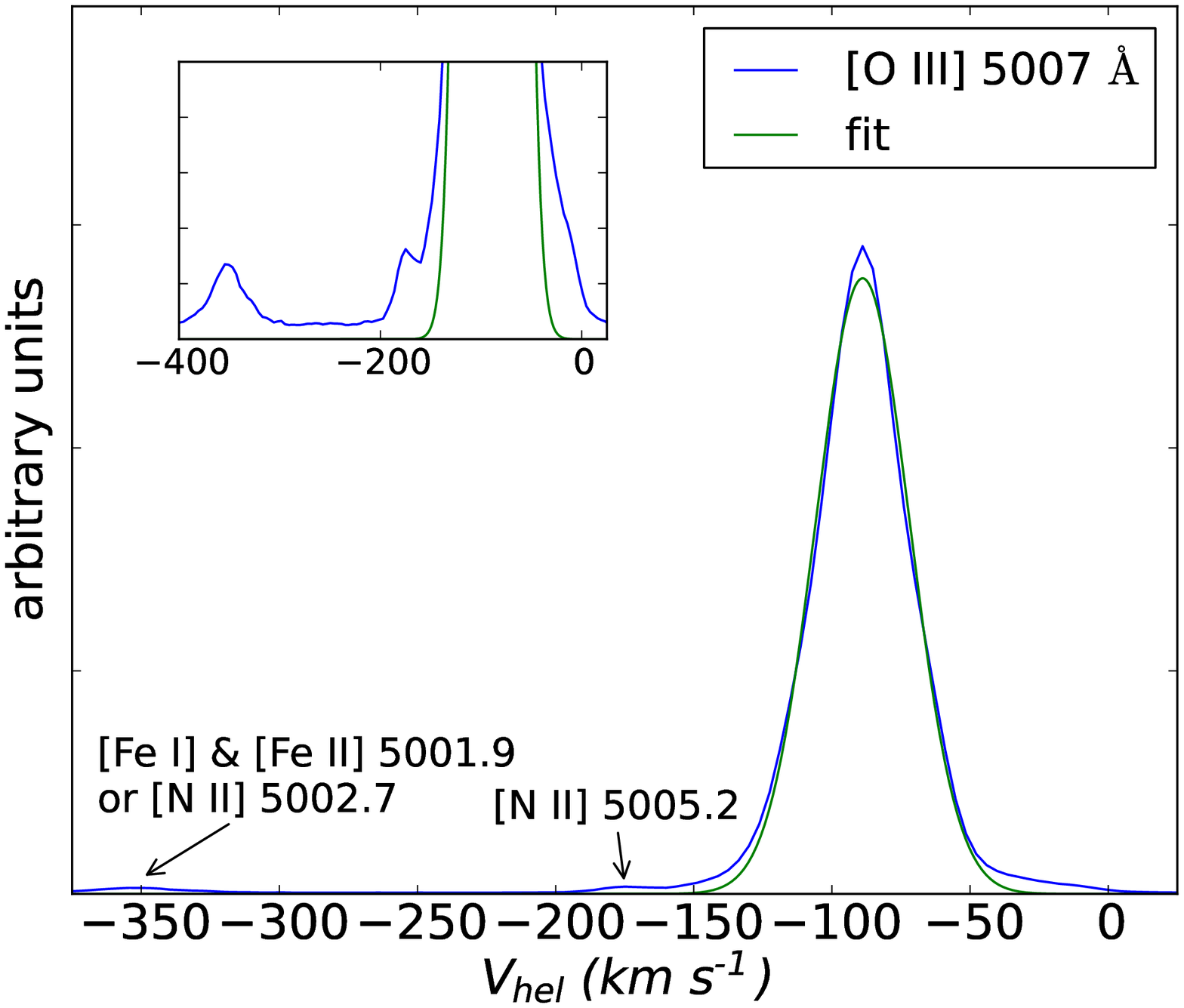}
\caption{The \ha, \nitrogen\ 6584~\AA\ and \oxygeniii\ 5007~\AA\ line profiles for slit position 2 
(P.A.=90$^{\circ}$) are shown in the left, middle and right panels, respectively. The blue and green lines 
correspond to the observed line profile and a single Gaussian best fit, respectively.}
\label{fig9}
\end{figure*} 

\begin{figure*}
\includegraphics[scale=0.25]{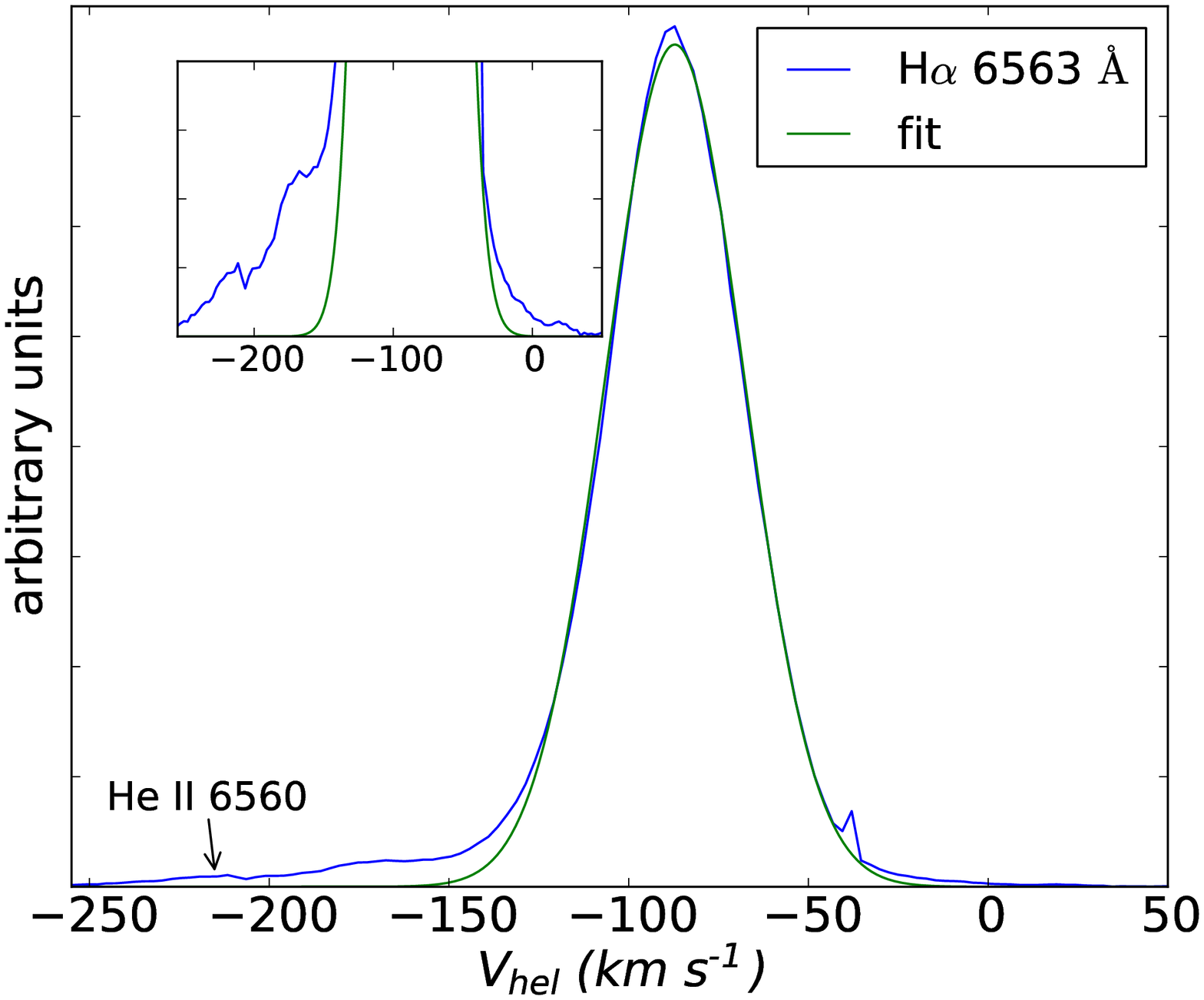}
\includegraphics[scale=0.25]{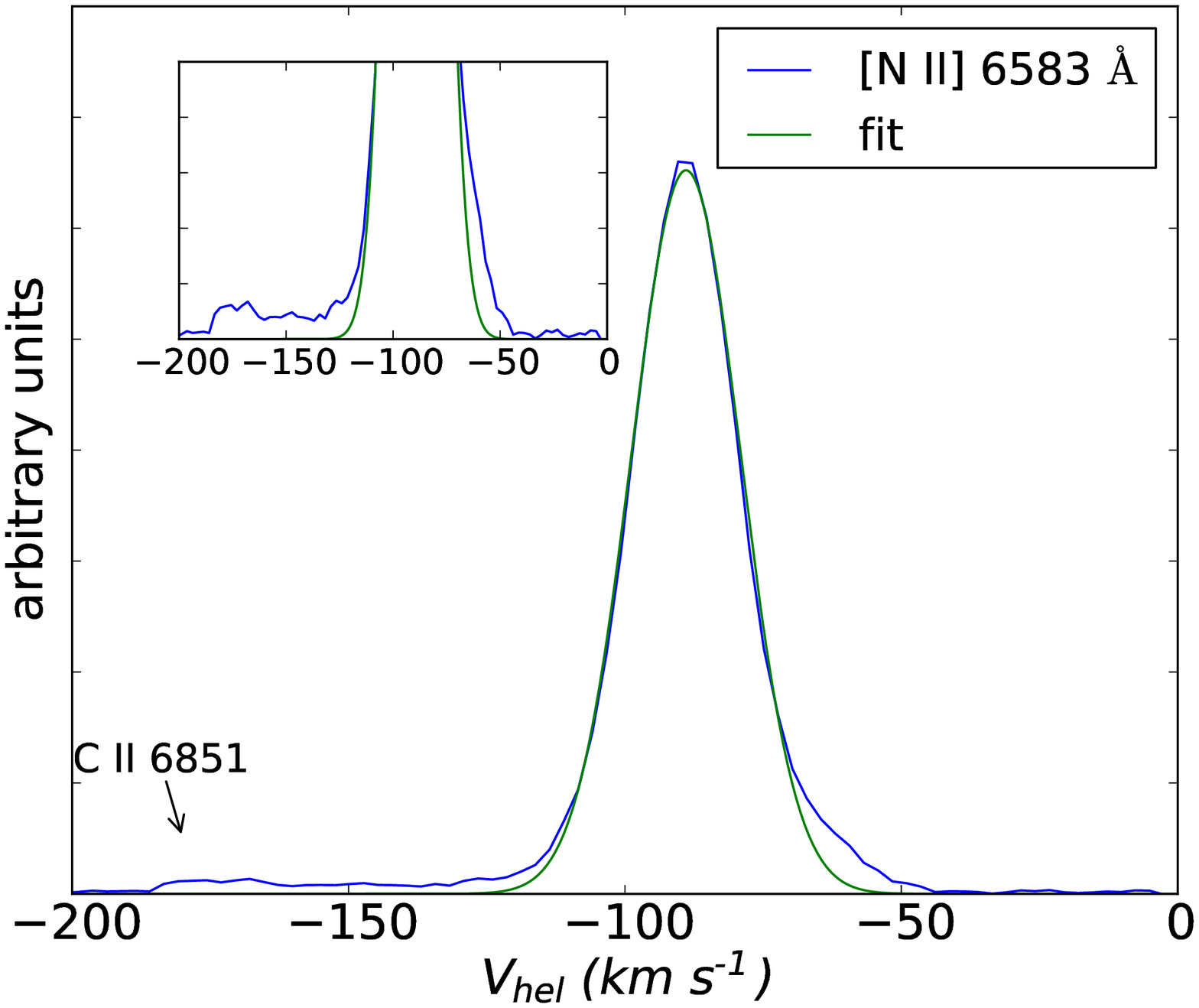}
\caption{The \ha\ and \nitrogen\ 6584~\AA\ line profiles for slit position 
3 (P.A.=0$^{\circ}$) are shown in the left and right panels, respectively. 
The blue and green lines correspond to the observed line profile and the 
single Gaussian best fit, respectively.}
\label{fig10}
\end{figure*}

\begin{figure*}
\includegraphics[scale=0.25]{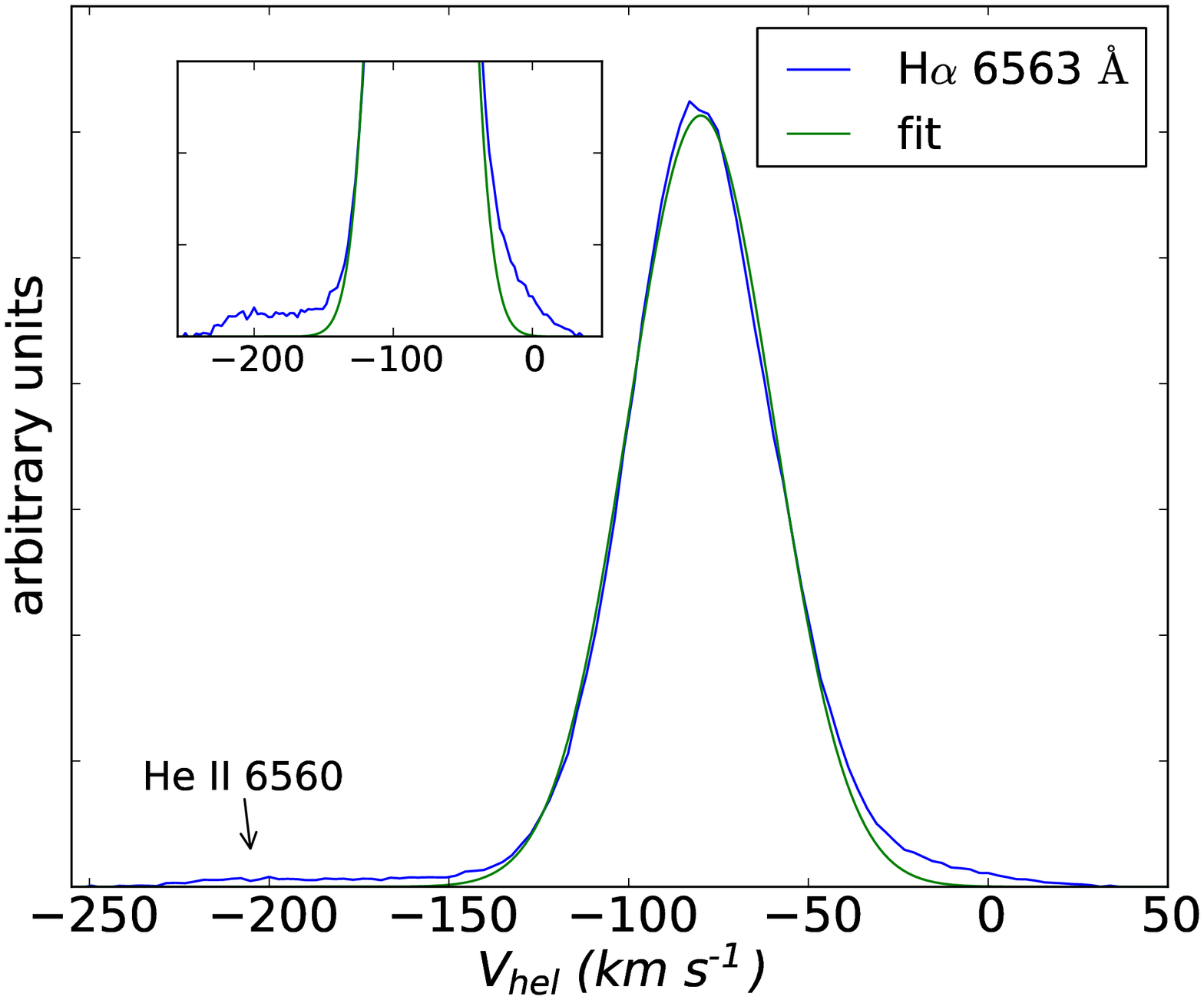}
\includegraphics[scale=0.25]{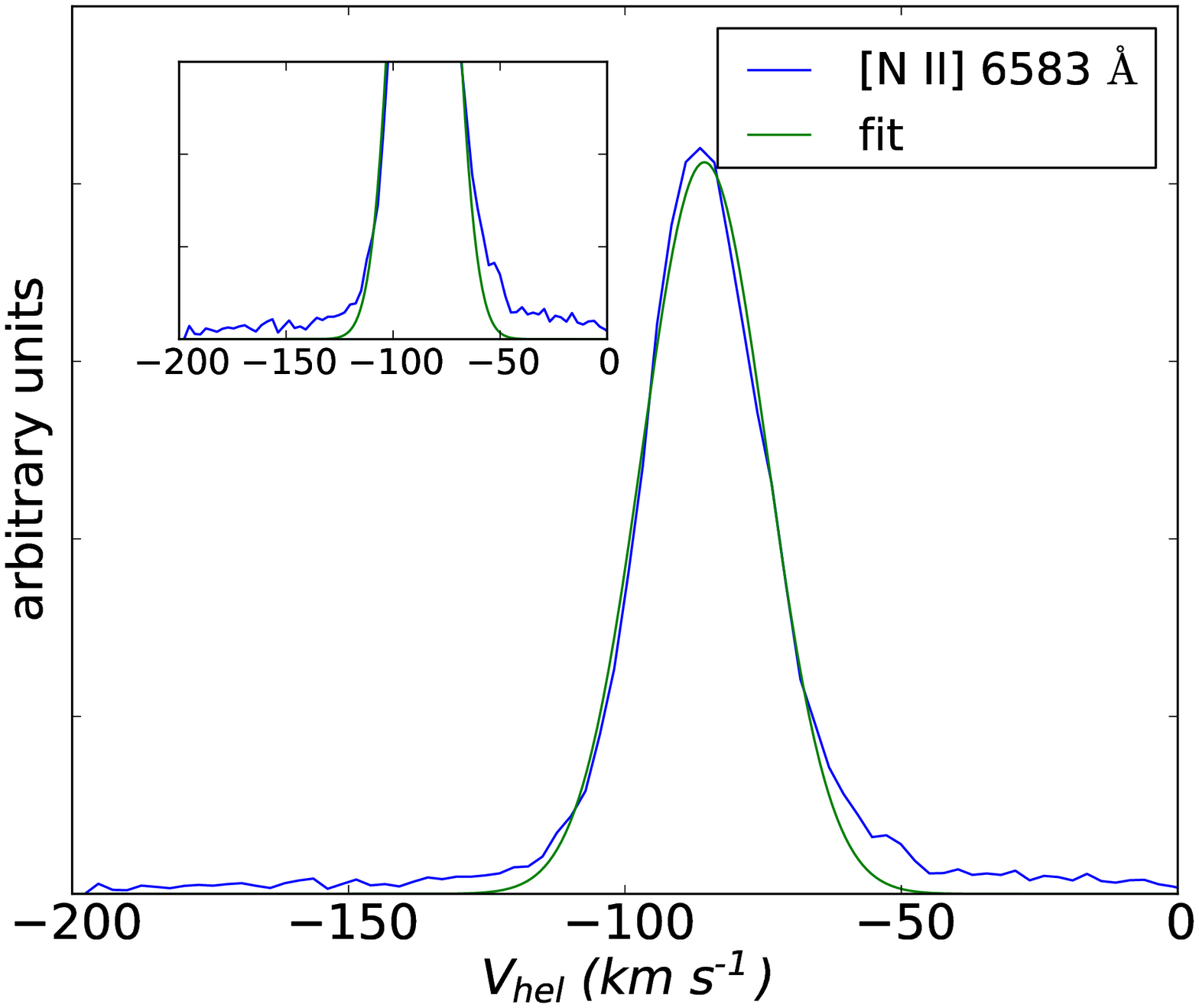}
\caption{The \ha\ and \nitrogen\ 6584~\AA\ line profiles for slit position 
4 (P.A.=55$^{\circ}$) are shown in the left and right panels, respectively. 
The blue and green lines correspond to the observed line profile and the 
single Gaussian best fit, respectively.}
\label{fig11}
\end{figure*}

\begin{figure*}
\includegraphics[scale=0.25]{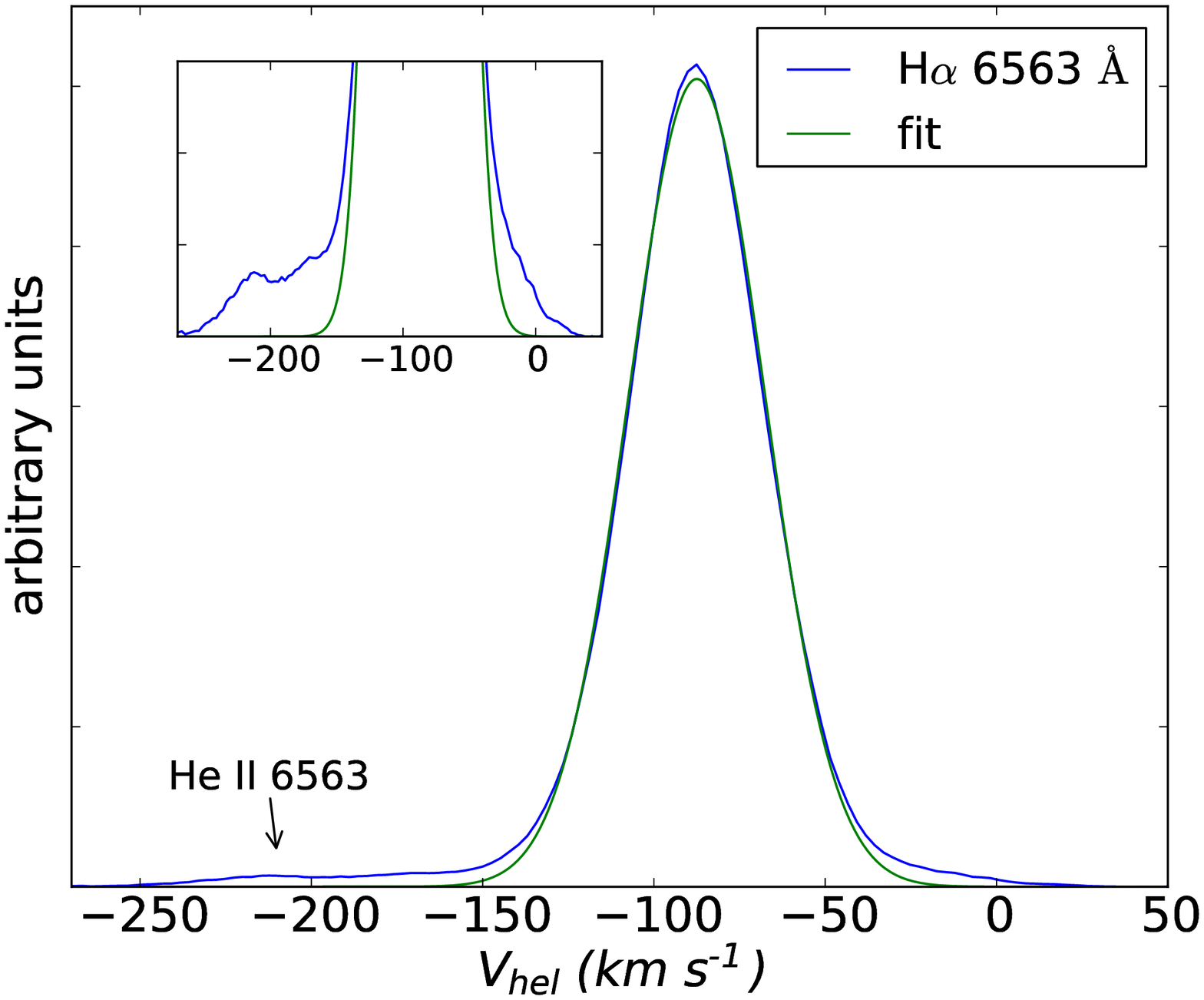}
\includegraphics[scale=0.25]{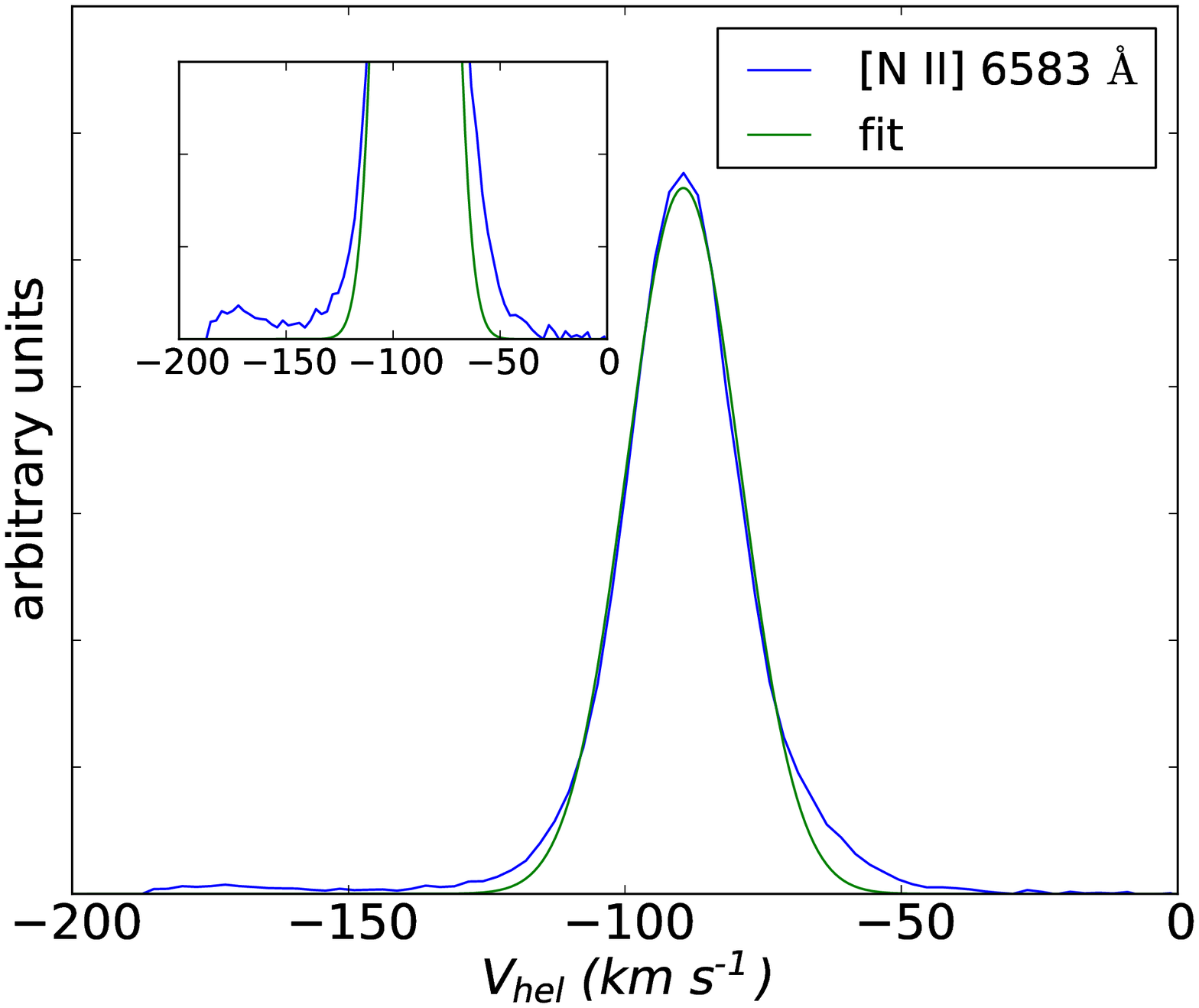}
\includegraphics[scale=0.25]{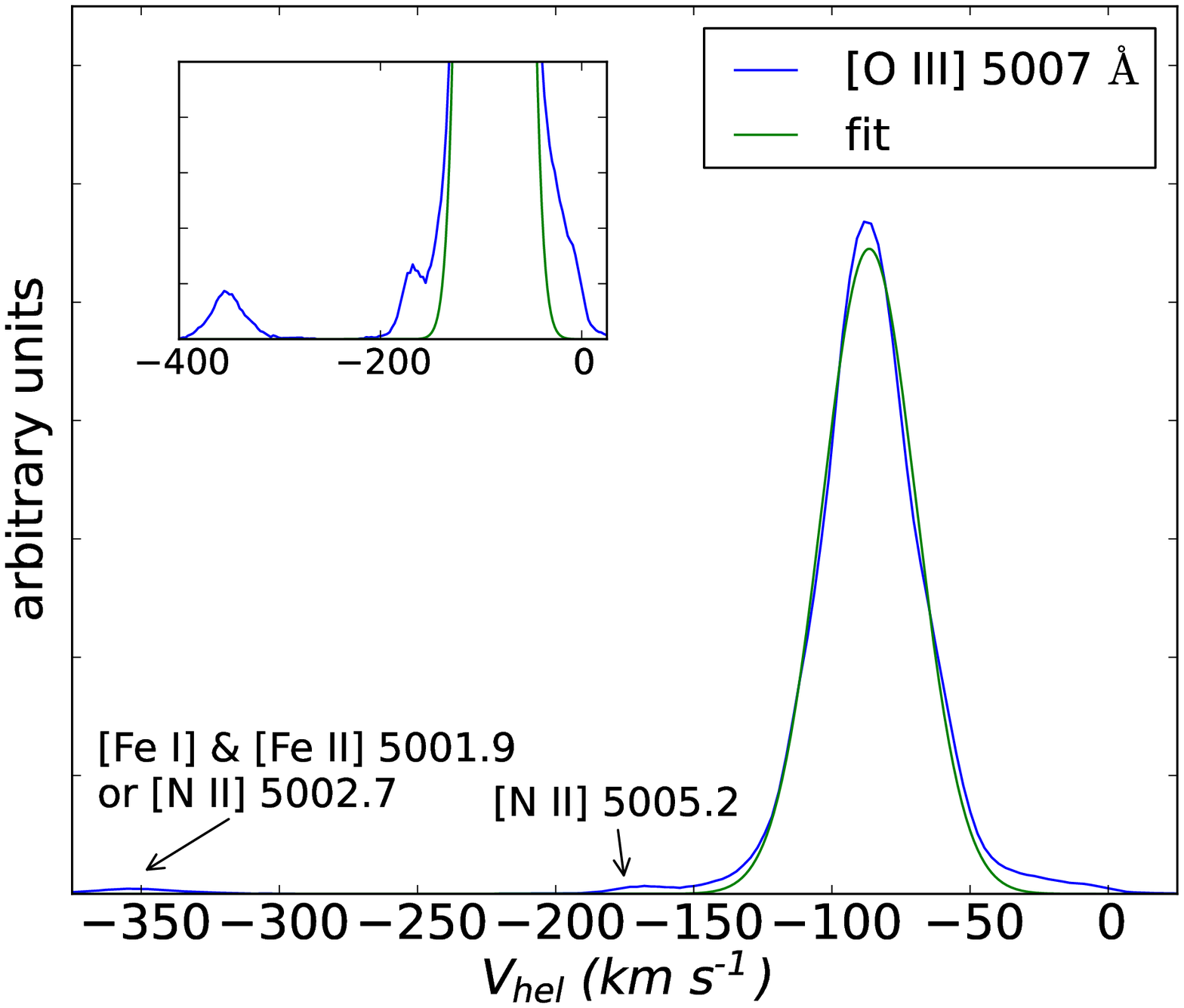}
\caption{The \ha, \nitrogen\ 6584~\AA\ and \oxygeniii\ 5007~\AA\ line profiles for slit position 
5 (P.A.=310$^{\circ}$) are shown in the left, middle and right panels, respectively.
The blue and green lines correspond to the observed line profile and the single 
Gaussian best fit, respectively.}
\label{fig12}
\end{figure*}

In Figures~4 to 8 we present the observed \ha\ 6563~\AA, \nitrogen\ 6584~\AA\ and \oxygeniii\ 
5007~\AA\ position--velocity (PV) diagrams along the slit positions 2 to 6. All slit positions are shown overlaid 
on the HST \oxygeniii\ image (Fig. 1, left panel). Besides the common bright \ha, \nitrogen\ and \oxygeniii\ emission 
lines, the \heliumb\ 6560~\AA\ and \helium\ 5015~\AA\ lines, as well as the much weaker lines, 
\carboniia\ 6581.2 \AA, \carboniib\ 6578~\AA, \nitrogen\ 4996.8~\AA, \ironi\ + \ironii\ 5001.9~\AA\ or \nitrogen\ 5002.7~\AA, 
\nitrogen\ 5005.2~\AA\ and \ironii\ 5018.2 or \helium\ 5017.2~\AA\ are also present (see Figs 4 to 8).

By conducting a literature search for these emission lines in well known objects, we found a small number of PNe 
(e.g  NGC~6543, Hyung et al. 2000; M~2--9, Lee H.--W. et al. 2001) and symbiotic stars (RR Telescopii and He~2--106,  
Lee H.--W. et al. 2001) that exhibit the 6581~\AA\ emission line. However, those objects also exhibit the 6545~\AA\ line,
which usually is attributed to the Raman scattered \heliumb\ line that is not detected in our data. 
Both the 6578~\AA\ and 6581~\AA\ lines have also been detected in the symbiotic star AE Circinus, and identified 
as carbon lines (Mennickent et al. 2008). The 5002 \AA\ line has been detected in the J320 (PNG190.3−17.7) nebula but 
no identification was provided (Harman et al. 2004).  

The PV diagrams clearly show the presence of a high--speed component (wide wings) with velocities up to 90--100~\kms, 
and a much slower component $\sim$20~\kms\ (Gaussian fit). The \ha\ 6563~\AA, \nitrogen\ 6584~\AA\ and \oxygeniii\ 5007~\AA\ emission 
line profiles, in Fig.~9 to 12, are from the slit positions 2 to 5, respectively. Single Gaussian fit to these line profiles 
gives turbulent widths (FWHM) of 19$\pm$3~\kms\ in the \ha, between (8--9)$\pm$2~\kms\ in the \nitrogen, and $18\pm$3~\kms\ 
in the \oxygeniii\ emission (see Table~2). These expansion velocities correspond to the bright, inner ring--like structure. 
Two separated blobs can be discerned in the \nitrogen\ 6584 \AA\ PV diagram from the slit position 4 (Fig. 6; right, lower panel), 
similar to those of a pole--on toroidal or ring--like structure, whilst high expansion velocities are not apparent, since this 
slit position covers only a small part of the bipolar lobes. All the widths are corrected: for the 20~\kms\ instrumental width; 
the 21.4~\kms\ thermal width of the \ha\ line at 10000~K (5.8~\kms and 5.4~\kms of the \nitrogen\ and \oxygeniii\ lines respectively); 
and the fine structural broadening of 6.4~\kms\ of the \ha\ line. Note that, unlike the \ha\ line, the \nitrogen\ and \oxygeniii\ 
lines have no fine structural components. The \heliumb\ expansion velocity is estimated (23--25)$\pm$4~\kms\ from the \heliumb\ 
6560~\AA\ recombination line, and is corrected for the instrumental broadening of 20~\kms, the thermal broadening of 10.7~\kms\ 
(10000~K), and the fine structural broadening from this line at $\sim$6~\kms\ (Meaburn et al. 2005). 

Additional to the previous low velocity components, the line profiles of Vy~1--2 also present broad wings (Figs. 9--12). 
This continuous high--velocity component likely corresponds to the expansion of the faint bipolar lobes seen in the HST images.
$\rm{V_{[O {\sc\ III}]}}$ is between 90$\pm$4~\kms and 100$\pm$4~\kms, measured at the base of line profiles for all the P.A. 
The \nitrogen\ emission line shows such high expansion velocity of 90--100~\kms\ only along the P.A.=310$^{\circ}$, whilst for the 
rest of the P.A.s, $\rm{V_{[N {\sc\ II}]}}$ is significantly lower than $\rm{V_{[O {\sc\ III}]}}$. This is in agreement 
with the Winberger's catalogue (Weinberger 1989). A similar velocity trend, $\rm{V_{[O {\sc\ III}]}}$ $>$ $\rm{V_{[N {\sc\ II}]}}$, 
has also been reported for a small group of PNe (see \S 4.1). Regarding the HST images and PV diagrams, we conclude that Vy 1-2 is 
seen almost pole--on with an inclination angle of 10$^{\circ}$ $\pm$2$^{\circ}$ to the line of sight.  
The average systemic velocity of the nebula is $V_{\rm{sys}}$=-85~\kms, with a standard deviation of 4~\kms, in good agreement 
with the previous value of -82.4$\pm$3.9\kms\ quoted by Schneider et al. (1983). 

Figure 8 shows the PV diagram obtained along P.A.=305$^{\circ}$ in \ha+\nitrogen, drawn from the recently 
released SPM Kinematic Catalogue of Galactic Planetary Nebulae (L\'opez et al. 2012). 
This spectrum covers, apart from the bright ring--like structure, the faint knot N (Fig. 1). The \ha\ line profile of this 
knot is centred on $V_{\rm{ HEL}}$=-80$\pm$4~\kms. This central value should be compared to the systemic heliocentric 
radial velocity of the whole nebula, $V_{\rm{sys}}$=-85$\pm$4~\kms.

\begin{table}
\centering
\caption[]{Observed and modelled emission line intensities for Vy~1--2, corrected for 
interstellar extinction. The lines outside the wavelength 
range 4700--6800~\AA\ were drawn from WLB05.}
\label{table4a}
\begin{tabular}{llllllllllll}
\hline
Line(\AA)    & I($\lambda$) &  error &   model &  WLB05\\ 
\hline
\oxygenii  \ 3726 &    $-$  &    $-$ &   17.69 &   20.04\\
\oxygenii  \ 3729 &    $-$  &    $-$ &    9.57 &    9.97\\
\neon      \ 3869 &    $-$  &    $-$ &   98.86 &   94.55\\
\neon      \ 3968 &    $-$  &    $-$ &   29.81 &   29.91\\
\helium    \ 4026 &    $-$  &    $-$ &    2.05 &    2.19\\
\sulfurt   \ 4070 &    $-$  &    $-$ &    1.49 &    1.21\\
\hc        \ 4100 &    $-$  &    $-$ &   25.89 &   24.44\\
\hdelta    \ 4340 &    $-$  &    $-$ &   46.81 &   45.30\\
\oxygeniii \ 4363 &    $-$  &    $-$ &    9.12 &    8.51\\
\helium    \ 4471 &    $-$  &    $-$ &    4.39 &    4.40\\
\heliumb   \ 4686 &    $-$  &    $-$ &   26.85 &   31.00\\ 
\helium    \ 4712 &    $-$  &    $-$ &    0.54 &    0.52\\
\argon     \ 4712 &    3.89 &   0.43 &    3.38 &    4.29\\
\argon     \ 4740 &    3.97 &   0.48 &    3.72 &    4.22\\
\hbeta     \ 4861 &  100.00 &      4 &  100.00 &  100.0 \\
\helium    \ 4924 &    1.38 &   0.34 &    1.18 &    1.15\\
\oxygeniii \ 4959 &  434.85 &     13 &  497.21 &  443.10\\
\nitrogen  \ 4997$^b$ &    $-$  &   $-$  &     $-$ &   $-$ \\
\ironii    \ 5002$^b$ &    $-$  &   $-$  &     $-$ &   $-$ \\
\nitrogen  \ 5003$^b$ &    $-$  &   $-$  &     $-$ &   $-$ \\
\nitrogen  \ 5005$^b$ &    $-$  &   $-$  &     $-$ &   $-$ \\
\oxygeniii \ 5007 & 1287.06 &     38 & 1496.59 & 1337.0\\
\helium    \ 5015$^b$ &    $-$  &   $-$  &     $-$ &   $-$ \\
\helium    \ 5017$^b$ &    $-$  &   $-$  &     $-$ &   $-$ \\
\ironii    \ 5018$^b$ &    $-$  &   $-$  &     $-$ &   $-$ \\
\silicon   \ 5040${^a}$ &  0.25 &    0.09 &     $-$ &   $-$\\
\nitrogena \ 5200 &    0.22 &   0.07 &    0.00 &   $-$\\
\heliumb   \ 5411 &    2.15 &   0.19 &    1.95 &   $-$\\
\chloro    \ 5517 &    0.56 &   0.11 &    0.52 &   $-$\\
\chloro    \ 5537 &    0.64 &   0.16 &    0.61 &   $-$\\
\oxygeni   \ 5577${^a}$ &    0.05 &  0.03 &    0.001 &   $-$\\
\nitrogen  \ 5678 &    0.13 &   0.04 &    0.00 &   $-$\\ 
\nitrogen  \ 5755 &    0.53 &   0.11 &    0.42 &   $-$\\
\helium    \ 5876 &   13.16 &   0.71 &   13.11 &   13.10\\
\kripto    \ 6101${^a}$ &    0.23 &   0.06  &    $-$  &   $-$\\
\heliumb   \ 6119 &    0.11 &   0.04 &    0.06 &   $-$\\
\heliumb   \ 6172 &    0.08 &   0.04 &    0.06 &   $-$\\
\heliumb   \ 6234 &    0.08 &   0.04 &    0.07 &   $-$\\
\oxygeni   \ 6300 &    2.02 &   0.14 &    0.06 &  1.70\\
\sulfur    \ 6312 &    1.45 &   0.15 &    5.09 &  1.30\\
\oxygeni   \ 6363 &    0.69 &   0.09 &    0.02 &  0.71\\
\heliumb   \ 6405 &    0.11 &   0.04 &    0.11 &   $-$\\
\argonV    \ 6435 &    0.17 &   0.06 &    0.43 &   $-$\\
\heliumb   \ 6525 &    0.17 &   0.06 &    0.13 &   0.5\\
\nitrogen  \ 6548 &    8.42 &   0.45 &    8.75 &   9.16 \\
\heliumb   \ 6560$^b$ &    $-$  &   $-$  &     $-$ &   $-$ \\
\ha        \ 6563 &  284.62 &     10 &  287.45 & 284.1\\
\carboniib  \ 6578$^b$ &    $-$  &   $-$  &     $-$ &   $-$ \\
\carboniia  \ 6581$^b$ &    $-$  &   $-$  &     $-$ &   $-$ \\
\nitrogen  \ 6584 &   26.23 &   1.32 &   25.81 &  22.68\\
\helium    \ 6678 &    3.79 &   0.31 &    3.59 &   3.89\\
\sulfurt   \ 6716 &    2.49 &   0.28 &    2.61 &   2.04\\
\sulfurt   \ 6731 &    4.13 &   0.32 &    4.69 &   2.47\\
\argonV    \ 7005 &    $-$  &    $-$ &    0.95 &   0.61\\
\helium    \ 7065 &    $-$  &    $-$ &    4.39 &   3.96\\
\argoniii  \ 7135 &    $-$  &    $-$ &   12.51 &  13.83\\
\helium    \ 7281 &    $-$  &    $-$ &    0.70 &   0.50\\
\argoniii  \ 7751 &    $-$  &    $-$ &    3.01 &   2.61\\       
\hline
logF(H$\beta$)    & -11.56  &  ----  &  -11.56& -11.53\\
\hline
\end{tabular}
\begin{flushleft}
\end{flushleft}
\end{table}

\begin{table}
\centering
\addtocounter{table}{-1}
\caption[]{Continued}
\label{table4a}
\begin{tabular}{llllllllllll}
\hline
Line(\AA)    & I($\lambda$) & error & model &  WLB05\\ 
\hline
\oxygeniii \ 5007/4363      & $-$   & $-$   & 164  & 157\\
\nitrogen  \ 6581/5755      & 49.5  & 10.4  & 61.4 & $-$\\
\oxygeni   \ 6300+6363/5577 & 54.2  & 28.1  & 80.0 & $-$ \\ 
\oxygenii  \ 3727/3729      & $-$   & $-$   & 1.85 & 2.01\\
\sulfurt   \ 6716/6731      & 0.60  & 0.12  & 0.56  & 0.83 \\
\argon     \ 4712/4740      & 0.98  & 0.27  & 0.92  & 1.01 \\    
\chloro    \ 5517/5537      & 0.88  & 0.36  & 0.85  & $-$ \\
\hline
J          \ 1.25$^c$      & 3.5$^d$ & 0.2$^d$ & 2.4    & $-$ \\
H          \ 1.65          & 2.0     & 0.1 & 2.3   & $-$ \\
K          \ 2.15          & 2.6     & 0.1 & 2.8   & $-$ \\
WISE       \ 3.35          & 3.4     & 0.1 & 2.5   & $-$ \\
WISE       \ 4.60          & 5.6     & 0.1 & 2.2   & $-$ \\
AKARI      \ 9.00          & 104     & 21  & 65    & $-$ \\
WISE       \ 11.6          & 164.    & 21  & 110   & $-$ \\
IRAS       \ 12.0          & 113     & 22  & 211   & $-$ \\
AKARI      \ 18.0          & 786     & 21  & 893   & $-$ \\
WISE       \ 22.1          & 1319    & 15  & 1804  & $-$ \\
IRAS       \ 25.0          & 1647    & 83  & 2361  & $-$ \\
IRAS       \ 60.0          & 2078    & 104 & 2122  & $-$ \\
AKARI      \ 65.0          & 1704    & $-^e$ & 1795  & $-$ \\
AKARI      \ 90.0          & 1184    & 58  & 796   & $-$ \\
IRAS       \ 100.0         & 1094    & 219 & 592   & $-$ \\
AKARI      \ 140.0         & 697     & $-^e$ & 218   & $-$ \\
\hline
\end{tabular}
\begin{flushleft} 
${^a}$ These lines were not used to constrain the model.\\
${^b}$ Emission lines detected in the high--dispersion spectra\\ 
${^c}$ In $\mu$m, ${^d}$ In mJy\\
${^e}$ No errors are provided for these measurements since the flux quality indicator is 1.\\
\end{flushleft}
\end{table}

\subsection{Ionisation structure}

\subsubsection{Nebular parameters and ICF abundances}
The panels in Figure 13 display the low--dispersion spectra of Vy~1--2 that cover the wavelength 
range of 4700--6800~\AA\ for exposure times of 3600 (upper panel) and 300~sec (lower panels), respectively. 
Several emission lines from low--to--high and very high ionisation states are present and some are 
detected for the first time in this nebula. Besides the common nebular lines, such as \ha, \hbeta, \helium, 
\heliumb, \oxygeniii, \nitrogen\ and \sulfurt\ lines, the \chloro\ 5517 and 5537~\AA, \carboniib\ 6461~\AA, 
\carboniv\ 5801 and 5812~\AA, \silicon\ 5040~\AA, \kripto\ 6101~\AA\ and \nitrogen\ 5679~\AA\ emission 
lines are also detected. 

It is worth mentioning here that the emission line at 5679~\AA\ has been 
identified either as the \nitrogen\ line by Liu et al. (2000; NGC~6135, 2001; M1--42 and M2--36) 
and Ercolano et al. (2004; NGC~1501), or as \ironvi\ line by Hyung \& Aller (1997). However, the latter 
classification requires, a very hot star of $\rm{T_{eff}}\geq$140~kK, which is not in agreement with the 
values previously published ($\rm{T_{eff}}$=119~kK, Stangellini et al. 2002; 75.4~kK$<\rm{T_{eff}}<$99~kK, Phillips 2003). 

All very faint emission lines presented here are only detected in our deep spectrum ($\rm{t_{exp}}$=3600~sec). 
Notice that WLB05 did not detect any of these lines due to their shallow spectroscopy.

In Table 3, we list the optical nebular emission line fluxes, corrected for 
atmospheric extinction and interstellar reddening. All line fluxes are normalized 
to F(\hbeta)=100. The interstellar extinction c(\hbeta) was derived 
from the Balmer \ha /\hbeta\ ratio (Eq. 1), using the interstellar
extinction law by Fitzpatrick (1999) and $R_v$=3.1,

\begin{equation}
c(H\beta)=\frac{1}{0.348}log\frac{F(H\alpha)/F(H\beta)}{2.85},
\end{equation}
 
\noindent where the 0.348 is the relative logarithmic extinction coefficient
for \hbeta /\ha. The observed reddening in magnitude E(B-V) was also 
calculated using the following relationship (Seaton 1979),

\begin{equation}
c(H\beta)=0.4X_{\beta}E(B-V),
\end{equation}

\begin{figure*}
\includegraphics[scale=0.48]{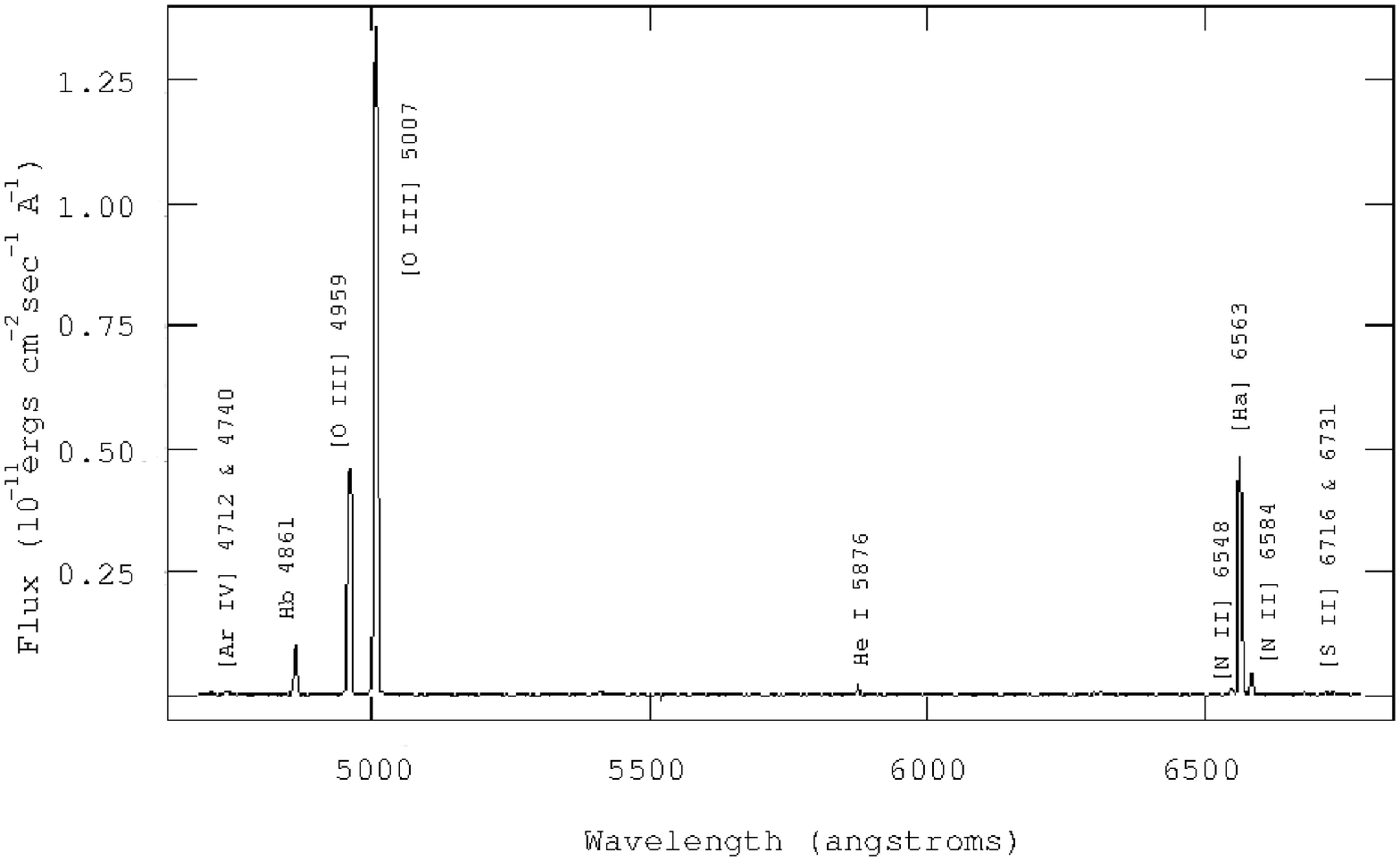}
\includegraphics[scale=0.48]{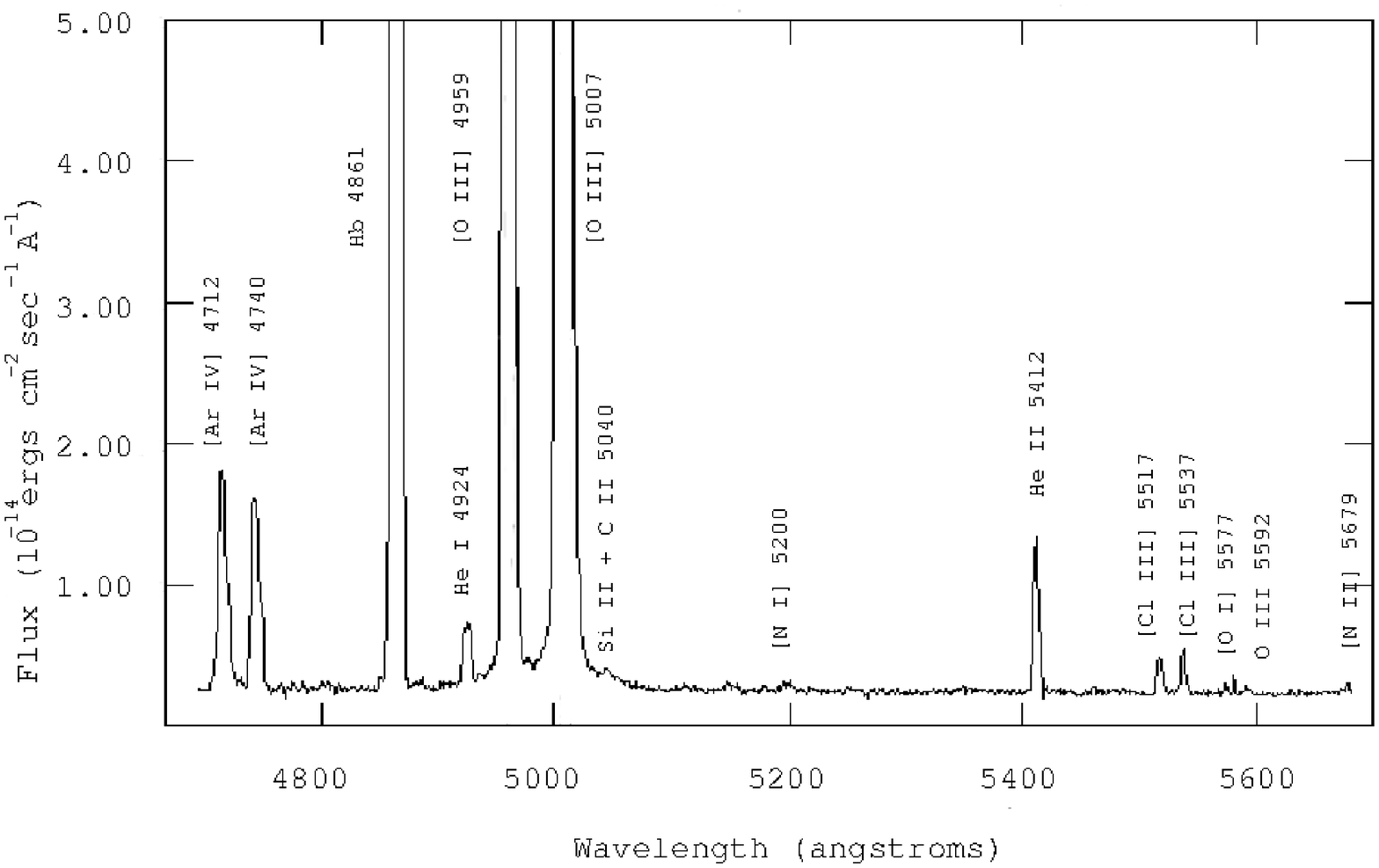}
\includegraphics[scale=0.48]{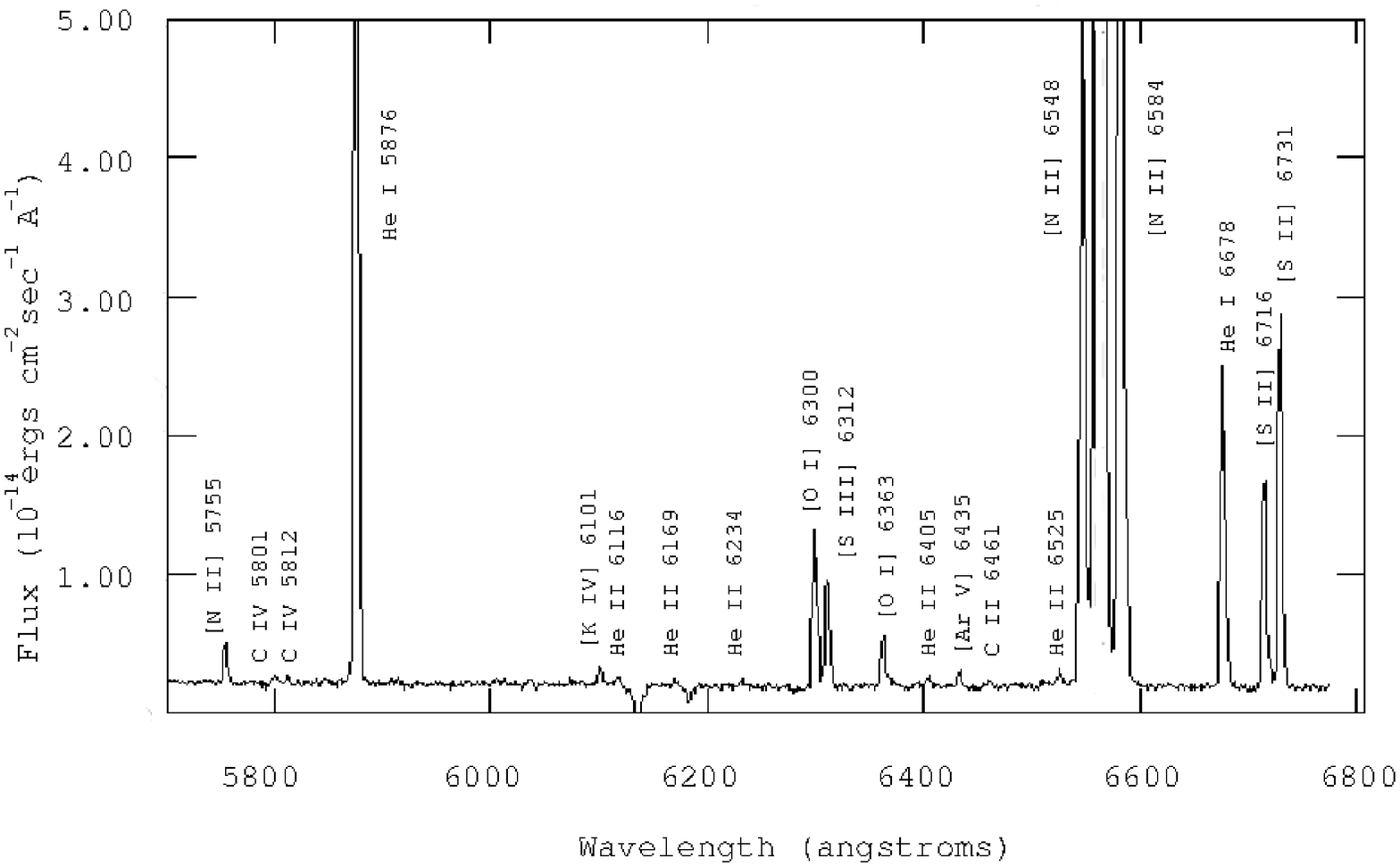}
\caption{Observed spectra of Vy~1--2 taken with the 1.3--m telescope. Upper panel 
displays the spectrum with an exposure time of 300s, covering the wavelength range of 
4700--6800~\AA. Lower panels display two enlargements of the spectrum (left: 4700--5700~\AA, 
right: 5700--6800~\AA) with an exposure time of 3600 sec, in order to highlight the 
weaker emission lines.}
\label{fig13}
\end{figure*}

\noindent where the extinction parameter $\rm{X_\beta}$ = 3.615 (Fitzpatrick 1999).
The errors of the extinction c(\hbeta) and E(B-V) are calculated through
standard error propagation of equations (1) and (2). We obtain 
c(\hbeta)=0.13$\pm$0.03, in agreement with the value quoted by WLB05 (c(\hbeta)=0.139)  
and E(B-V)=0.09$\pm$0.02. Our logarithm \hbeta\ flux -11.56$\pm$0.03 is also in very good 
agreement with previous studies (-11.53, WLB05; -11.52, Bar87a) and the recently published catalogue 
of integrated \ha\ fluxes of Galactic PNe (Frew et al. 2013). Notice that the interstellar 
extinction c(\hbeta) of Vy~1--2, given by Frew et al. (2013) is significantly lower than the previous values.

\begin{table}
\centering
\caption[]{Electron temperature (in K) and electron density (in $\rm{cm^{-3}}$) of Vy~1--2.}
\label{table5}
\begin{tabular}{llllllllllll}
\hline 
                         & this work           & WLB05   & Bar78a &McNabb$^{\dagger}$\\   
\hline
$\rm{T_e}$({\oxygeniii}) &  $-$                    & 10400  &  9800  &$-$\\ 
$\rm{T_e}$(\nitrogen)    & 10850$\pm$2100         & $-$    &   $-$   &$-$\\   
$\rm{T_e}$(\oxygenii)    &  $-$                    & 16179  &   $-$  &7943\\  
$\rm{T_e}$(\oxygeni)     & 10400$\pm$5400          & $-$    &   $-$  &$-$\\    
$\rm{N_e}$(\sulfurt)     & 4500$\pm$650            & 1160   &  5000  &$-$\\ 
$\rm{N_e}$(\chloro)      & 4330$\pm$1600           & $-$    &   $-$  &$-$\\ 
$\rm{N_e}$(\argon)       & 4550$\pm$1300$^{\ddagger}$ & 3300   &   $-$ &$-$\\ 
$\rm{N_e}$(\oxygenii)    &  $-$                    & 4100   &   $-$  &4466\\ 
\hline
\end{tabular}
\medskip{}
\begin{flushleft}
${\dagger }$  T$_e$\oxygenii\ and N$_e$\oxygenii\ estimated by McNabb et al. are obtained using 
the \rm{O~{\sc ii}} optical recombination lines.\\
${\ddagger }$  The contribution of \helium\ 4712 \AA\ emission line to \argon\ 4711 \AA\ 
line, was calculated based on the work of  Benjamin et al. (1999).\\
\end{flushleft}
\end{table}

\begin{table}
\centering
\caption[]{Ionic and total (${\rm}$ $\epsilon$(X)=log(n(X)/n(H))+12) 
abundances derived from the observed fluxes of Vy~1--2.}
\label{table5}
\begin{tabular}{llllllllllll}
\hline
ionic                        &this work         & this work & WLB05     & Bar78b \\
abundances                    &  (ICF)           & (model) &            &\\
$\rm{He^{+}/H^{+}}$             & 0.087$\pm$0.008 & 0.083   &  0.084    & 0.076 \\
$\rm{He^{++}/H^{+}}$            & 0.022$\pm$0.004 & 0.024   &  0.023    & 0.019 \\
ICF(He)                        & 1.00             & 1.00    &  1.00     & 1.0   \\
$\rm{\epsilon(He)}$            & 11.04$\pm$0.03   & 11.03   & 11.03     & 10.98 \\
\\
$10^5\times\rm{O^{+}/H^{+}}$     & 1.55$\pm$0.17$^a$ & 2.35   &   1.36    & 2.51 \\
$10^5\times\rm{O^{2+}/H^{+}}$    & 37.74$\pm$2.78    & 66.11 &  41.55    & 47.1\\
ICF(O)                         & 1.17              & 1.19  &   1.18    & 1.25 \\ 
$\rm{\epsilon(O)}$             & 8.66$\pm$0.04     & 8.91  &   8.70    & 8.79   \\ 
\\
$10^5\times\rm{N^{+}/H^{+}}$    & 0.47$\pm$0.05     & 0.62  &   0.395   & 0.435 \\
$10^5\times\rm{N^{2+}/H^{+}}$   & $-$               & $-$   &   9.300   & $-$    \\
ICF(N)                         & 29.28            & 41.45 &   1.206   & 24.8   \\
$\rm{\epsilon(N)}$             & 8.13$\pm$0.05    & 8.43  &   8.06    & 8.03  \\ 
\\
$10^6\times\rm{S^{+}/H^{+}}$     & 0.24$\pm$0.04   & 0.38   & 0.144     & 0.385 \\
$10^6\times\rm{S^{2+}/H^{+}}$    & 2.71$\pm$0.37   & 12.19  & 2.467     & $-$  \\
ICF(S)                         & 2.17            & 1.86    & 2.334     & 24.8 \\
$\rm{\epsilon(S)}$             & 6.80$\pm$0.08   & 7.37   &  6.78     & 6.98 \\ 
\\
$10^8\times\rm{Cl^{2+}/H^{+}}$   & 7.09$\pm$1.97   & 10.07  &  $-$      & $-$ \\
ICF(Cl)                        & 2.35            & 2.03    &  $-$      & $-$ \\
$\rm{\epsilon(Cl)}$            & 5.22$\pm$0.12   & 5.31   &  $-$      & $-$ \\ 
\\
$10^7\times\rm{Ar^{2+}/H^{+}}$  & 11.23$\pm$1.35$^a$ & 13.03 &   9.370   & $-$ \\
$10^7\times\rm{Ar^{3+}/H^{+}}$  & 6.26$\pm$1.43      & 10.34 &   6.902   & $-$  \\
$10^7\times\rm{Ar^{4+}/H^{+}}$  & 0.75$\pm$0.26      & 1.09  &   1.005   & $-$ \\
ICF(Ar)                        & 1.04              & 1.01   &   1.035   & $-$ \\
$\rm{\epsilon(Ar)}$            & 6.28$\pm$0.06     & 6.39  &   6.25    & $-$ \\ 
\hline
\end{tabular}
\medskip{}
\begin{flushleft}
${^a }$ The ionic abundances of these ions were calculated using the intensities from WLB05, 
considering an error of 10\%, since no errors are given in WLB05.\\ 
\end{flushleft}
\end{table}

The nebular electron temperature ($\rm{T_e}$) and density ($\rm{N_e}$) are derived using the {\sc TEMDEN} task in {\sc IRAF} 
(Shaw \& Dufour 1994). From the \nitrogen\ (6584+6548)/5755 and \oxygeni\ (6300+6363)/5577 diagnostic line ratios, we obtain 
$\rm{T_e}$= 10850$\pm$2100~K and 10400$\pm$5400~K, respectively, whilst $\rm{N_e}$ are calculated as 4500$\pm$650~$\rm{cm^{-3}}$, 
4330$\pm$1600~$\rm{cm^{-3}}$ and 4550$\pm$1300~$\rm{cm^{-3}}$ from the \sulfurt\ 6716/6731, \chloro\ 5517/5537 and \argon\ 4711/4740 
diagnostic line ratios, respectively. No evidence of density variation with the ionisation potential is found (see also Fig. 14). 
For direct comparison, Table 4 lists the $\rm{N_e}$ and $\rm{T_e}$ together with those derived in previous studies from collisionally 
excited and recombination lines.

Regarding the N$_e$[Ar], it should be noticed that the \argon\ 4711~\AA\ line is blended 
with the \helium\ 4712~\AA, \neoniv\ 4724~\AA\ and 4726~\AA\ lines. Therefore, to estimate the correct $\rm{N_e}$, 
the contribution of the latter to \argon\ 4711~\AA\ was calculated and subtracted. The theoretical value of \helium\ 
4712~\AA /\helium\ 4471~\AA\ is 0.146 for $\rm{T_e}$=10000~K, $\rm{N_e}$=10000~$\rm{cm^{-3}}$ 
and 0.105 for $\rm{T_e}$=10000~K, $\rm{N_e}$=1000~$\rm{cm^{-3}}$ (Benjamin et al. 1999). The observed intensity 
of \helium\ 4471~\AA\ is 4.403 (relative to $\rm{H_{\beta}}$=100; WLB05). This value gives I(4712) equal 
to 0.64 and 0.46 for $\rm{N_e}$=10000~$\rm{cm^{-3}}$ and $\rm{N_e}$=1000~$\rm{cm^{-3}}$, respectively. 
The observed intensity of \helium\ 4712 \AA\ is 0.519 (WLB05) and  lies between the 
theoretical lower and upper limits, as it is expected, given that $\rm{N_e}$= 4330~$\rm{cm^{-3}}$ also lies 
between the equivalent lower (1000~$\rm{cm^{-3}}$) and upper (10000~$\rm{cm^{-3}}$) limits. Our best--fit photo--ionisation 
model predicts I(4712)=0.53 (see \S 4.2). The contribution of \neoniv\ was considered 
negligible since only a  star with $\rm{T_{eff}}$ $>$ 100~kK can significantly ionise
this line. WLB05 found the \neoniv/\hbeta\ equal to 0.216 or 5\% of the \argon\ 4711~\AA\ line.

The recently released software, namely PyNeb (Luridiana et al. 2011) was used to 
construct the $\rm{T_e}$--$\rm{log(N_e)}$ diagnostic plot for Vy~1--2 from our 
data and those from WLB05 (Fig. 14). This diagram illustrates the whole picture of the physical 
conditions in Vy~1--2 where a good solution for $\rm{T_e}$--$\rm{log(N_e)}$ is found; 
$\rm{T_e} \sim$11000~K and $\rm{log(N_e)} \sim$ 3.6~$\rm{cm^{-3}}$. Note that 
both high T$_e$\oxygenii\ (16200~K or 17110~K corrected for recombination excitation on the nebular lines) 
and low N$_e$\sulfurt\ (1160~$\rm{cm^{-3}}$) derived by WLB05 do not agree with this solution. 

Most of the emission lines in our spectra are in very good agreement with those reported by WLB05, except the 
\sulfurt\ 6731~\AA\ emission line, which surprisingly differs by a factor of 1.5 (see Table 3) and 
reflects on the significant N$_e$ difference. Our estimation of $\rm{N_e}$(\sulfurt)=4500~$\rm{cm^{-3}}$ agrees with 
the value reported by Baker (1978b) as well as the recent $\rm{N_e}$ determinations from the 
\rm{O~{\sc ii}} optical recombination lines (ORLs) by Mc Nabb et al. (2013). 
It is worth mentioning the high difference of order 2 in T$_e$ derived from the
\rm{O~{\sc ii}} optical recombination lines by WLB05 and McNabb et al. (2013).

The ionic abundances of N, O, S, Cl and Ar, relative to H$^{+}$, are derived 
using the ionic task in IRAF (Shaw \& Dufour 1994)\footnote{Transition probabilities: 
N$^+$, O$^+$ and O$^{2+}$, Wiese et al. (1996); S$^+$, Verner et al. (1996); S$^{+2}$, Mendoza 
\& Zeippen (1982); Cl$^{2+}$, Mendoza (1983); Ar$^{4+}$, Mendoza \& Zeippen (1982a); Ar$^{4+}$, 
Mendoza \& Zeippen (1982b). Collisional strengths:  N$^+$, Lennon \& Burke (1994); O$^+$, 
Pradhan (1976); O$^{2+}$, Lennon \& Burke (1994); S$^+$, Ramsbottom et al. (1996); S$^{+2}$, 
Galavis et al. (1995); Cl$^{2+}$, Butler \& Zeippen (1989); Ar$^{3+}$, Zeippen et al. (1987); 
Ar$^{4+}$, Galavis et al. (1995).}. $\rm{T_e}$=10600~K and $\rm{N_e}$= 4500~$\rm{cm^{-3}}$ are 
used for both low and high--ionisation regions. 

The $\rm{He^+/H^+}$ and $\rm{He^{++}/H^{+}}$ ionic abundance ratios are calculated 
based on the work of Benjamin et al. (1999). For the total chemical abundances we used the ionisation 
correction factor in order to correct for the unobserved states of ionisation (ICF; Kingsburgh \& Barlow 1994). 
Since the aforementioned authors do not provide an ICF for the Cl, we used the equivalent equation from Liu et al. (2000). 
In Table 5, we list the observed ionic and total abundances derived in this work, together with those derived by WLB05 
and Bar78b. For a direct comparison the ionic abundances from our best--fit model are also included in Table 5. 
The ICFs computed from our model are listed in Table~5. One can see that N shows the largest discrepancy in ICFs 
between the empirical method (Kingsburgh \& Barlow 1994) and our photo--ionisation model. This difference in the ICFs 
is not such high to significantly alter the N/O abundance ratio and the classification of the nebula as non-Type I, like 
in NGC 1501 (Ercolano et al. 2004).

\begin{figure}
\centering
\includegraphics[scale=0.45]{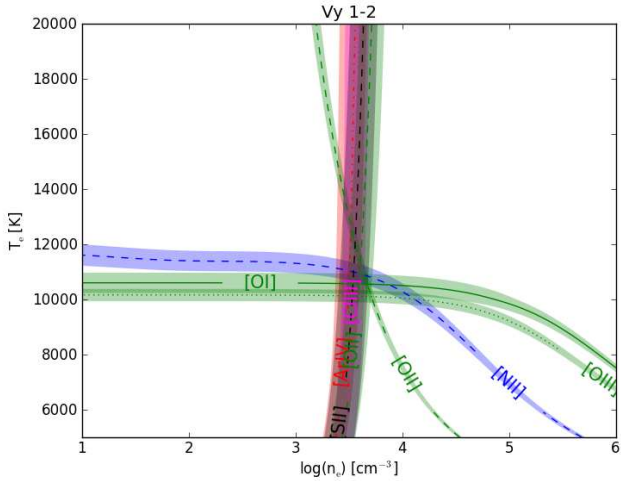}
\caption[]{$\rm{T_e}$--$\rm{log(N_e)}$ diagnostic plot for Vy~1--2.} 
\label{fig14}
\end{figure}

\subsubsection{Photo--ionisation modelling}

In order to perform a more assiduous study on the ionisation structure of Vy~1--2, we constructed a 
photo--ionisation model using Cloudy (version 13.02; Ferland et al. 2013). Several atomic, ionic 
and molecular databases are used within Cloudy (e.g. Dere at al. 1997; Badnell et al. 2003; Badnell 2006, Landi et al. 
2012; among others). The complete information can be found in Linkins et al. (2013) and Ferland et al. (2013).

Cloudy code uses as input parameters, (a) the energy distribution of the ionising radiation, (b) the luminosity of the central 
star and (c) some additional constraints that are usually estimated from the observed spectrum, which should as 
well be reproduced by the model such as the $\rm{T_e}$ and $\rm{N_e}$. Both parameters are required for determining 
the thermal balance of PNe. Because the configuration of the spectrograph does not include the lines of 
\oxygenii\ 3726~\AA\ and 3729~\AA, \neon\ 3869~\AA\ and 3968~\AA, \heliumb\ 4686~\AA, \argoniii\ 
7135~\AA, they were drawn from WLB05.

The model of Vy~1--2 was developed using the simplest set of assumptions (i) a blackbody approximation for the 
central star and (ii) a spherically symmetric nebula with constant density. The initial abundances of He, N, 
O, S, Ar and Cl were taken from our empirical analysis whereas Ne and C abundances from WLB05, and all of them were 
kept as free parameters, except C. Dust grains were also included in the model since they play an important role in 
the thermal balance of PNe. However, there is no information regarding the grain composition (no ISO spectra 
are available) and a mixture of Polycyclic Aromatic Hydrocarbons (PAHs; optical constants from Desert et al. 1990, 
Schutte et al. 1993, Li \& Draine 2001; for more detail see Ferland et al. 2013) and astronomical graphite and 
silicate (Martin \& Rouleau 1991; Laor \& Draine 1993) with a constant dust--to--gas ratio, was adopted. 
The amount of PAH, graphite and silicate, grains were also kept as free parameters. Furthermore, the filling factor was kept 
as a free parameter in order to reproduce the observed $\rm{H\beta}$ flux.

\begin{table*}
\centering
\caption[]{Total abundances, ${\rm}$ $\epsilon$(X)=log(n(X)/n(H))+12, of Vy~1--2.}
\label{table5}
\begin{tabular}{llllllllllll}
\hline
$\epsilon$(X)        &this work      & this work      & WLB05    & Bar78b      & Stan06 & Solar$^a$\\
               & (ICF)         & (model)        &          &             &        &          \\  
\hline
He & 11.04         & 11.03           & 11.03   & 10.98       & 11.23 & 10.93\\
C  & $-$           &  8.08$^b$       & 8.08$^b$ & $-$        & $-$    & 8.43\\
O  & 8.66          &  8.91           & 8.70    & 8.79        & 8.46  & 8.69\\
N  & 8.13          &  8.42           & 8.06    & 8.03        & 7.90  & 7.83\\  
S  & 6.80          &  7.37           & 6.78    & 6.98        & $-$   & 7.12\\
Ne & $-$           &  8.11           & 8.00    & 8.19        & 7.81  & 7.93\\
Ar & 6.28          &  6.39           & 6.25    & $-$         & 5.84  & 6.40\\
Cl & 5.22          &  5.31           & $-$     & $-$         & $-$   & 5.50\\
\hline
log(N/O)       & -0.53         & -0.49           & -0.64   & -0.76       & -0.56 & -0.86\\
log(S/O)       & -1.86         & -1.54           & -1.92   & -1.81       &  $-$  & -1.57\\
log(Ne/O)      &  $-$          & -0.80           & -0.70   & -0.60       & -0.65 & -0.76\\
log(Ar/O)      & -2.38         & -2.52           & -2.45   &  $-$        & -2.62 & -2.29\\
log(Cl/O)      & -3.44         & -3.60           &  $-$    &  $-$        &  $-$  & -3.19\\
\hline
\end{tabular}
\medskip{}
\begin{flushleft}
$^a$ Asplund et al. 2009 \\
$^b$ C abundance was derived from collisionally excited lines. A higher C abundance of $\epsilon$(C)=8.98 
was also derived from the optical recombination lines (WLB05). Vy 1--2 is a nebula with moderate high abundance 
discrepancy factors.\\
\end{flushleft}
\end{table*}

Given that C, O and N are the most important coolant elements in PNe, and they play a significant role in 
the thermal balance, small changes in their abundances can significantly affect the intensities of 
the emission lines of all elements. Hence, in order to get the best fit between the modelled 
and observed spectrum, we first varied the N and O abundances until the intensities 
of \nitrogen\ 6584~\AA\ and \oxygeniii\ 5007~\AA\ emission lines were fitted. 
Then, regarding the other elements, we varied the abundances of the rest of elements (He, S, Ar and Cl) up to the point 
that the intensities of \heliumb\ 5876~\AA, \sulfurt\ 6716, 6731~\AA, \argon\ 4712, 4740~\AA\ and \chloro\ 5517, 5538~\AA\ lines, 
were also fitted. The modelled spectrum of Vy 1-2 is presented in Table 3 together with the observed 
spectrum. The mean difference between the observed and modelled intensities is $\sim$10\% for the bright 
lines (S/N$>$5) and $\sim$20--30\% for the fainter lines (S/N$<5$). The \kripto\ 6101~\AA\ and \silicon\ 5040~\AA\ 
lines, as well as the stellar \carboniib\ 6461~\AA, \carboniv\ 5801 and 5812~\AA\ and the \oxygeniiib\ 5592~\AA\ lines, were 
not used to constrain the model.

As for the carbon, we assumed a constant abundance of log(C/H)+12=8.08, which was derived from the collisionally excited lines 
but it was not possible to get a good match of the CEL and ORL at the same time (very weak ORL), while the ionisation balance of C was far from the 
observed value. Assuming a higher C abundance (of 9.0), the model better predicts the ORL but overestimates the CEL.
In addition to the blackbody stellar model, a H--deficient stellar atmosphere model for the ionizing source was also used. 
A good match between the modelled and observed spectrum could be obtained only for the case of high C abundance ($\epsilon$(C)$>$9). 

In conclusion, whatever stellar atmosphere model was chosen for Vy 1-2, we did not fully reproduce the 
observed properties (e.g. the ionisation balance of the elements) to a satisfactory degree. 
In particular, the ionisation balance of all the elements is overestimated by the low C abundance model, 
whilst the higher C abundance model seems to produce an ionisation balances that matches better the observed ones.
Also, our simple spherically symmetric shell geometry seems to be far from the real structure of the nebula, which as 
discussed in \S~3.1 seems to have a bipolar structure or at least a significant deviation from a spherically symmetric morphology.

\section{Discussion}

\subsection{Morpho--kinematic analysis}

Vy~1--2 shows a high--speed component expanding with velocities up to 90--100~\kms, which is likely associated with the material 
in the faint bipolar lobes, with the eastern and western being blue-- and red--shifted, respectively.
Considering that these lobes are inclined by 10$^{\circ}$ with respect to the line of sight and their projected distance on 
the sky is 3.6\arcsec\ (measured from the eastern to the western lobe), we get a rough estimation of the nebula's size along the 
polar axis of 20.4\arcsec\ (r=10.2\arcsec). Given that the bipolar lobes have a de--projected radial expansion velocity 
of 90--100~\kms, their kinematic age is found to be 5200$\pm$750~years, for D=9.7~kpc.  

A pair of knots is marginally detected in the HST \oxygeniii\ image, located at 11.5\arcsec\ (knot S) and 
14.5\arcsec\ (knot N) from the central star. The projected expansion velocity of knot N differs from the $V_{\rm{sys}}$ 
of the nebula by 5~\kms. This converts to a de--projected radial velocity in a direction away from the central 
star of V=5/cos($\theta$)~\kms, where $\theta$ is the inclination angle of the knot to the line of sight. 
If we consider that the pair of knots is the result from the same ejection that formed the bipolar nebula (same age), 
its de--projected expansion velocity should be up to $\sim$130~\kms. Taking into account the projected velocity 
of knot N (5~\kms), as measured from the PV diagrams (Fig. 8), the inclination angle of the knots 
is found 89$^{\circ}$ with the respect to the line of sight. This suggests a possible precession of the putative central 
star system and multiple ejection events leading to the formation of a bipolar nebula.

A single Gaussian was fitted to the emission line profiles of \ha, \heliumb, \nitrogen\ and \oxygeniii\ and gave expansion 
velocities for the unresolved ring--like structure of 19$\pm$3~\kms, (23--25)$\pm$4~\kms, (8--9)$\pm$2~\kms\ and 18$\pm$3~\kms, respectively. 
For V$_{\rm{exp}}$=19~\kms and r=1.4\arcsec, we estimate the kinematic age of the ring--like structure $\sim$3500$\pm$500~years.
This means that the ring--like structure was formed more recently than the bipolar lobes, which is not feasible. 
If, although, the inclination angle of the bipolar lobes is slightly higher of 15$^{\circ}$, we get the same dynamical age for 
both component, ring--like and bipolar lobes, of 3500 years. 

The velocity trend, on which the low--ionisation gas (e.g. \nitrogen) expand slower than the high--ionisation gas--e.g. \oxygeniii, \heliumb, 
is not consistent with a typical homologous expansion law, where their expansion velocity increases linearly with the distance 
from the central star. Despite the fact that $\rm{V_{[O {\sc\ III}]}}$ and $\rm{V_{He {\sc\ II}}}$ could be roughly considered 
the same within the uncertainties, the much lower $\rm{V_{[N {\sc\ II}]}}$ provides strong evidence for a non--homologous expansion. 
A small group of PNe (e.g. BD +30\degree3639, Akras \& Steffen 2012; see also Medina et al. 2006) has been found to show this velocity trend. 
The central stars of those PNe are either [WR] or WEL--type stars (Medina et al. 2006). The acceleration of the inner zones of 
the nebula might be the result of strong, fast winds from a [WR] or WEL nucleus (Gesicki \& Zijlstra 2003). The detection of the \carboniib\ 
6461~\AA, \carboniv\ 5801 and 5812~\AA\ and \oxygeniiib\ 5592~\AA\ stellar lines in Vy~1--2 supports this 
scenario (see \S 4.3.2). Another possible scenario responsible for the acceleration of the inner nebular regions 
is a nova--type eruption of a close binary system. Akras \& Steffen (2012) propose this mechanism in order to 
explain the morpho--kinematic characteristics of the young BD +30\degree3639 nebula.

\subsection{Chemical analysis}

\begin{figure*}
\centering
\includegraphics[scale=0.35]{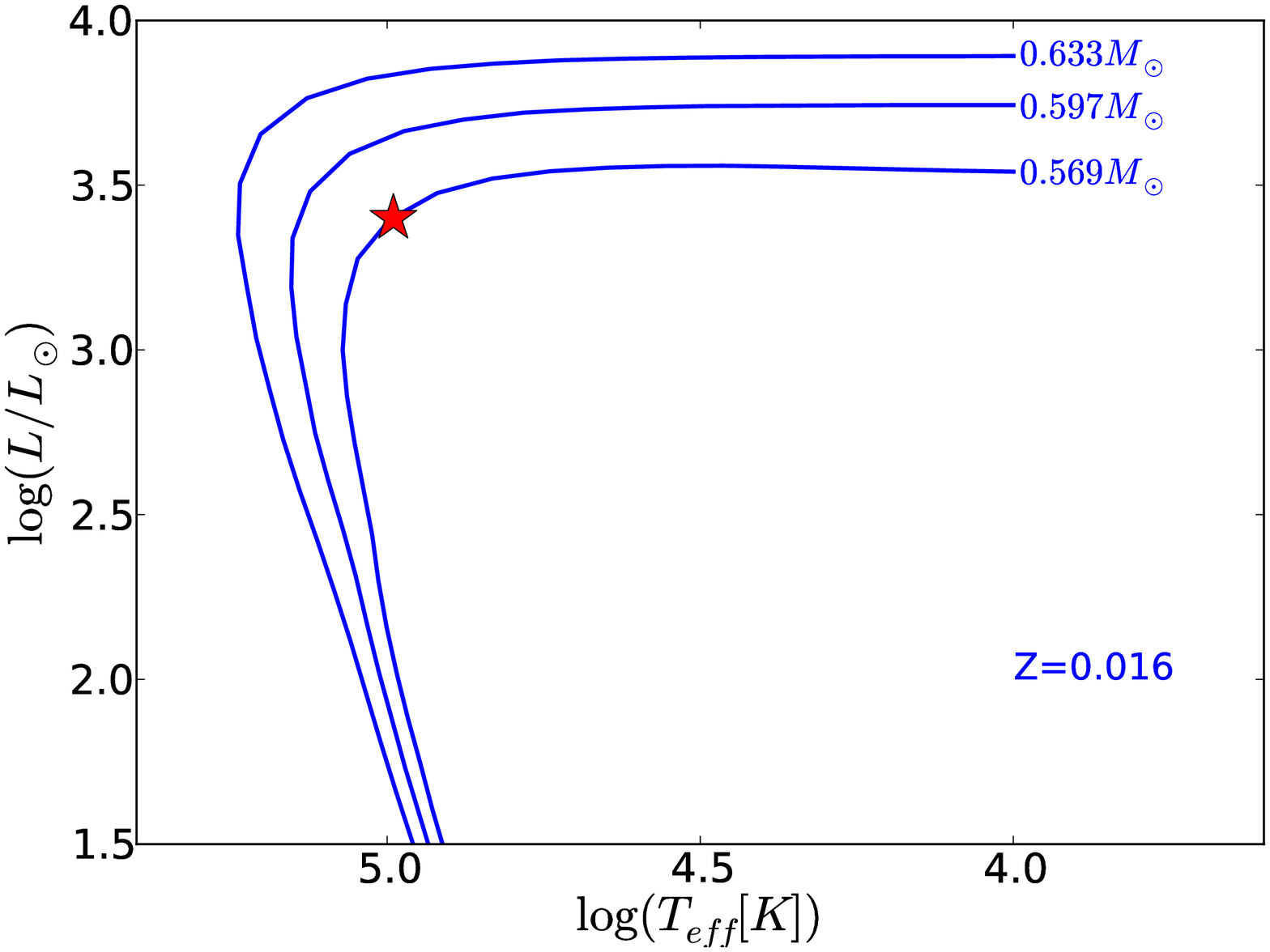}
\includegraphics[scale=0.35]{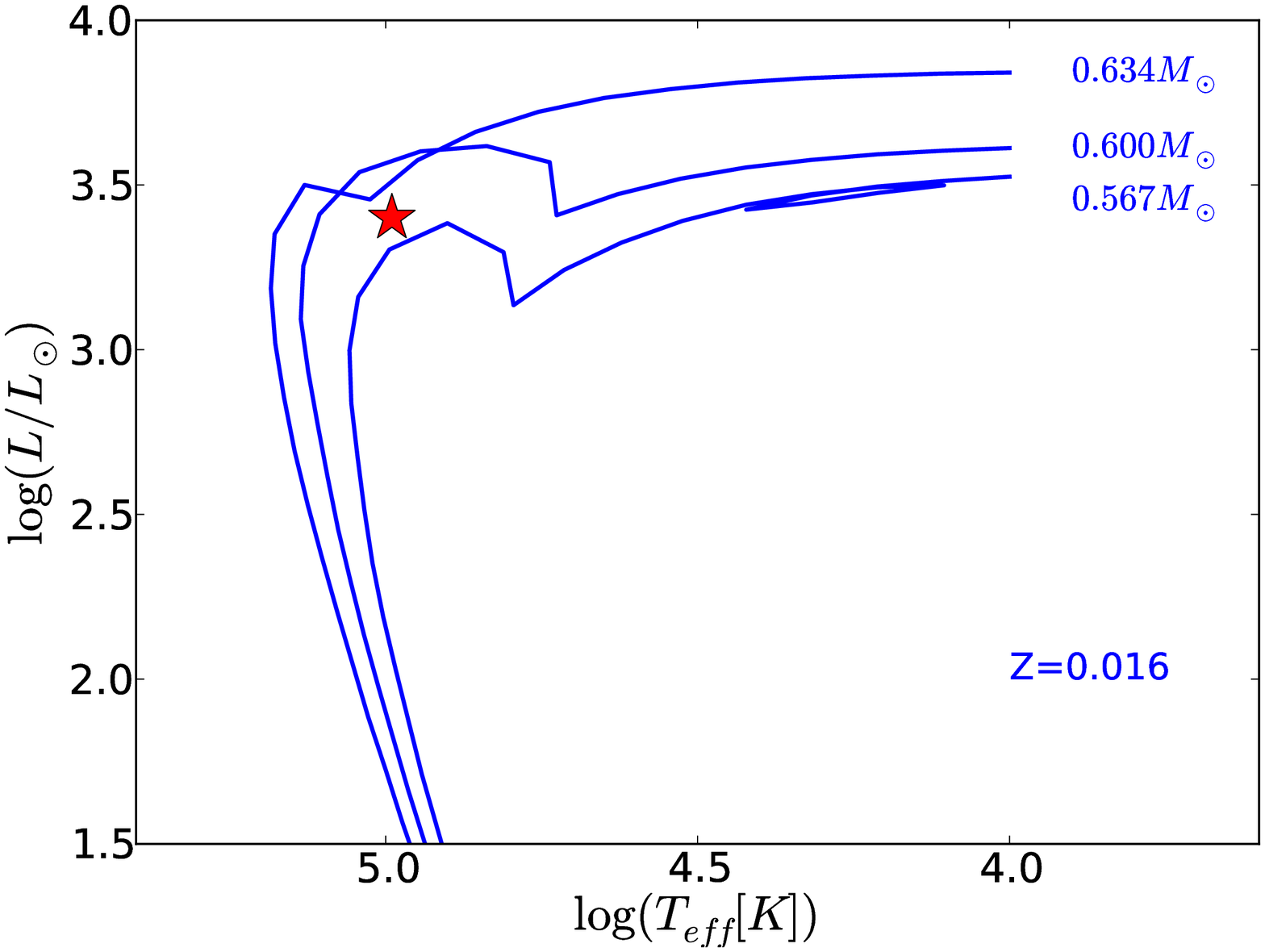}
\includegraphics[scale=0.35]{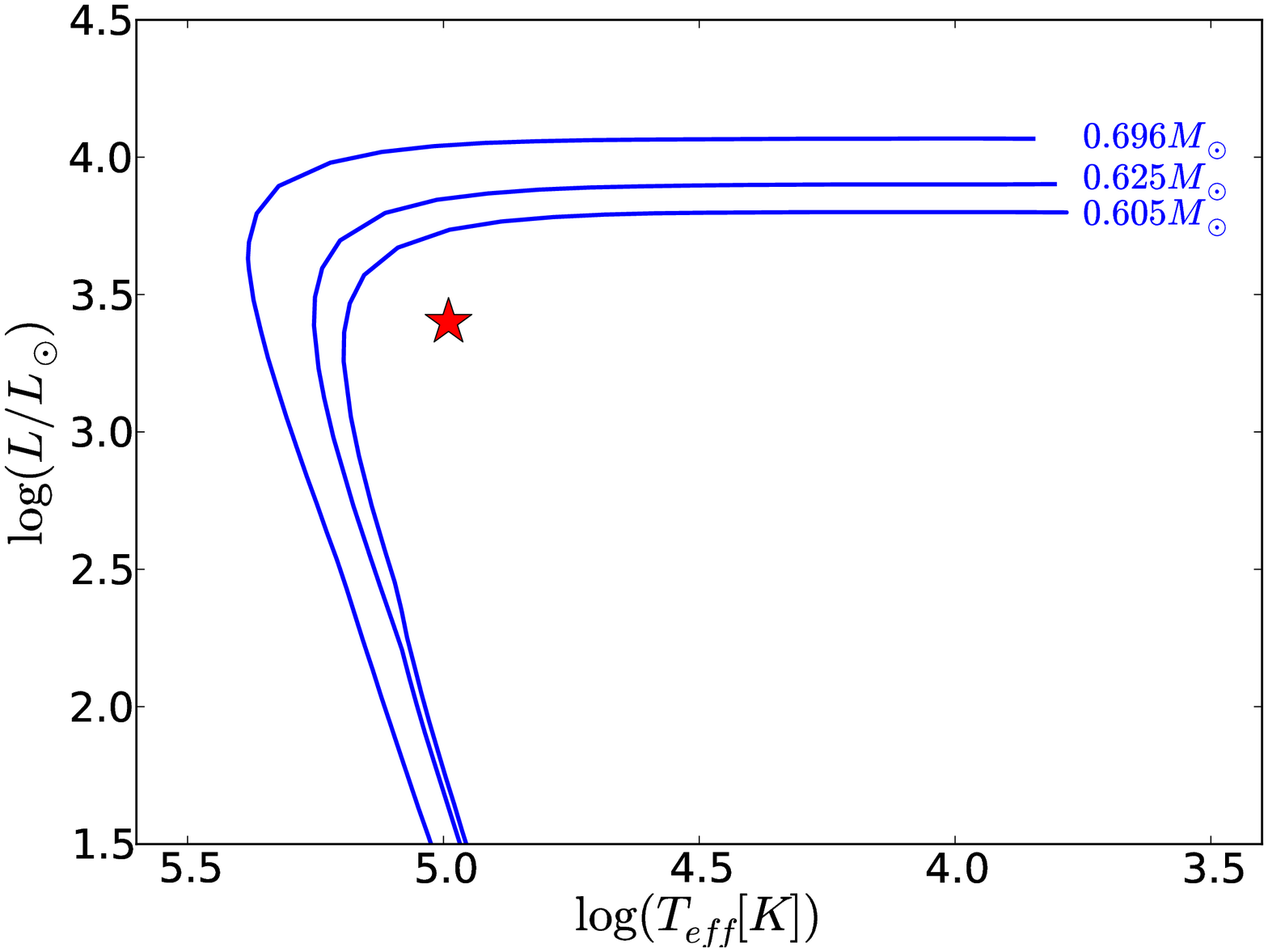}
\includegraphics[scale=0.35]{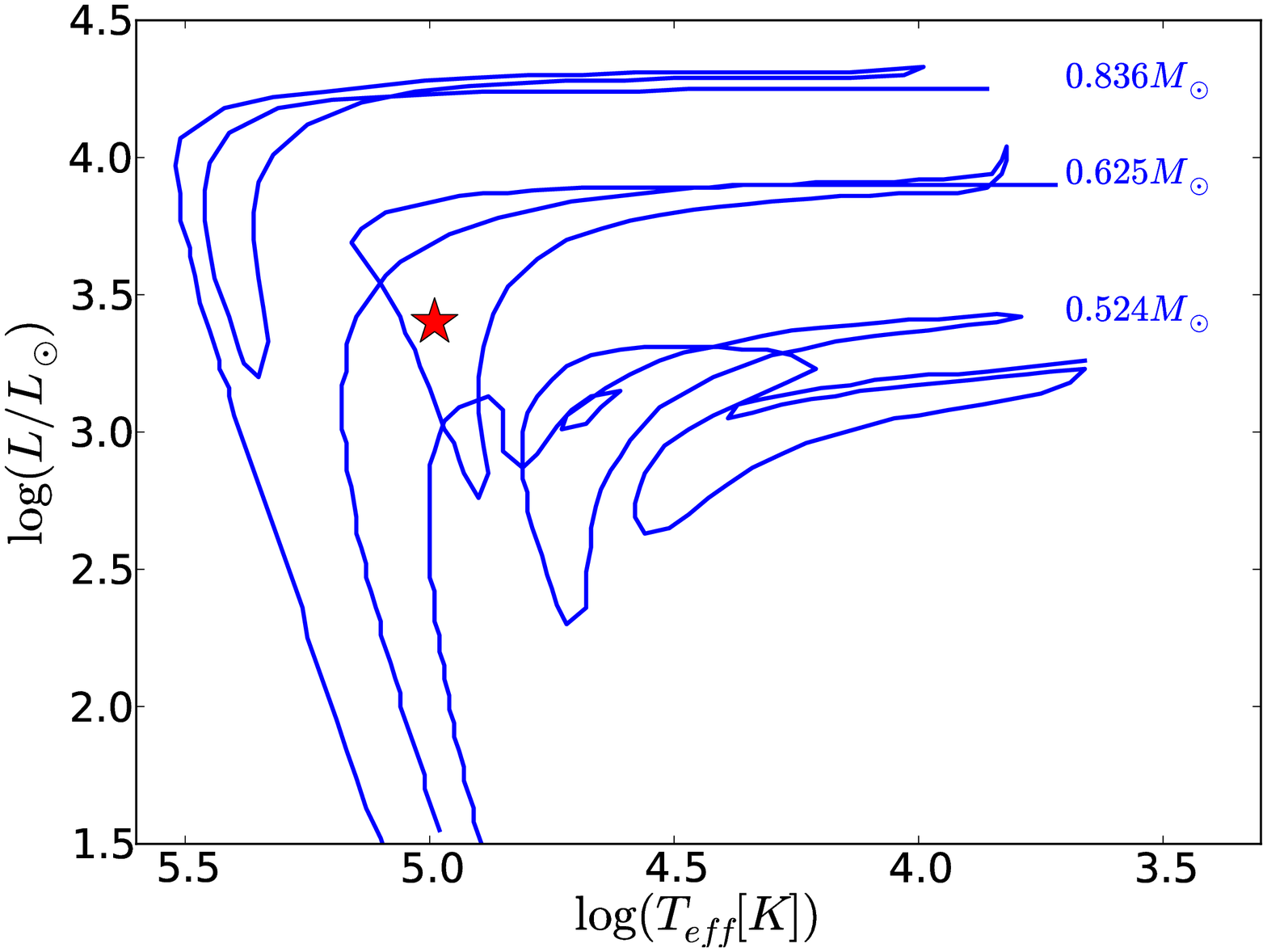}
\caption[]{Hertzprung--Russell diagrams for hydrogen--burning (left panels) and helium--burning (right panels) models.
Upper panels display the evolutionary tracks from Vassiliadis \& Wood (1994) produced by ($\rm{M_{initial}}$, $\rm{M_{final}}$) = 
(1$\rm{M_{\odot}}$, 0.569$\rm{M_{\odot}}$), (1.5$\rm{M_{\odot}}$, 0.597$\rm{M_{\odot}}$) and (2$\rm{M_{\odot}}$, 0.633$\rm{M_{\odot}}$) 
for the hydrogen--burning model (left panel) and (1$\rm{M_{\odot}}$, 0.567$\rm{M_{\odot}}$), (1.5$\rm{M_{\odot}}$, 0.600$\rm{M_{\odot}}$) 
and (2$\rm{M_{\odot}}$, 0.534$\rm{M_{\odot}}$) for the helium--burning model (right panel).  
Lower panels display the evolutionary tracks from Bl\"ocker (1995) produced by ($\rm{M_{initial}}$, $\rm{M_{final}}$) = 
(3$\rm{M_{\odot}}$, 0.605$\rm{M_{\odot}}$), (3$\rm{M_{\odot}}$, 0.625$\rm{M_{\odot}}$) and (4$\rm{M_{\odot}}$, 0.696$\rm{M_{\odot}}$) 
for the hydrogen--burning model (left panel) and (1$\rm{M_{\odot}}$, 0.524$\rm{M_{\odot}}$), (3$\rm{M_{\odot}}$, 0.625$\rm{M_{\odot}}$) 
and (5$\rm{M_{\odot}}$, 0.836$\rm{M_{\odot}}$) for the helium--burning model (right panel).
The position of the central star of Vy~1--2 for a distance of 9.7~kpc is marked with a red star symbol.}
\label{fig15}
\end{figure*}

The plasma parameters of Vy~1--2 such as T$_{\rm{e}}$, N$_{\rm{e}}$ and chemical abundances were derived by means of the empirical 
methods and photo--ionisation modelling. Our best--fitting model provides a good approximation of the observed spectrum, except for the 
\oxygeniii\ 4959, 5007 \AA\ doublet, and adds some constraints to the nebular and stellar properties of Vy~1--2. The predicted \hbeta\ flux is in a good agreement 
with the observed flux (see Table 3) with a filling factor equal to 0.44. The mean difference between 
the observed and modelled lines varies from 10\% for the bright lines (S/N$>5$; the discrepancy in the \oxygeniii\ 4959, 5007~\AA\ doublet is about 14\%) 
to $\sim$20--30\% for the fainter lines, S/N$<$5. Moreover the intensities of \oxygeni\ and \nitrogena\ neutral lines are underestimated.
There are two possible scenarios for this: i) there is a large amount of neutral material around the nebula and/or ii) shock excitation of neutral 
lines close to the ionisation front may be significant. We can actually rule out the first scenario since the model predicts a density--bounded nebula 
whereas high expansion velocities up to 90--100~\kms\ could favour the second scenario.

The total chemical abundances of Vy~1--2, for a blackbody stellar flux distribution, are listed in Table 6, which also includes those 
derived from previous studies, as well as the solar abundances, for direct comparison. The low He and N/O abundances imply
a non Type--I nebula based on both Peimbert \& Torres--Peimbert's (1983) and Kingsburgh \& Barlow 
(1994) classifications. The O abundance derived from the empirical ICF is found to be almost solar and close to the 
average value of non--Type I PNe (Bohigas 2008). Our best--fitting model predicts a somewhat higher value by 0.25 dex meaning that the 
star has not undergone the ON--cycle burning during the second dredge--up phase. 

Given that Vy~1--2 is an O--rich nebula -- log(C/O)=-0.62, WLB05; or log(C/O)=-0.83; our model -- and the C abundance 
($\epsilon$(C)=8.08) is lower than half of the solar (Asplund et al. 2009) as well as the average non--type I PNe values, 
the N enrichment must have occurred by the conversion of C to N, via the hot--bottom burning (HBB; CN--cycle). It is known from theoretical 
models of stellar nucleosynthesis that only the progenitor stars with masses $\geq$3~$\rm{M_{\odot}}$ undergo the $\rm{2^{nd}}$ dredge--up and 
HBB events (Karakas et al. 2009). This result is inconsistent with the low N/O abundance ratio (-0.53 (ICF), -0.48 (model)) and a low--mass 
progenitor star ($\leq$1~$\rm{M_{\odot}}$).

Unlike the single star evolutionary models, a close binary system can give such low N/O and C/O abundance ratios via 
the mass--transfer exchange (Iben \& Tutukov 1993, De Marco 2009). Hf 2--2, NGC 2346 and Hb 12\footnote{Despite that Hsia 
et al. (2006) presented the detection of periodic variabilities, the binarity of the central star of Hb 12 still remains 
under debate (De Marco et al. 2008)} are PNe that have confirmed close binary systems and low N/O and C/O values 
(see De Marco 2009). In a common envelope evolution, the AGB--phase may be significantly shorter than in a single star 
case resulting in the formation of an O--rich envelope (Izzard et al. 2006).

A comparison of the chemical abundance of Vy~1--2 with those derived from PNe with different types of nucleus (normal, WEL and [WR]) 
shows that the He abundance and N/O ratio better match with those from WEL--type nuclei while the O abundance with those from 
WEL-- and [WR] types (G\'orny et al. 2009). As for the other elements, Ne matches those from WEL--type nuclei, 
Ar those from WEL-- and normal--type nuclei, while Cl abundance is found to be significantly lower than those values derived by Gorny et al. 
(2009) for all types of nuclei, but closer to the solar value. In particular, the average abundance of Cl varies from 6.12 to 6.48, while 
we find $\epsilon$(Cl)=5.22 (ICF method) and 5.31 (photo--ionisation model). S abundance seems to match all types of nuclei 
due to the large difference between the ICF method ($\epsilon$(S)=6.80) and the photo--ionisation model ($\epsilon$(S)=7.37). This may be a consequence of the poor fit of the 
\sulfur\ 6312~\AA\ emission line. The abundance uncertainties are $\sim$10\% for the He and $\sim$20--30\% for the rest of the elements.

It is worth mentioning here that Vy~1--2 has also been considered as a candidate halo PN due to its large distance from 
the Galactic plane ($\sim$500) and its high systemic velocity (-85$\pm$4~\kms) (Quireza et al. 2007). However, a comparison of 
its chemical abundances with those from Galactic disk and halo PNe (Otsuka et al. 2009) does not support this classification. 

The best-fitting model of Vy~1--2, presented here, fails to reproduce the \oxygeniii\ 5007~\AA\ and 4959~\AA\ emission lines. The difference 
between the observed and modelled values is about 14\%, similar to another PN with a WR--type central star as well as large abundance discrepancy 
factors (ADFs), NGC 1501 (Ercolano et al. 2004). We have to point out, though, that the emission line ratio \oxygeniii\ 5007/4363, which is an 
indicator of $\rm{T_e}$ of the nebula, is predicted with an accuracy better than 5\%. The main causes of the model's inability to reproduce the observed 
ionization balance and predict the bright \oxygeniii\ lines may be: i) the structure of the nebula, which is far from the spherically symmetric assumption, 
ii) the stellar continuum, where neither the blackbody nor an H--deficient stellar atmosphere model were able to reproduce all the observed data and/or 
iii) the dust composition, which seems to play a crucial role in this objects, but no information is available for Vy~1-2. 

The large ADFs found in Vy~1--2 for O$^{++}$, N$^{++}$, C$^{3+}$, C$^{++}$ and Ne$^{++}$ are of 6.17, 11.8, 14.0, 9.27 and 13.8, respectively 
(see  WLB05) and could indicate the possible presence of cold H--deficient clumps in the nebula, which may also hinder the results of 
our model This possibility is, however, beyond the scope of this paper.

\subsection{Central star}

\subsubsection{Evolutionary status}

Our simple blackbody stellar model predicts a luminosity $\rm{\log(L/L_\odot)}$=3.40 and an effective temperature 
$\rm{T_{eff}}$=98~kK, for D=9.7~kpc. For a distance of 7.8~kpc, we obtain $\rm{\log(L/L_\odot)}$=3.21 in good agreement 
with those values reported by Stangellini et al. (2002) and Phillips (2003). Figure 15 portrays the positions of Vy~1--2's 
central star in the HR diagram, for a distance of 9.7~kpc. The theoretical evolutionary tracks plotted are the hydrogen--burning 
(left panel) and helium--burning (right panel) models from Vassiliadis \& Wood (1994; upper panels) and (Bl\"ocker 1995; 
lower panels). In Vassiliadis and Wood's models, the central star corresponds to a progenitor mass of 1~$\rm{M_{\odot}}$ and a core mass of 
0.569~$\rm{M_{\odot}}$ (hydrogen--burning ) and 0.567~$\rm{M_{\odot}}$ (helium--burning). The timecales from these evolutionary tracks are 
significantly higher than the kinematic age of the nebula with 26000~years and 41000~years for the hydrogen-- and helium--burning 
respectively.

In Bl\"ocker's models, the evolutionary track for a very late thermal pulse (VLTP) (lower, right panel) 
implies a progenitor star of 3~$\rm{M_{\odot}}$ and a core mass of 0.625~$\rm{M_{\odot}}$. 
Even though the timescale of the former evolutionary track is still higher than the kinematic age of the nebula, by a factor of $\sim$2, 
it is closer to the observed value. The scenario of a close companion and mass--transfer exchange would be able to 
explain the rapid evolution of the star. Unlike the  1~$\rm{M_{\odot}}$ low mass star, a 3 $\rm{M_{\odot}}$ 
progenitor star can go through the HBB effect during the 2$^{\rm{nd}}$ dredge--up phase resulting in enhancement 
of N and depletion of C. The detection of the \carboniib\ 6461~\AA, the doublet \carboniv $\lambda$ $\lambda$ 5801, 
5812 and the \oxygeniiib\ 5592~\AA\ stellar lines emitted from a hydrogen--deficient star supports the VLTP scenario.

Our new distance value indicates that Vy~1--2 is a distant nebula (9.7~kpc) and provides a much better agreement between the 
kinematic age of the nebula and the evolutionary age of the central star. 
 
\subsubsection{[WR] or WEL type nucleus?}

Apart from the detection of the common nebular emission lines, weak stellar emission lines 
such as \carboniib\ 6461~\AA, \carboniv\ 5801 and 5812~\AA\ and  \oxygeniiib\ 5592~\AA\ were also detected
for the first time in this nebula in our very deep spectra (Table 7). Their line profiles 
are shown in Fig. 16. There are two groups of stars with emission lines of highly ionised He, C, O 
and N: [WR] (Crowther 2008) and WELS stars (Tylenda et al. 1993). 
Although, both classes exhibit the same emission lines (e.g \carboniv\ doublet 5801 and 5812~\AA\ and 
\carboniii\ 5696~\AA), the former show much broader lines profiles and stronger intensities than 
the latter due to the fast and strong stellar winds.

Based on the classification scheme of Acker \& Neiner (2003) and the \carboniib/\carboniv\ and 
\oxygeniiib/\carboniv\ line ratios, the central star of Vy~1--2 is classified as a late--type star 
([WRL]). On the other hand, the absence of the \carboniii\ 5696~\AA\ line (see WLB05) in conjunction with 
the detection of the \carboniv\ doublet, {N~{\sc iii}} and {C~{\sc iii--iv}} 4650~\AA\ lines, also suggest 
a WEL--type nucleus. Therefore, new high--resolution observations are required in order to confirm the 
spectral type of this star.

\begin{figure}
\centering
\includegraphics[scale=0.20]{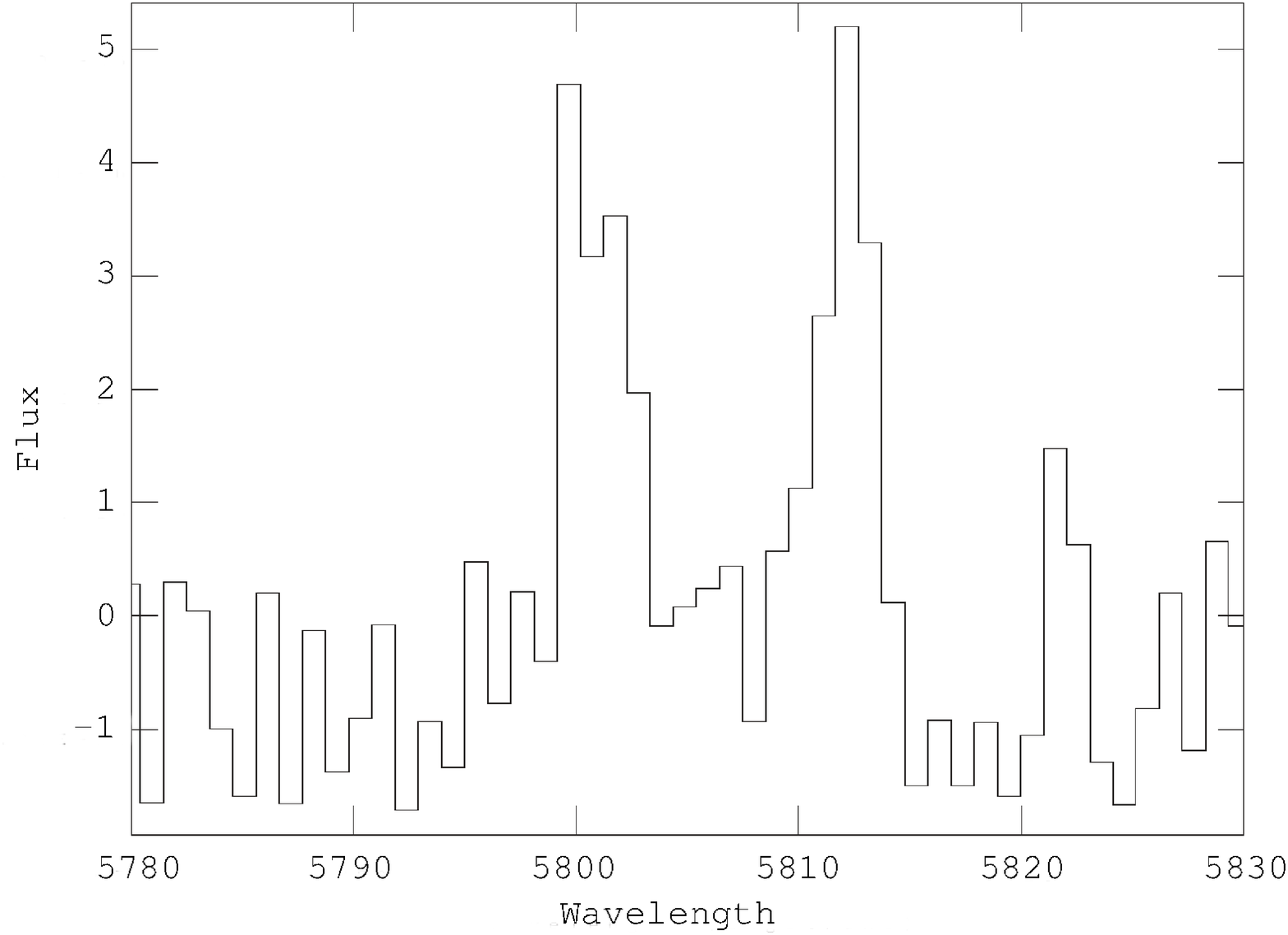}
\includegraphics[scale=0.20]{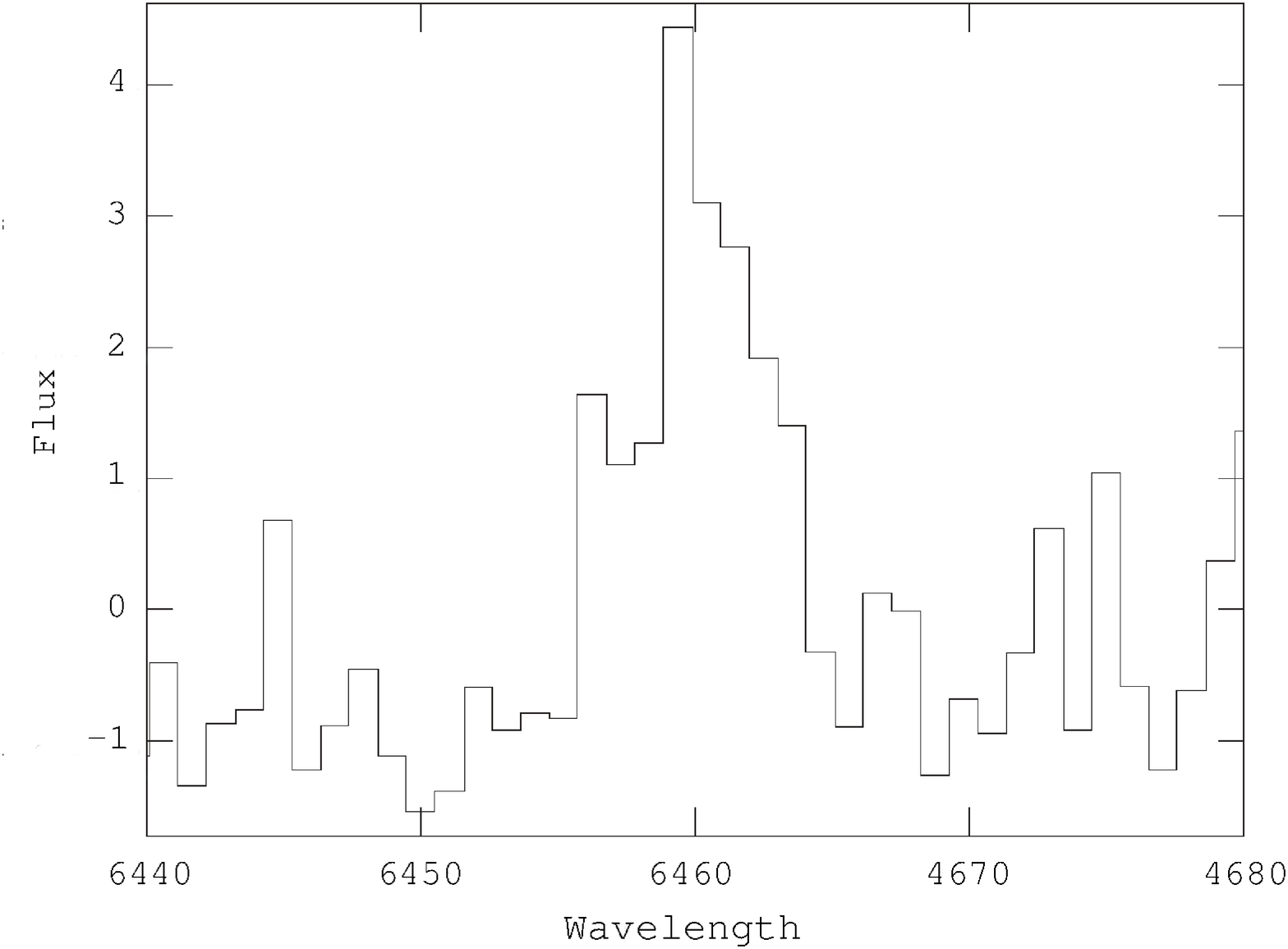}
\includegraphics[scale=0.20]{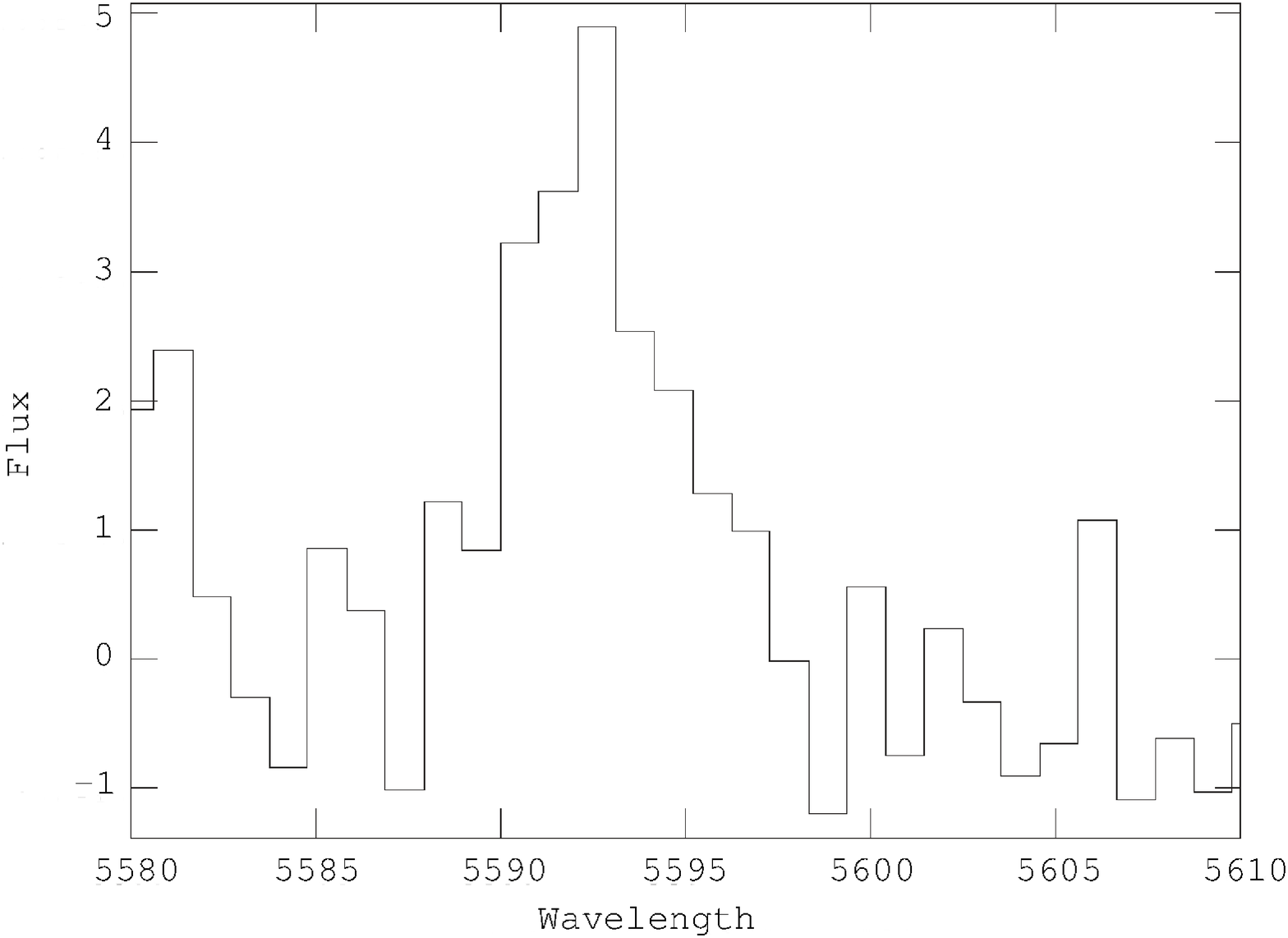}
\caption[]{Part of the spectrum of Vy~1--2, in which the faint stellar lines \carboniv\ $\lambda$ ~$\lambda$ 5801, 5812~\AA, 
\carboniib\ 6461~\AA\ and \oxygeniiib\ 5592~\AA\ are seen (in units of $10^{-14}$ erg $\rm{s^{-1}}$ $\rm{cm^{-2}}$).}
\label{fig16}
\end{figure}

\subsection{Infrared properties and SED}
Vy 1-2 is also a strong infrared emitter indicating the presence of significant amount of dust. 
Thus far, no one had studied the infrared emission and dust properties of this compact nebula. 
For this study, we used infrared data from 2MASS, IRAS (IRAS 17524+2800), AKARI and WISE data archives. 
The data cover a wavelength range from the near-- to the far--infrared.

\begin{table}
\centering
\caption[]{Stellar emission lines of Vy~1--2}
\label{table4a}
\begin{tabular}{llllllllllll}
\hline
\multicolumn{2}{c|}{} \\
{\em \rm{Line}} (\AA) & F($\lambda$)$^a$ & S/N \\
\hline
\oxygeniiib\ 5592 &    0.08 &    3.0 \\
\carboniv  \ 5801 &    0.12 &    3.0 \\
\carboniv  \ 5812 &    0.11 &    3.0 \\
\carboniib  \ 6461 &    0.09 &    3.0 \\
\hline
\end{tabular}
\medskip{}
\begin{flushleft}
${^a}$ Relative to the nebular \hbeta=100
\end{flushleft}
\end{table}

The spectral energy distribution (SED) of Vy 1-2 is presented in Fig. 17 and displays a double--peaked profile. 
One peak lies at the optical wavelength due to the hot central star (100,000~K) of the nebula and the second one 
at ~30--40~$\mu$m implying the presence of a cold dusty circumstellar shell with a temperature of $\sim$100--150~K. 
Vy~1--2 may also exhibit a small K--band excess, associated either with a hot dust component or the emission from a cool 
($\sim$2000~K) unresolved companion.

The all--sky Wide--Field Infrared Survey Explorer (WISE) observed Vy 1-2 in four photometric bands at 3.35~$\mu$m, 4.6~$\mu$m, 
11.6~$\mu$m and 22.1~$\mu$m (W1 through W4, respectively; Fig. 18). Because of the low angular resolution in these bands 
(6.1\arcsec, 6.4\arcsec, 6.5\arcsec and 12\arcsec, respectively; Wright et al. 2010), Vy 1-2 remains unresolved. 
It is worth mentioning here that W1 and W2--band images unveil the presence of a seemingly collimated 
outflow in the same direction as the NW knot (PA=310$^{\circ}$). An arrow in Figure 18 indicates the position of the collimated structure. 
Despite the W3--band having a similar angular resolution to those of W1 and W2 bands, the collimated structure can not be discerned.

By scrutinizing the DSS images for Vy 1-2, we discover the presence of a very faint starlike object in 
the same direction as the outflow and in the same position as the NW knot ($\sim$14\arcsec\ away from the central star). 
The detection of this field star in the DSS IR band, as well as in the W1 and W2-bands but not in the W3-band, implies a  
very cold star with a temperature of $\sim$2000~K. Note that the seemingly halo observed in the W3 and W4 band 
images (panels (c) and (d) in Fig. 18) is not a real structure since it is also apparent around the bright field stars.

WISE bands, encompass the dust continuum and several features attributed to PAH features at 3.3~$\mu$m, 8.6~$\mu$m, 
11.3~$\mu$m, 12.7~$\mu$m and  16.4~$\mu$m, silicate dust features at 9.7~$\mu$m and 18~$\mu$m as well as high excitation nebular lines 
(e.g. \sulfurvi, \neoniv\ and \neonv). However, ISO spectra are not available, prohibiting any further investigation on the dust composition.
Nevertheless, the detection of single ionised recombination C lines (\carboniib\ 6578~\AA\ and 6581~\AA) may suggest the presence of 
C--rich dust. The large excess in the W3 and W4 bands may also be associated with the presence of silicate dust 
features. The SwSt 1 nebula exhibits these C lines as well as the silicate dust feature at 10 $\mu$m (De Marco et al. 2001). 
A double--dust chemistry (O-- and C--rich) is observed in PNe with [WR] or WEL type central stars, probably due to the VLTP 
that turn the central stars from O--rich to C--rich.

In order to constrain the dust emission, we used the aforementioned broad--band infrared fluxes. 
The best--fitting SED to the infrared data was possible only assuming a C--rich dust model. Figure 17 displays the 
modelled SED of Vy 1--2. The PAHs are $\sim$5.7 times more abundant (by mass) than the silicates and the dust-to-gas ratio equal to 
1.05$\times$10$^{-2}$. The total mass of the nebula and the dust mass are ~0.08~M$\odot$ and 8.4 10$^{-4}$~M$\odot$, respectively. 
The high uncertainty of Vy 1--2's distance implies as well high uncertainties in these masses.
This result is consistent with the scenario of a hydrogen deficient star and dual--dust chemistry.
Interestingly, the ionized gas indicates an O-rich nebula (C/O$<$1), whilst the dust composition is C-rich. 
One possible explanation, as recently proposed by Guzm\'an--Ramirez et al. (2014), is the formation of PAHs from the photo--dissociation of 
CO in a torus, such as the bright ring--like structure in Vy 1-2, whereas the dust in the main nebula remains O--rich. 
Moreover, a recent change of the stellar surface chemistry due to the second dredge--up of C, as has been 
proposed to explain the dual--dust chemistry in the young PN BD +3036 (Guzm\'an--Ramirez et al. 2015), can also provide a plausible 
explanation for Vy~1--2 as well. We have to stress out though there is no available infrared spectroscopic data for Vy~1--2 and 
our analysis is based on the results from our best-fitting photo--ionisation model using the broad--band infrared fluxes.

In summary, a careful look at the modelled SED reveals a lower flux in the infrared H--band, indicating that this measurement 
may be underestimated and the seemingly K-band excess is not really an excess associated with the presence of a cold companion 
as previously mentioned.

\begin{figure}
\centering
\includegraphics[scale=0.155]{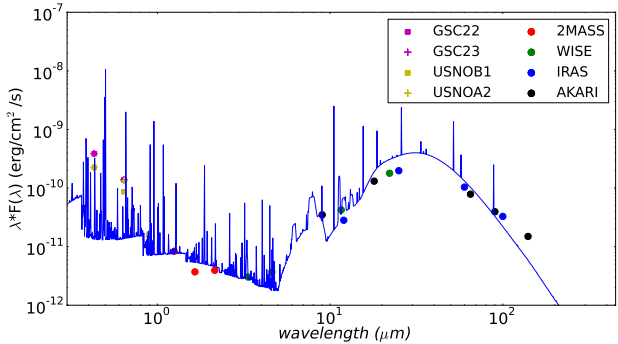}
\caption[]{The observed and modelled SED of Vy~1--2 in the wavelength range from 0.43 to 140~$\mu$m. The data are from 
GSC2.2, GSC2.3, USNOB1, USNOB2, 2MASS, AKARI and WISE surveys.}
\label{fig17}
\end{figure}

\begin{figure}
\centering
\includegraphics[scale=0.325]{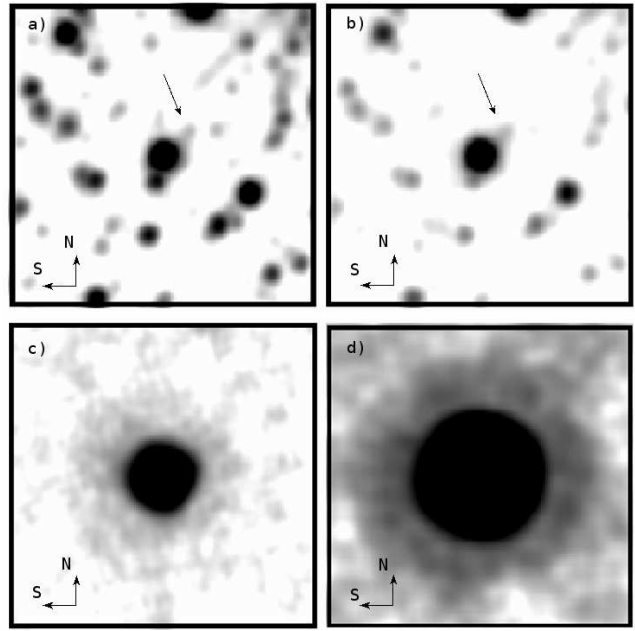}
\caption[]{WISE images of Vy~1--2 in four bands: a) 3.35~$\mu$m, b) 4.6~$\mu$m, c) 11.6~$\mu$m and d) 22.1~$\mu$m. 
The arrows indicate the position of the field star. The field of view is 180\arcsec$\times$180\arcsec. 
North is up and East is to the left.}
\label{fig18}
\end{figure}

\section{Conclusions} 
The main conclusions of this work are summarized as follow: 

1) Vy~1--2 presents a bipolar structure, seen almost pole--on. The bipolar lobes expand with velocities 
up to 90--100~\kms, whereas the bright, inner ring--like structure shows lower expansion velocities of V$_{\rm{H\alpha}}$ =19~\kms, 
V$_{\rm{He {\sc\ II}}}$ =23--25~\kms, $\rm{V_{[N {\sc\ II}]}}$=8--9~\kms\ and $\rm{V_{[O {\sc\ III}]}}$=18~\kms. 
These velocities indicates a possible acceleration process of the inner nebular regions, which may be associated with
the strong/fast stellar wind from a [WR] or WEL nucleus. The kinematic age of the nebula is $\sim$3500~years for a distance of 9.7~kpc.

2) A pair of knots is also apparent along P.A.=305$^{\circ}$, located at 11.5\arcsec\ and 14.5\arcsec\ and moving perpendicular to the 
line of sight. Assuming that they were formed approximately at the same time as the nebula, we estimated their velocities up to 
a value of 130$\pm$15~\kms.

3) Vy~1--2 is apparently an O--rich, non--type I nebula (log(C/O)=-0.58 (ICF) and -0.83 (model), -0.53$\leq$log(N/O)$\leq$-0.49). 
Compared to solar abundances, O was found to be solar, C subsolar and it is enriched in N. The HBB CN--cycle must 
have occurred in order to explain these abundances, implying a progenitor star with initial mass $\geq$3~M$_{\odot}$.

4) None of the stellar ionizing continua used in our photo--ionisation model (blackbody or H--deficient) 
were able to reproduce the observed ionisation balance of the elements. The spherical symmetry asssumed for the nebula 
may be far from the real structure of Vy~1--2, and a more sophisticated 3--D bi--abundance model is required for the 
proper study of its ionisation structure.

5) The low C/O and N/O abundance ratios of Vy~1--2 can not be attained by single star evolutionary models, and a post--common envelope 
binary system seems a more plausible scenario. The N enhancement can be explained by the binary interaction via mass--transfer exchange. 

6) A very late thermal pulse may also have occurred resulting in the formation of a H--deficient star. The detection of 
the \carboniib\ 6461~\AA, the \carboniv $\lambda$ $\lambda$ 5801, 5812 doublet and the \oxygeniiib\ 5592~\AA\ stellar lines 
supports this scenario. The timescale of VLTP evolutionary track for a 3 M$_{\odot}$ progenitor star is consistent with the kinematic 
age of the nebula.

7) Vy~1--2 is found to be very bright at the mid--IR indicating large amount of dust. Its SED shows a double--peaked profile with one peak 
at 30--40~$\mu$m which implies a circumstellar envelope with a temperature of $\sim$100~K. The C/O$<$1 abundance ratio of the ionised gas 
indicates a O--rich nebula, whilst our best--fitting model predicts a C--rich dust, with PAHs being almost 5 times more abundant (by mass) 
than silicates. A possible interpretation of this result may be the formation of PAHs in a torus due to the photo--dissociation of 
CO and the formation of silicates in the main nebula or a recent change of stellar chemical composition due to the second dredge--up 
of C.


\section*{Acknowledgments} 

SA is supported by the Brasilian CAPES post-doctoral fellowship 'Young Talents Attraction' - Science Without Borders, 
A035/2013. Based on observations made with the NASA/ESA Hubble Space Telescope, obtained from the
data archive at the Space Telescope Institute. We would like to thank the staff at the SPM and Skinakas Observatory for their 
excellent support during these observations. Moreover, we thank the anonymous referee for his/her
valuable comments that greatly improved the presentations of this paper.

\bibliographystyle{mnras}

\begin{thebibliography}{99}

\bibitem[\protect\citeauthoryear{Acker}{2003}]{Ac03}
Acker A., Neiner C., 2003, A\&A, 403, 659.

\bibitem[\protect\citeauthoryear{Akras}{2012}]{Ak2012}
Akras S., Steffen W., 2012, MNRAS, 423, 925.

\bibitem[\protect\citeauthoryear{Asplund}{2009}]{Asp09}
Asplund M., Grevesse N., Sauval A. J., Scott P., 2009, ARA\&A, 47, 481.

\bibitem[\protect\citeauthoryear{Badnell}{2003}]{Bad03}
Badnell N. R., O’Mullane M. G., Summers H. P., Altun Z., Bautista M. A., Colgan J., Gorczyca T. W., Mitnik D. M.,
Pindzola M. S., Zatsarinny O., 2003, A\&A, 406, 1151.

\bibitem[\protect\citeauthoryear{Badnell}{2006}]{Bad06}
Badnell N. R., 2006, ApJS, 167, 334.

\bibitem[\protect\citeauthoryear{Barker}{1978}]{Bar78a}
Barker T., 1978a, AJ, 219, 914.

\bibitem[\protect\citeauthoryear{Barker}{1978}]{Bar78b}
Barker T., 1978b, AJ, 220, 193.

\bibitem[\protect\citeauthoryear{Benjamin}{1999}]{Ben99}
Benjamin R. A., Skillman E. D., Smits D. P., 1999, ApJ, 514, 307.

\bibitem[\protect\citeauthoryear{Blocker}{1995}]{Blok95}
Bl\"ocker T., 1995, A\&A, 299, 755.

\bibitem[\protect\citeauthoryear{Bohigas}{2001}]{Boh01}
Bohigas J., 2001, RMxAA, 37, 237.

\bibitem[\protect\citeauthoryear{Bohigas}{2008}]{Boh08}
Bohigas J., 2008, ApJ, 674, 954.

\bibitem[\protect\citeauthoryear{Bohigas}{2015}]{Boh15}
Bohigas J., Escalante V., Rodr\'iguez M., Dufour R. J., 2015, MNRAS, 447, 817B.

\bibitem[\protect\citeauthoryear{Boumis}{2006}]{Boum06}
Boumis P., Akras S., Xilouris E. M., et al., 2006, MNRAS, 367, 1551B.

\bibitem[\protect\citeauthoryear{Boumis}{2006}]{Boum03}
Boumis P., Paleologou E. V., Mavromatakis F., Papamastorakis J., 2003, MNRAS, 339, 735B.

\bibitem[\protect\citeauthoryear{Bryce}{1999}]{Bryc99}
Bryce M., Mellema G., 1999, MNRAS, 309, 731.

\bibitem[\protect\citeauthoryear{Butler}{1989}]{But89}
Butler K., Zeippen C. J., 1989, A\&A, 208, 337.

\bibitem[\protect\citeauthoryear{Cahn}{1971}]{Cah71}
Cahn J. H., Kaler J. B., 1971, ApJS, 22, 319C.

\bibitem[\protect\citeauthoryear{Cahn}{1992}]{Cah92}
Cahn J. H., Kaler J. B., Stanghellini L., 1992, A\&AS, 94, 399.

\bibitem[\protect\citeauthoryear{Calavis}{1995}]{Cal95}
Galavis M. E., Mendoza C., Zeippen C. J., 1995, A\&A Supp., 111, 347.

\bibitem[\protect\citeauthoryear{Condon}{1998}]{Con98}
Condon J. J., Kaplan D. L, 1998, ApJS, 117, 361.

\bibitem[\protect\citeauthoryear{Crowther}{2008}]{Crow08}
Crowther P. A., 2008, ASPC, 391, 83.

\bibitem[\protect\citeauthoryear{Dere}{1997}]{Der97}
Dere K. P., Landi E., Mason H. E., Monsignori Fossi B. C., Young P. R., 1997, A\&AS, 125, 149.

\bibitem[\protect\citeauthoryear{DeMarco}{2001}]{Mar09}
De Marco O., Crowther P. A., Barlow M. J., et al., 2001, MNRAS, 328, 527D.

\bibitem[\protect\citeauthoryear{DeMarco}{2008}]{Mar08}
De Marco O., Hillwig Todd C., Smith A. J., 2008, AJ, 136, 323D.

\bibitem[\protect\citeauthoryear{DeMarco}{2009}]{Mar09}
De Marco O, 2009, PASP, 121, 316D.

\bibitem[\protect\citeauthoryear{Depew}{2011}]{Dep11}
Depew K., Parker Q. A., Miszalski B., De Marco O., Frew D. J., 
Acker A., Kovacevic A. V., Sharp R. G., 2011, MNRAS, 414, 2812.

\bibitem[\protect\citeauthoryear{Desert}{1990}]{Des90}
Desert F.-X., Boulanger F., Puget J. L. 1990, A\&A, 237, 215.

\bibitem[\protect\citeauthoryear{Ercolano}{2004}]{Erc04}
Ercolano B., Wesson R., Zhang Y., Barlow M. J., De Marco O.,
Rauch T. and X.-W. Liu X. -W., 2004, MNRAS, 354, 558.

\bibitem[\protect\citeauthoryear{Ferland}{2013}]{Fer13}
Ferland G. J., Porter R. L., van Hoof P. A. M., Williams R. J. R., 
Abel N. P., Lykins, M. L., Shaw, G., Henney, W. J., Stancil, P. C., 
2013, RMxAA, 49, 137

\bibitem[\protect\citeauthoryear{Flores-Dur\'an}{2014}]{Flor14}
Flores--Dur\'an S. N., Pe\~na M., Hern\'andez--Mart\'inez L., Garc\'ia--Rojas J., 
Ruiz M. T., 2014, A\&A, 568A, 82F.

\bibitem[\protect\citeauthoryear{Fitzpatrick}{1999}]{Fitz99}
Fitzpatrick E. L., 1999, PASP, 111, 63.

\bibitem[\protect\citeauthoryear{Frew}{2013}]{Frew13}	
Frew D. J., Boji\v ci\'c I. S., Parker Q. A., 2013, MNRAS, 431, 2F.

\bibitem[\protect\citeauthoryear{Frew}{2010}]{Frew10}
Frew D. J., Parker Q. A., 2010, APN5 Conf. Proc., Ebrary, 33, 
arXiv:1010.5003.

\bibitem[\protect\citeauthoryear{Gesicki}{2003}]{Ges03}
Gesicki K., Zijlstra A. A., 2003, MNRAS, 338, 347.

\bibitem[\protect\citeauthoryear{Girard}{2007}]{Gir07}
Girard P, Koppen J., Acker A,. 2007, 463, 265.

\bibitem[\protect\citeauthoryear{Gorny}{2009}]{Gor09}
G\'orny S. K., Chiappini C., Stasi\'nska G., Cuisinier F., 2009, A\&A, 500, 1089.


\bibitem[\protect\citeauthoryear{Gurzadian}{1988}]{Gurz88}
Gurzadian G. A., 1988, Ap\&SS, 149, 343.

\bibitem[\protect\citeauthoryear{Guzman}{2014}]{Guz14}
Guzman--Ramirez L., Lagadec E., Jones D., Zijlstra A. A., Gesicki K., 2014, MNRAS, 441, 364G.

\bibitem[\protect\citeauthoryear{Guzman}{2015}]{Guz15}
Guzman--Ramirez L., Lagadec E., Wesson R., et al. 2015, MNRAS, 451L, 1G.

\bibitem[\protect\citeauthoryear{Hamuy}{1992}]{Ham92}
Hamuy M., Walker A. R., Suntzeff N. B., Gigoux P., Heathcote S. R., 
Phillips M. M., 1992, PASP, 104, 533.

\bibitem[\protect\citeauthoryear{Harman}{2004}]{Har04}
Harman D. J., Bryce M.,  L\'opez J. A., Meaburn J., Holloway A. J., MNRAS, 2004, 348, 1047

\bibitem[\protect\citeauthoryear{Hyung}{1997}]{Hyu97}
Hyung S., Aller H. L., 1997, MNRAS, 292, 71.

\bibitem[\protect\citeauthoryear{Hyung}{2000}]{Hyu00}
Hyung, S., Aller, L. H., Feibelman, W. A., Lee, W. B., de Koter, A. 2000, MNRAS, 318, 77.

\bibitem[\protect\citeauthoryear{Jones}{2014}]{Jones14}
Jones D., Boffin H. M. J., Miszalski B., Wesson R., Corradi R. L. M., Tyndall A. A.,
2014, A\&A, 562A, 89J.

\bibitem[\protect\citeauthoryear{Iben}{1993}]{Iben93}
Iben I., Tutukov V. A., 1993, ApJ, 418, 343.

\bibitem[\protect\citeauthoryear{Izzard}{2006}]{Izz06}
Izzard R. G., Dray L. M., Karakas A. I., Lugaro M., Tout 
C. A., 2006, A\&A, 460, 565.

\bibitem[\protect\citeauthoryear{Karakas}{2009}]{Kar09}
Karakas A. I., van Raai M. A., Lugaro M., Sterling N. C., Dinerstein
H. L., 2009, ApJ, 690, 1130.

\bibitem[\protect\citeauthoryear{King}{1994}]{King94}
Kingsburgh R. L., Barlow M. J., 1994, MNRAS, 271, 257.

\bibitem[\protect\citeauthoryear{Krist}{2011}]{Kris11}
Krist J. E., Hook R. N., Stoehr F., 2011, SPIE, 8127E, 0JK.

\bibitem[\protect\citeauthoryear{Kwok}{2010}]{Kwok10}
Kwok S., 2010, PASA, 27, 174.

\bibitem[\protect\citeauthoryear{Lagadec}{2011}]{Lag11}
Lagadec E., et al. 2011, MNRAS, 417, 32.

\bibitem[\protect\citeauthoryear{Laor}{1993}]{Lao93}
Laor A., Draine B. T., 1993, ApJ, 402, 441.

\bibitem[\protect\citeauthoryear{Landi}{2012}]{Lan12}
Landi E., Del Zanna G., Young P. R., Dere K. P., Mason H. E., 2012, ApJ, 744, 99.

\bibitem[\protect\citeauthoryear{Lee}{2001}]{Lee01}
Lee H.--W., Kang Y.--W., Byun Y.--I., 2001, ApJ, 551L, 121.

\bibitem[\protect\citeauthoryear{Lennon}{1994}]{Len94}
Lennon D. J., Burke V. M., 1994, A\&AS, 103, 273.

\bibitem[\protect\citeauthoryear{Li}{2001}]{Li01}
Li A., Draine B. T., 2001, ApJ, 554, 778.

\bibitem[\protect\citeauthoryear{Liu}{2000}]{Liu00}
Liu X.-W., Storey P. J., Barlow M. J., Danziger I. J., Cohen M. and Bryce M., 
2000, MNRAS, 312, 585.

\bibitem[\protect\citeauthoryear{Liu}{2001}]{Liu01}
Liu X.-W., Luo S.-G., Barlow M. J., Danziger I. J. and Storey P. J., 2001, 
MNRAS, 327, 141.

\bibitem[\protect\citeauthoryear{Lopez}{2012}]{Lop12}
L\'opez J. A., Richer M. G., Garc\'ia--D\'iaz Ma. T., Clark D. M., 
Meaburn J., Riesgo H., Steffen W., \& Lloyd, M., 2012, RevMexAA, 48, 3.

\bibitem[\protect\citeauthoryear{Luridiana}{2012}]{Lur12}
Luridiana V., Morisset C., Shaw R. A., 2012, IAUS, 283, 422.

\bibitem[\protect\citeauthoryear{Lykins}{2013}]{Lyk}
Lykins M. L., Ferland G. J., Porter R. L. van Hoof P. A. M., Williams R. J. R., 
Gnat O., 2013, MNRAS, 429, 3133L.

\bibitem[\protect\citeauthoryear{Maciel}{1984}]{MAc84}
Maciel, W. J., 1984, A\&AS, 55, 253.

\bibitem[\protect\citeauthoryear{Manchado}{1996}]{Man96}
Manchado A., Guerrero M. A., Stanghellini L., \& Serra--Ricart M. 1996, The
IAC morphological catalogue of northern Galactic planetary nebulae.

\bibitem[\protect\citeauthoryear{Manchado}{2011}]{Man11}
Manchado A., Garc\'ia--Hern\'andez D. A., Villaver E., Guironnet de Massas J., 
2011, ASPC, 445, 161.

\bibitem[\protect\citeauthoryear{Martin}{1991}]{Mar91}
Martin P. G., Rouleau F. 1991, in Extreme Ultraviolet Astronomy, ed. R. F. Malina 
\& S. Bowyer (New York: Pergamon), 341.

\bibitem[\protect\citeauthoryear{McKenna}{1996}]{McK96}
McKenna F. C., Keenan F. P., Kaler J. B., Wickstead A. W., Bell 
K. L., Aggarwal K. M., 1996, PASP, 108, 610.

\bibitem[\protect\citeauthoryear{Meaburn}{2003}]{Meab03}
Meaburn J., L\'opez J. A., Guti\'errez L., Quir\'oz F.; Murillo J. M., 
Vald\'ez, J., Pedrayez M., 2003, RMxAA, 39, 185.	

\bibitem[\protect\citeauthoryear{Meaburn}{2005}]{Meab05}
Meaburn J., Boumis P., L\'opez J. A., Harman D. J., Bryce M., Redman M. P., Mavromatakis F., 
2005, MNRAS, 360, 963M.

\bibitem[\protect\citeauthoryear{Medina}{2006}]{Med06}
Medina S., Pe\~na M., Morisset C., Stasi\'nska G.,2006, RMxAA, 42, 53.

\bibitem[\protect\citeauthoryear{Meixner}{1999}]{Meix99}
Meixner M., et al., 1999, ApJS, 122, 221.

\bibitem[\protect\citeauthoryear{Mendoza}{1982}]{Men82}
Mendoza C., Zeippen C. J., 1982, MNRAS, 198, 127a.

\bibitem[\protect\citeauthoryear{Mendoza}{1982}]{Men82}
Mendoza C., Zeippen C. J.,1982, MNRAS, 199, 1025b.

\bibitem[\protect\citeauthoryear{Mendoza}{1983}]{Men83}
Mendoza C., 1983, "Planetary Nebulae", 143.

\bibitem[\protect\citeauthoryear{Mennickent}{2008}]{Men08}
Mennickent R., Greiner J., Arenas J., Tovmassian G., Mason E., Tappert C., Papadaki C., MNRAS, 2008, 383, 845.

\bibitem[\protect\citeauthoryear{Miszalski}{2009}]{Mis09}
Miszalski B., Acker A., Parker Q. A., Moffat A. F. J., 2009, A\&A, 505, 249.

\bibitem[\protect\citeauthoryear{Miszalski}{2008}]{Mis08}
Miszalski B., Parker Q. A., Acker A, Birkby J. L., Frew D. J., Kovacevic A., 2008, MNRAS,
384, 525.

\bibitem[\protect\citeauthoryear{McNabb}{2013}]{McN13}
McNabb I. A., Fang X., Liu X.--W., Bastin R. J., Storey P. J., 2013, MNRAS, 428, 3443M.


\bibitem[\protect\citeauthoryear{Otsuka}{2009}]{Ots09}
Otsuka M., Hyung S., Lee S.--J., Izumiura H., Tajitsu A.,  2009, ApJ, 705, 509O.

\bibitem[\protect\citeauthoryear{Parker}{2006}]{par06}
Parker Q. A., et al., 2006, MNRAS, 373, 79.

\bibitem[\protect\citeauthoryear{Peimbert}{1983}]{Peim83}
Peimbert M., Torres--Peimbert S., 1983, in Flower D. R., ed., IAU Symp. 
103, 233.

 
\bibitem[\protect\citeauthoryear{Phillips}{1998}]{phi98}
Phillips J. P., A\&A, 1998, 340, 527.

\bibitem[\protect\citeauthoryear{Phillips}{2003}]{Phil03}
Phillips J. P., 2003, MNRAS, 344, 501.

\bibitem[\protect\citeauthoryear{Phillips}{2011}]{phi11}
Phillips J. P.,  M\'arquez--Lugo R. A., 2011, RMxAA, 47, 83.

\bibitem[\protect\citeauthoryear{Pradhan}{1976}]{Pra76}
Pradhan A. K., 1976, MNRAS, 177, 31.

\bibitem[\protect\citeauthoryear{Quireza}{2007}]{Qui07}
Quireza C., Rocha--Pinto H. J., Maciel W. J., 2007, A\&A, 475, 217Q.

\bibitem[\protect\citeauthoryear{Lario}{2012}]{Lar12}
Ramos--Larios G., Guerrero M. A., V\'azquez R., Phillips J. P., 2012, MNRAS, 420, 1977.

\bibitem[\protect\citeauthoryear{Ramsbottom}{1996}]{Ram96}
Ramsbottom C. A., Bell K. L., Stafford R. P., 1996, ADNDT, 63, 57.

\bibitem[\protect\citeauthoryear{Sabadin}{1984}]{Sabb84}
Sabbadin F., Gratton R. G., Bianchini A., Ortolani S., 1984, A\&A, 136, 181S.

\bibitem[\protect\citeauthoryear{Sabin}{2014}]{Sab14}
Sabin L., Parker Q. A., Corradi R. L. M., et al., 2014, MNRAS, 443, 3388S.

\bibitem[\protect\citeauthoryear{Sahai}{1999}]{Sah99}
Sahai R., Zijlstra A., Bujarrabal V., Hekkert Te Lintel P., AJ, 1999, 117, 1408. 

\bibitem[\protect\citeauthoryear{Sahai}{2011}]{Sah11}
Sahai R., Morris M. R., Villar G. G., 2011, AJ, 141, 134.

\bibitem[\protect\citeauthoryear{Schneider}{1983}]{Sch83}
Schneider S. E., Terzian Y., Purgathofer A., Perinotto M., 1983, ApJS, 52, 399.

\bibitem[\protect\citeauthoryear{Schutte}{1993}]{Schu93}
Schutte W. A., Tielens A. G. G. M., Allamandola L. J. 1993, ApJ, 415, 397.

\bibitem[\protect\citeauthoryear{Shaw}{1994}]{Shaw}
Shaw, R. A., Dufour, R. J., 1994, ASPC, 61, 327.

\bibitem[\protect\citeauthoryear{Seaton}{1979}]{Seat79}
Seaton M. J., 1979, MNRAS, 187, 73.

\bibitem[\protect\citeauthoryear{Stanghellini}{2006}]{Sta06}
Stanghellini L., Guerrero M A., Cunha K, Manchado A., Villaver E., 2006, ApJ, 651, 898.

\bibitem[\protect\citeauthoryear{Stanghellini}{2002}]{stan02}
Stanghellini L., Villaver E., Manchado A., Guerrero M. A., 2002, ApJ, 576, 285.

\bibitem[\protect\citeauthoryear{Steffen}{2011}]{Ste11}
Steffen W., Koning N., Wenger S., Morisset  C., Magnor M., 2011, IEEE
Transactions on Visualization and Computer Graphics, 17, 454.

\bibitem[\protect\citeauthoryear{Soker}{1994}]{Sok94}
Soker N., Livio M., 1994, ApJ, 421, 219.

\bibitem[\protect\citeauthoryear{Tylenda}{1993}]{Tyl93}
Tylenda R., Acker A., Stenholm B., 1993, A\&AS, 102, 595.

\bibitem[\protect\citeauthoryear{Vassiliadis}{1994}]{Vas94}
Vassiliadis E., Wood P. R., 1994, ApJS, 92, 125.

\bibitem[\protect\citeauthoryear{Vaytet}{2009}]{Vayt09}
Vaytet N. M. H., Rushton A. P., Lloyd M., L\'opez J. A., Meaburn J., O\' Brien T. J., 
Mitchell D. L., Pollacco D., 2009, MNRAS, 398, 385.

\bibitem[\protect\citeauthoryear{Verner}{1996}]{Ver96}
Verner D. A., Verner E. M., Ferland G. J., 1996, ADNDT, 64, 1.

\bibitem[\protect\citeauthoryear{Weinberger}{1989}]{Wei89}
Weinberger R., 1989, A\&AS, 78, 301.

\bibitem[\protect\citeauthoryear{Wesson}{2005}]{Wes05}
Wesson R., Liu X.-W., Barlow M. J., 2005, MNRAS, 362, 424.


\bibitem[\protect\citeauthoryear{Wiese}{1996}]{Wie96}
Wiese W. L., Fuhr J. R., Deters T. M., 1996, JPCRD, Monograph 7.

\bibitem[\protect\citeauthoryear{Wright}{2010}]{Wri10}
Wright E. L., Eisenhardt P. R. M., Mainzer A. K., 2010, AJ, 140, 1868W.


\bibitem[\protect\citeauthoryear{Zeippen}{1987}]{Zei87}
Zeippen C. J., Butler K., Le Bourlot J., 1987, A\&A, 188, 251.



\end{thebibliography}

\end{document}